\documentclass[10pt]{article}
\usepackage{amsmath}
\usepackage{latexsym}
\usepackage{amssymb}
\usepackage{amsfonts}
\usepackage{amsthm}
\title{ALGEBRAIC ANALYSIS \\  AND \\  MATHEMATICAL PHYSICS}
\author{J.-F. Pommaret \\ CERMICS, Ecole des Ponts ParisTech,\\ 6/8 Av. Blaise Pascal, 77455 Marne-la-Vall\'ee Cedex 02, France \\
E-mail: jean-francois.pommaret@wanadoo.fr\\
URL: http://cermics.enpc.fr/$\sim$pommaret/home.html }
\date{  }
\begin{document}
\maketitle

\noindent
{\bf ABSTRACT}   \\

This paper, which is largely self-contained with many explicit examples, is an extended written version of a series of lectures (10 h) given during a short intensive course organized by the Center of Theoretical Physics (CPT) in Marseille (may 23-25, 2016) under the title " Algebraic Analysis, Double Duality and Applications in Mathematical Physics ". It aims to revisit the mathematical foundations of both General Relativity and Electromagnetism after one century, in the light of the formal theory of systems of partial differential equations and Lie pseudogroups (D.C. Spencer, 1970) or Algebraic Analysis, namely a mixture of differential geometry and homological algebra (M. Kashiwara, 1970). Among the new results obtained, we may quote:  \\    
\noindent
1) In dimension $4$ {\it only}, the $9$ Bianchi identities that must be satisfied by the $10$ components of the Weyl tensor are described by a {\it second order operator} and have thus {\it nothing to do} with the $20$ {\it first order} Bianchi identities for the $20$ components of the Riemann tensor. This result has been recently confirmed by A. Quadrat (INRIA) using new computer algebra packages.  \\
\noindent
2) The {\it Ricci tensor} $ R=(R_{ij})$ is a section of the {\it Ricci bundle} $S_2T^*$ of symmetric covariant $2$-tensors which is the kernel of the canonical projection of the {\it Riemann bundle} onto the {\it Weyl bundle}, induced by the canonical inclusion of the {\it classical Killing system} (Poincar\'e group) into the {\it conformal Killing system} (Conformal group). {\it It has only to do with the second order jets} ${\hat{g}}_2$ ({\it elations}) of the conformal Killing system because we have the isomorphisms $T^*\otimes  {\hat{g}}_2\simeq T^*\otimes T^*\simeq S_2T^* \oplus {\wedge}^2T^*=(R,F)$ with a direct sum where the electromagnetic field $F=(F_{ij})$ is a section of the vector bundle ${\wedge}^2T^*$ of skewsymmetric covariant $2$-tensors. It follows therefore that {\it electromagnetism and gravitation have only to do with second order jets}.  \\
\noindent 
3) The $10$ linearized second order Einstein equations are parametrizing the $4$ first order Cauchy stress equations but cannot be parametrized themselves. As a byproduct of this negative result, these $4$ Cauchy stress equations have {\it nothing to do} with the $4$ divergence-type equations usually obtained from the $20$ Bianchi identities by contraction of indices.   \\

\
\vspace{2cm}

\noindent
{\bf KEY WORDS}\\
 
General relativity, Riemann tensor, Weyl tensor, Ricci tensor, Einstein equations, Lie groups, Lie pseudogroups, Differential sequences, Spencer operator, Janet sequence, Spencer sequence, Differential modules, Split exact sequence, Algebraic analysis, Homological algebra, Extension modules,

\newpage

\noindent
{\bf 1)  INTRODUCTION}  \\

The first motivation for studying the methods used in this paper has been a $1000 \$ $ challenge proposed in $1970$ by J. Wheeler in the physics department of Princeton University while the author of this paper was a student of D.C. Spencer in the closeby mathematics department:  \\ 

{\it Is it possible to express the generic solutions of Einstein equations in vacuum by means of the derivatives of a certain number of arbitrary functions like the potentials for Maxwell equations ?}.\\

Then, being already in contact with M.P. Malliavin as I gave a seminar on the "{\it Deformation Theory of Algebraic and Geometric Structures} " [55], I presented in 1995 a seminar at IHP in Paris, proving the impossibility to parametrize Einstein equations, a result I just found. Meeting with the participants in a caf\'e after the seminar, one of them called my attention on a recently published translation from japanese of the 1970 master thesis of M. Kashiwara that he just saw on display in the library of the Institute [20]. This has been a real "shock" and the true starting of the story. In the meantime, following U. Oberst [34,35], a few persons were trying to adapt these methods to control theory and, thanks to J.L. Lions, I have been able to advertise about this new approach in a european course, held with succes during 6 years [41] and continued for 5 other years in a slightly different form [45]. However, we may say that "{\it  the battle died down because there has been nobody left to carry on the fight} " [67]. By chance I met A. Quadrat, a good PhD student interested by control and computer algebra and we have been staying alone because the specialists of Algebraic Analysis were pure mathematicians, not interested at all by applications.\\

Now, let us start with a completely different approach. Indeed, looking at any textbook of mechanics and using the well known Newton formula, the movement of a body of mass $m$ falling freely in the constant gravitational field $\vec{g}$ is described by $\vec{f}=m\vec{\gamma}$ with $\vec{f}=m\vec{g}$ and $\vec{\gamma}=\frac{d\vec{v}}{dt}$, that is by the $2$ {\it purely geometrical equations} $\frac{d\vec{x}}{dt}=\vec{v} , \frac{d\vec{v}}{dt}=\vec{g}$ and by $\frac{\partial \vec{g}}{\partial x}=0$, that we may rewrite as:  \\
\[  \left\{  \begin{tabular}{cl}
$ \frac{d\vec{x}}{dt}- \vec{v} = 0  $ &\hspace{15mm} derivative of a zero order jet - first order jet  \\
$ \frac{d\vec{v}}{dt}-\vec{g}=0 $ & \hspace{16mm}derivative of a first order jet - second order jet  \\
$ \frac{\partial \vec{g}}{\partial x} - 0=0 $& \hspace{15mm} derivative of a second order jet - third order jet 
\end{tabular}  \right.  \]
It is only after following the course of D.C. Spencer on jet theory that we understood this was just one way to describe the {\it Spencer operator}, namely to identify the speed with a first order jet (Lorentz rotation) and the gravity with a second order jet. Accordingly, the accelerometers on a gyroscopic platform in a rocket are thus only able to measure the three components of the Spencer operator described by the middle line. This comment has been the main physical motivation for using the conformal group of space-time with vanishing third order jets, adopting a quite different philosophy and framework compared to G. Nordstr\"{o}m [31] and H. Weyl [66]. With more details, a section of a {\it jet bundle} of order $q$ can be represented locally by functions $(f^k(x), f^k_i(x), f^k_{ij}(x), ... )$ transforming like the respective derivatives $(f^k(x),{\partial}_if^k(x), {\partial}_{ij}f^k(x),...)$ up to order $q$ but in such a way that ${\partial}_if^k(x)-f^k_i(x)\neq 0, {\partial}_if^k_j(x) - f^k_{ij}(x)\neq 0, ...$ and so on.       \\

A second physical motivation has been to understand the " {\it analogies} " described by E. Mach [1,27,28], G. Lippmann [1,24,25] and H. von Helmholtz [9,44] at the end of the $19^{th}$ century while discovering that they were implicitly used in the finite element approach to the following tabular dealing with variational calculus along the Poincar\'e duality scheme {\it geometry $\leftrightarrow$ physics} [37] but precise definitions will be given in the forthcoming sections (See [43] and [44], p 740-786, for more details):  \\

\[ \renewcommand{\arraystretch}{1.5}
\begin{tabular}{|c|c|c|}
\multicolumn{3}{c}{THEORY}  \\ [1mm]
\hline 
 ELASTICITY & HEAT & ELECTROMAGNETISM \\  [3pt]
 \hline  
 \multicolumn{3}{c}{GEOMETRY }  \\ [1mm]
\hline   
 DISPLACEMENT & TEMPERATURE & POTENTIAL \\  [1mm]
 DEFORMATION & GRADIENT &  FIELD  \\  [1mm]
 DEFORMATION EQUATIONS & CURL &FIELD EQUATIONS   \\ [1mm]
 \hline
 \multicolumn{3}{ c } {PHYSICS}  \\  [1mm]
 \hline
 STRESS & HEAT FLUX & INDUCTION \\   [1mm]
 STRESS EQUATIONS & HEAT EQUATION  & INDUCTION EQUATIONS  \\   [1mm]
 \hline
 \multicolumn{3}{c}{COUPLINGS}  \\  [1mm]
 \hline
 HOOKE LAW  & FOURIER LAW  &  MINKOWSKI LAW  \\   [1mm]
\hline
\multicolumn{3}{|c|}  { DIAGONAL  \{ Photoelasticity, Piezzoelectricity, Thermoelectricity, ...\} }\\  [1mm]
\hline
\end{tabular}  \]

\[   \renewcommand{\arraystretch}{1.5}
\begin{tabular}{|c|c|c|}
\multicolumn{3}{c}{GROUP THEORY $\rightarrow$ SPENCER SEQUENCE  }\\
\hline
ELASTICITY& HEAT& \begin{tabular}{c}ELECTROMAGNETISM\\ $\oplus$ GRAVITATION  \end{tabular} \\
\hline
 $\{{\xi}^k(x),{\xi}^k_i(x)\mid k\neq i \}$& ${\xi}^r_r(x) $& ${\xi}^r_{ri}=A_i $  \\ [3mm]

 $\left\{\begin{array}{rl}{\partial}_i{\xi}^k-{\xi}^k_i&=X^k_{,i}\\{\partial}_i{\xi}^k_j-{\xi}^k_{ij}&=X^k_{j,i} \end{array} \right.$&${\partial}_i{\xi}^r_r-{\xi}^r_{ri}=X_i$ &  $\left\{  \begin{array}{c}\hspace{2mm}{\partial}_i{\xi}^r_{r,j}-{\partial}_j{\xi}^r_{r,i}=F_{ij}\\\frac{1}{2}({\partial}_i{\xi}^r_{rj}+{\partial}_j{\xi}^r_{ri})=R_{ij} \end{array} \right.$ \\ [5mm]
 
 $\left\{  \begin{array}{c}{\partial}_iX^k_{,j}-{\partial}_jX^k_{,i}+X^k_{i,j}-X^k_{j,i}=0 \\ {\partial}_iX^k_{l,j}-{\partial}_jX^k_{l,i}+X^k_{li,j}-X^k_{lj,i}=0  \end{array}  \right. $ 
&${\partial}_iX_j-{\partial}_jX_i +F_{ij}=0 $   &${\partial}_iX_{l,j}-{\partial}_jX_{l,i}=0 $   \\ [5mm]
\hline
\end{tabular}   \]
where the rows are successively describing POTENTIAL, FIELD and FIELD EQUATIONS like in the previous tabular. As we shall see in Section $4$, $\oplus$ is the direct sum $T^*\otimes T^*\simeq S_2T^* \oplus {\wedge}^2T^*$ with standard notations and, using the fact that the third order jets vanish, we have set:  \\
\[X^r_{r,i}=X_i, \hspace{1cm} X^r_{rj,i}=X_{j,i}=R_{ij}+\frac{1}{2}F_{ij}={\partial}_i{\xi}^r_{rj}-{\xi}^r_{rij}={\partial}_i{\xi}^r_{rj}\neq X_{i,j}\hspace{5mm}  \forall n\geq 3\] 
Accordingly, the {\it field} is a section of ${\hat{C}}_1$ parametrized by the first Spencer operator $D_1$ and thus killed by $D_2$ in the initial part ${\hat{C}}_0 \stackrel{D_1}{\longrightarrow} {\hat{C}}_1 \stackrel{D_2}{\longrightarrow} {\hat{C}}_2$ of the Spencer sequence with ${\hat{C}}_r={\wedge}^rT^*\otimes   {\hat{R}}_2$ and $dim({\hat{R}}_2)=15$ in the case of the conformal Killing equations on space-time. {\it It is essential to notice that the field is a $1$-form with value in a Lie equation}. Of course, the metric and Riemann tensor cannot exist in this scheme because we shall see that ... {\it they are in another differential sequence and only the Ricci part is surprisingly left} as we shall explain in Section $4$. {However, such a result, which is coherent with the ideas of both the Cosserat brothers [13] (first row), Weyl [66] and Nordstr\"om [31] (second and third row), cannot be explained by standard tensorial methods and could not have been found before $1975$ because of the lack of any convenient mathematical framework for dealing with second and third order jets [50, 51].  \\

 \newpage

 \noindent
{\bf EXAMPLE 1.1} : {\it Cosserat Elasticity Theory} \\
If we restrict our study to the group of isometries of the euclidean metric $\omega$ in dimension $n\geq 2$, exhibiting the Janet and the Spencer sequences is not easy at all because the corresponding Killing operator ${\cal{D}}\xi={\cal{L}}(\xi)\omega=\Omega  \in S_2T^*$, which involves the Lie derivative ${\cal{L}}$ and provides twice the so-called infinitesimal deformation tensor of continuum mechanics, is not involutive. In order to overcome this problem, one must differentiate once by considering also the Christoffel symbols $\gamma$ and add the operator ${\cal{L}}(\xi)\gamma=\Gamma \in S_2T^*\otimes T$ with the well known Levi-Civita isomorphism $j_1(\omega)=(\omega,{\partial}_x\omega)\simeq (\omega,\gamma)$. Introducing the bundle ${\wedge}^rT^*$ of completely skewsymmetric covariant tensors or $r$-forms and the exterior derivative $d$ with $d^2=d\circ d\equiv 0$, we have the Poincar\'{e} sequence:\\
 \[   {\wedge}^0T^* \stackrel{d}{\longrightarrow} {\wedge}^1T^* \stackrel{d}{\longrightarrow} {\wedge}^2T^* \stackrel{d}{\longrightarrow}  ...  \stackrel{d}{\longrightarrow} {\wedge}^nT^* \longrightarrow 0  \]
{\it For Lie groups of transformations, we shall prove that the Spencer sequence is locally isomorphic to the tensor product of the Poincar\'{e} sequence by the Lie algebra of the underlying Lie group}. Hence, the bigger is the group involved, the bigger are the dimensions of the Spencer bundles, contrary to what happens in the Janet sequence where the first Janet bundle has only to do with differential invariants. This rather philosophical comment, namely to {\it replace the Janet sequence by the Spencer sequence}, must be considered as the crucial key for understanding the work of the brothers E. and F. Cosserat in 1909 [13, 42], the best picture being that of Janet and Spencer playing at see-saw. Also, contrary to what happens in the Janet sequence with ${\cal{D}}$, the formal adjoint of the Spencer operator $D_1$ brings as many dual equations as the number of parameters [68].\\

When $n=2$, one has $n(n+1)/2=3$ parameters, namely $2$ translations and $1$ rotation with infinitesimal generators ${\partial}_1, {\partial}_2$ and $x^1{\partial}_2 - x^2 {\partial}_1$. The following commutative diagram only depends on the left commutative square and each operator generates the {\it compatibility conditions} (CC) of the previous one with 
$j_2(\xi)(x)=({\xi}^k(x), {\partial}_i{\xi}^k(x), {\partial}_{ij}{\xi}^k(x))$ and ${\cal{D}}={\Phi}_0 \circ j_2$: \\
  
  \[  \begin{array}{rccccccccccccr}
 &&&&& 0 &&0&&0&  & \\
 &&&&& \downarrow && \downarrow && \downarrow  &\\
  & 0& \longrightarrow& \Theta &\stackrel{j_2}{\longrightarrow}&  3 &\stackrel{D_1}{\longrightarrow}& 6 &\stackrel{D_2}{\longrightarrow} & 3 &\longrightarrow  0 & \hspace{3mm}Spencer \hspace{2mm}  \\
  &&&&& \downarrow & & \downarrow & & \downarrow & &    & \\
   & 0 & \longrightarrow & 2 & \stackrel{j_2}{\longrightarrow} & 12& \stackrel{D_1}{\longrightarrow} & 16 &\stackrel{D_2}{\longrightarrow} & 6 &   \longrightarrow 0 &\\
   & & & \parallel && \hspace{5mm}\downarrow {\Phi}_0 & &\hspace{5mm} \downarrow {\Phi}_1 & & \hspace{5mm}\downarrow {\Phi}_2 &  &\\
   0 \longrightarrow & \Theta &\longrightarrow & 2 & \stackrel{\cal{D}}{\longrightarrow} & 9 & \stackrel{{\cal{D}}_1}{\longrightarrow} & 10 & \stackrel{{\cal{D}}_2}{\longrightarrow} & 3& \longrightarrow  0 & \hspace{7mm} Janet \hspace{2mm} \\
   &&&&& \downarrow & & \downarrow & & \downarrow &      &\\
   &&&&& 0 && 0 && 0  &  &
   \end{array}     \]
Even in this elementary case, the reader will fast discover that only working out the middle row is at the limit of what can be done by hand 
(exercise) and that it finally seems quite "{\it magical} " that the induced upper row has to do with the Poincar\'{e} sequence $1 \stackrel{d}{\longrightarrow} 2 \stackrel{d}{\longrightarrow} 1 \rightarrow 0$. \\

More generally, for $n\geq 2$ arbitrary, the adjoint of the first Spencer operator $D_1$ provides the Cosserat equations which can be parametrized by the adjoint of the second Spencer operator $D_2$ because it is well known that the Poincar\'{e} sequence is self-adjoint up to sign. A delicate theorem of homological algebra on the vanishing of the so-called {\it extension modules} (Section $3$) finally proves that the adjoint of the Lie operator ${\cal{D}}$ (stress equations) can also be parametrized by the adjoint of its compatibility conditions ${\cal{D}}_1$. As a byproduct, the following result does not seem to be known:  \\
$\bullet${\it The parametrization of the Cosserat couple-stress equations is first order}.\\
$\bullet${\it The parametrization of the Cauchy stress equations} (Airy [2] when $n=2$, Beltrami [5] and Maxwell [29] or Morera [30] when $n=3$, Einstein [50, 54] when $n=4$) {\it is second order}.  \\

When $n=2$, the Killing system brings ${\xi}^1_1=0, {\xi}^1_2+{\xi}^2_1=0,{\xi}^2_2=0, {\xi}^r_{ij}=0$ and the adjoint of $D_1$ provides the Cosserat couple-stress equations (Compare to [64]!). Indeed, lowering the upper indices by means of the (constant) euclidean metric, we just need to look for the factors of 
${\xi}_1,{\xi}_2$ and ${\xi}_{1,2}$ in the integration by parts of the sum:\\
 \[ {\sigma}^{11}({\partial}_1{\xi}_1-{\xi}_{1,1})+{\sigma}^{12}({\partial}_2{\xi}_1-{\xi}_{1,2})+{\sigma}^{21}({\partial}_1{\xi}_2-{\xi}_{2,1})+{\sigma}^{22}({\partial}_2{\xi}_2-{\xi}_{2,2})+{\mu}^{r}({\partial}_r{\xi}_{1,2}-{\xi}_{1,2r}) \]
 in order to obtain the {\it force} $f=(f^1,f^2)$ and the {\it momentum} $m$ by the formulas ([13], p 137): \\
 
 \[ {\partial}_1{\sigma}^{11}+{\partial}_2{\sigma}^{12}=f^1, {\partial}_1{\sigma}^{21}+{\partial}_2{\sigma}^{22}=f^2, \hspace{5mm}{\partial}_1{\mu}^1+{\partial}_2{\mu}^2+{\sigma}^{12}-{\sigma}^{21}=m  \]

 Finally, we obtain the nontrivial {\it first order} parametrization ${\sigma}^{11}={\partial}_2{\phi}^1, {\sigma}^{12}=-{\partial}_1{\phi}^1, {\sigma}^{21}=-{\partial}_2{\phi}^2, {\sigma}^{22}={\partial}_1{\phi}^2, {\mu}^{1}={\partial}_2{\phi}^3+{\phi}^1, {\mu}^{2}=-{\partial}_1{\phi}^3-{\phi}^2$ by means of the three arbitrary functions ${\phi}^1,{\phi}^2,{\phi}^3$, in a coherent way with the Airy {\it second order} parametrization obtained if we set  ${\phi}^1={\partial}_2{\phi}, {\phi}^2={\partial}_1{\phi}, {\phi}^3=-\phi$ when 
 ${\mu}^1=0,{\mu}^2=0$ [48] . \\
 
The adjoint of the second order {\it Riemann operator} ${\cal{D}}_1: \Omega \rightarrow R={\partial}_{11}{\Omega}_{22}+{\partial}_{22}{\Omega}_{11}-2{\partial}_{12}{\Omega}_{12}$ is nothing else but the {\it second order parametrization} ${\sigma}^{11}={\partial}_{22}\phi, {\sigma}^{22}={\partial}_{11}\phi, {\sigma}^{12}={\sigma}^{21}=-{\partial}_{12}\phi$ of the classical Cauchy stress equations by means of the single Airy function $\phi$ which has therefore {\it nothing to do} with any metric.\\

More generally, using the conformal Killing system with ${\xi}^1_1= ... ={\xi}^n_n=(1/n){\xi}^r_r$ and $n=4$, we may similarly 
introduce $tr(\sigma)={\omega}_{ij}{\sigma}^{ij}=-{\sigma}^{44}\sim \rho$ as usual in relativistic mechanics and obtain:  \\ 
\[  {\sigma}^{ij}{\xi}_{i,j}=({\sigma}^{11}{\xi}_{1,1} + ... )+ ({\sigma}^{12}{\xi}_{1,2} + {\sigma}^{21}{\xi}_{2,1}+ ...)= \frac{1}{n}tr(\sigma) {\xi}^r_r  + {\sum}_{i<j}  ({\sigma}^{ij}-{\sigma}^{ji}){\xi}_{i,j}      \]
in the conformal case, when ${\sigma}$ is arbitrary. Integrating now by parts the summation: \\
\[  n{\sigma}^{ij} ({\partial}_i {\xi}^j - {\xi}_{j,i}) + g^i({\partial}_i{\xi}^r_r - {\xi}^r_{ri}) + g^{ij}( {\partial}_i{\xi}^r_{rj} - 0)  \]
in order to find the adjoint of $D_1$, we obtain therefore  the so-called {\it virial equations} [53]: \\
\[   {\partial}_i {\sigma}^{ij} = 0, \hspace{5mm} {\sigma}^{ij} - {\sigma}^{ji}=0 \hspace{5mm} {\partial}_jg^j + tr(\sigma)=0, \hspace{5mm}{\partial}_ig^{ij} +g^j=0  \]
and thus ${\partial}_{ij}g^{ij}=tr(\sigma)\sim \rho$. If $(g^{ij}=g^{ji})$ is the symmetric tensor density dualizing $(R_{ij}=R_{ji})$ with $g^{ij}=\psi {\omega}^{ij}$ by isotropy, we get the Newton law ${\omega}^{ij}{\partial}_{ij}\psi\sim\rho$. Finally, if $(g^{ij}= - g^{ji})$ is the EM induction dualizing the EM field $(F_{ij}= - F_{ji})$, we obtain at once:  \\
\[{\partial}_ig^{ij}+g^j=0 \hspace{5mm}\Rightarrow \hspace{5mm}{\partial}_{ij}g^{ij}=0 \hspace{5mm}\Rightarrow \hspace{5mm}{\partial}_jg^j=0 \hspace{5mm}\Rightarrow\hspace{5mm}tr(\sigma)=0 \] 
as conjectured by Weyl in [66]. Accordingly, {\it there is no conceptual difference between the Cosserat couple-stress equations, the Newton equations of gravitation and the Maxwell equations of EM}, in a coherent way with the preceding tabular. Therefore, the main problem left 
and solved in section $4$ is to understand why {\it only the Ricci tensor is appearing in this scheme}, with a unique reference to the splitting 
$T^*\otimes T^*\simeq S_2T^*\oplus {\wedge}^2T^*$ but {\it without any reference to the Riemann or Weyl tensors}.  \\

\noindent
{\bf EXAMPLE 1.2}: {\it Classical Elasticity Theory} \\
In classical elasticity, the {\it stress tensor density} $\sigma=({\sigma}^{ij}={\sigma}^{ji})$ existing inside an elastic body is a symmetric $2$-tensor density introduced by A. Cauchy in 1822. The corresponding Cauchy {\it stress equations} can be written as ${\partial}_r{\sigma}^{ir}=f^i$ where the right member describes the local density of forces applied to the body, for example gravitation. With zero second member, we study the possibility to "{\it parametrize} " the system of PD equations ${\partial}_r{\sigma}^{ir}=0$, namely to express its general solution by means of a certain number of arbitrary functions or {\it potentials}, called {\it stress functions}. Of course, the problem is to know about the number of such functions and the order of the parametrizing operator. In what follows, the space has $n$ local coordinates $x=(x^i)=(x^1, ... , x^n)$. For $n=1,2,3$ one may introduce the Euclidean metric $\omega=({\omega}_{ij}={\omega}_{ji})$ while, for $n=4$, one may consider the Minkowski metric. A few definitions used thereafter will be provided later on.\\

\newpage

\noindent
$\bullet \hspace{3mm}n=2$: The stress equations become ${\partial}_1{\sigma}^{11}+{\partial}_2{\sigma}^{12}=0, {\partial}_1{\sigma}^{21}+{\partial}_2{\sigma}^{22}=0$. Their second order parametrization ${\sigma}^{11}={\partial}_{22}\phi, {\sigma}^{12}={\sigma}^{21}=-{\partial}_{12}\phi, {\sigma}^{22}={\partial}_{11}\phi$ has been provided by George Biddell Airy (1801-1892) in 1863 [2]. It can be simply recovered in the following manner: \\
\[ \begin{array}{rcl}
{\partial}_1{\sigma}^{11}- {\partial}_2( - {\sigma}^{12})= 0 \hspace{5mm} & \Rightarrow & \hspace{5mm} \exists \varphi,\, {\sigma}^{11}={\partial}_2\varphi, {\sigma}^{12}= 
- {\partial}_1\varphi \\  
{\partial}_2{\sigma}^{22}- {\partial}_1( - {\sigma}^{21})=0  \hspace{5mm} & \Rightarrow & \hspace{5mm}  \exists \psi, \,{\sigma}^{22}={\partial}_1\psi, {\sigma}^{21}= - {\partial}_2\psi \\
   {\sigma}^{12}={\sigma}^{21} \Rightarrow {\partial}_1 \varphi - {\partial}_2\psi =0 \hspace{5mm}& \Rightarrow & \hspace{5mm}\exists \phi, \,\varphi={\partial}_2\phi, \psi={\partial}_1\phi  
   \end{array}Ê \]
We get the second order system:  \\
\[ \left\{  \begin{array}{rll}
{\sigma}^{11} & \equiv {\partial}_{22}\phi =0 \\
-{\sigma}^{12} & \equiv {\partial}_{12}\phi =0 \\
{\sigma}^{22} & \equiv {\partial}_{11}\phi=0
\end{array}
\right. \fbox{ $ \begin{array}{ll}
1 & 2   \\
1 & \bullet \\  
1 & \bullet  
\end{array} $ } \]
which is involutive with one equation of class $2$, $2$ equations of class $1$ and it is easy to check that the $2$ corresponding first order CC are just the stress equations.\\

\noindent
$\bullet \hspace{3mm} n=3$: Things become quite more delicate when we try to parametrize the $3$ PD equations: \\
\[ {\partial}_1{\sigma}^{11}+{\partial}_2{\sigma}^{12}+{\partial}_3{\sigma}^{13}=0,\hspace{3mm} {\partial}_1{\sigma}^{21}+{\partial}_2{\sigma}^{22}+{\partial}_3{\sigma}^{23}=0, \hspace{3mm} {\partial}_1{\sigma}^{31}+{\partial}_2{\sigma}^{32}+{\partial}_3{\sigma}^{33}=0 \]

A direct computational approach has been provided by Eugenio Beltrami (1835-1900) in 1892 [5], James Clerk Maxwell (1831-1879) in 1870 [29] and Giacinto Morera (1856-1909) in 1892 [30] by introducing the $6$ {\it stress functions} ${\phi}_{ij}={\phi}_{ji}$ in the {\it Beltrami parametrization} obtained by considering:\\
\[{\sigma}^{11} = {\partial}_{33}{\phi}_{22}+{\partial}_{22}{\phi}_{33}-2{\partial}_{23}{\phi}_{23} \]
\[{\sigma}^{12}={\sigma}^{21} = {\partial}_{13}{\phi}_{23}+{\partial}_{23}{\phi}_{13}-{\partial}_{33}{\phi}_{12}-{\partial}_{12}{\phi}_{33}  \]
and the additional $4$ relations obtained by using a cyclic permutation of $(1,2,3)$. The system:\\
\[   \left\{  \begin{array}{rll}
{\sigma}^{11} \equiv& {\partial}_{33}{\phi}_{22}+{\partial}_{22}{\phi}_{33}-2{\partial}_{23}{\phi}_{23}=0  \\
-{\sigma}^{12}\equiv & {\partial}_{33}{\phi}_{12}+{\partial}_{12}{\phi}_{33}-{\partial}_{13}{\phi}_{23}-{\partial}_{23}{\phi}_{13}=0  \\
 {\sigma}^{22}\equiv & {\partial}_{33}{\phi}_{11}+{\partial}_{11}{\phi}_{33}-2{\partial}_{13}{\phi}_{13}=0  \\
{\sigma}^{13}\equiv & {\partial}_{23}{\phi}_{12}+{\partial}_{12}{\phi}_{23}-{\partial}_{22}{\phi}_{13}-{\partial}_{13}{\phi}_{22} =0 \\
-{\sigma}^{23}\equiv & {\partial}_{23}{\phi}_{11}+{\partial}_{11}{\phi}_{23}-{\partial}_{12}{\phi}_{13}-{\partial}_{13}{\phi}_{12} =0 \\
{\sigma}^{33}\equiv & {\partial}_{22}{\phi}_{11}+{\partial}_{11}{\phi}_{22}-2{\partial}_{12}{\phi}_{12}=0
\end{array}
\right. \fbox{ $ \begin{array}{lll}
1 & 2 & 3   \\
1 & 2 & 3  \\
1 & 2 & 3  \\
1 & 2 &  \bullet  \\
1 & 2 & \bullet  \\
1 & 2 & \bullet
\end{array} $ } \]
is involutive with $3$ equations of class $3$, $3$ equations of class $2$ and no equation of class $1$. The $3$ CC are describing the stress equations which admit therefore a parametrization ... but without any geometric framework, in particular without any possibility to imagine that the above second order operator is {\it nothing else but} the {\it formal adjoint} of the {\it Riemann operator}, namely the (linearized) Riemann tensor with $n^2(n^2-1)/2=6$ independent components when $n=3$ [54].\\

{\it Surprisingly}, the Maxwell parametrization is obtained by keeping ${\phi}_{11}=A, {\phi}_{22}=B, {\phi}_{33}=C$ while setting ${\phi}_{12}={\phi}_{23}={\phi}_{31}=0$. However, the fact that this system is involutive can only be found after effecting the linear change of coordinates $x^1 \rightarrow x^1+x^3, x^2\rightarrow x^2+x^3, x^3\rightarrow x^3$ and it is easy to check that the $3$ CC obtained just amount to the desired $3$ stress equations when coming back to the original system of coordinates. Again, {\it if there is a geometrical background, this change of local coordinates is hidding it totally}. The Morera parametrization is obtained similarly by keeping now ${\phi}_{23}=L, {\phi}_{13}=M, {\phi}_{12}=N$ while setting ${\phi}_{11}={\phi}_{22}={\phi}_{33}=0$.\\

\noindent
$\bullet \hspace{3mm}n\geq 4$: As a direct computational way cannot be applied, we don't know if a parametrization may exist and in any case no analogy with the previous situations $n=1,2,3$ could be used. Moreover, no known differential geometric background could be used at first sight in order to provide a hint towards the solution. Now, if $n=4$, $\omega$ is the Minkowski metric and $\phi=GM/r$ is the gravitational potential, then $\phi/c^2\ll 1$ and a perturbation $\Omega\in S_2T^*$  of $\omega$ may satisfy in vacuum the $10$ second order (linearized) {\it Einstein equations} for the $10$ $ \Omega$:  \\
\[  2 E_{ij}\equiv {\omega}^{rs}(d_{ij}{\Omega}_{rs}+d_{rs}{\Omega}_{ij}-d_{ri}{\Omega}_{sj}-d_{sj}{\Omega}_{ri})-{\omega}_{ij}({\omega}^{rs}{\omega}^{uv}d_{rs}{\Omega}_{uv}
-{\omega}^{ru}{\omega}^{sv}d_{rs}{\Omega}_{uv})=0  \]
by introducing the corresponding second order {\it Einstein} operator $S_2T^* \stackrel{Einstein}{\longrightarrow} S_2T^*:\Omega \rightarrow E$ with $E_{ij}=R_{ij}- \frac{1}{2} {\omega}_{ij}tr(R)$ and $tr(R)={\omega}^{ij}R_{ij}$ when $n=4$ [50,51]. For $n\geq 4$, this is a second order involutive system with $n(n-1)/2$ equations of class $n$ and thus $\alpha=n(n+1)/2 - n(n-1)/2=n$ equations of class $n-1$ providing the well known $n$ $div$ first order involutive CC induced from the Bianchi identities. The " {\it founding stone} " of General relativity (GR) is that {\it the Einstein operator is parametrizing the Cauchy stress equations}. However, by analogy with the Maxwell equations of electromagnetism (EM), the challenge of parametrizing Einstein equations themselves has been proposed in 1970 by J. Wheeler for 1000 \$ and solved {\it negatively} in 1995 by the author who only received 1 \$. We shall see that, {\it exactly as before and though it is quite striking}, the key ingredient will be to use the linearized Riemann tensor considered as a second order operator [49,50,54]. As an {\it even more striking fact}, we shall discover that the condition $n\geq 4$ has only to do with the Spencer cohomology for the symbol of the 
{\it classical and conformal Killing equations}.\\

The next tricky example will prove that the possibility to exhibit different parametrizations of the stress equations that we have presented has surely nothing to do with the proper mathematical background of elasticity theory !.  \\

\noindent
{\bf EXAMPLE 1.3}: {\it PD Control Theory}  \\
Let us consider the (trivially involutive) inhomogeneous first order PD equations with two independent variables $(x^1,x^2)$, two unknown functions $({\eta}^1, {\eta}^2)$ and a second member $\zeta$: \\
\[ {\partial}_2{\eta}^1 - {\partial}_1{\eta}^2 + x^2 {\eta}^2=\zeta   \hspace{1cm} \Leftrightarrow \hspace{1cm}  {\cal{D}}_1 \eta = \zeta \]
Multiplying on the left by a test function $\lambda$ and integrating by parts, the corresponding adjoint system of PD 
equations is:  \\
\[  \left\{ \begin{array}{rcll}
{\eta}^1 & \rightarrow  & - {\partial}_2 \lambda  & ={\mu}^1  \\
{\eta}^2 & \rightarrow & \hspace{3mm}{\partial}_1 \lambda + x^2 \lambda  & ={\mu}^2
\end{array} \right.  \hspace{1cm} \Leftrightarrow \hspace{1cm}  ad({\cal{D}}_1)\lambda=\mu\]
Using crossed derivatives, we get $\lambda={\partial}_2{\mu}^2+{\partial}_1{\mu}^1+x^2{\mu}^1$ and substituting, we get the two CC:  \\
\[  \left  \{  \begin{array}{lcl}
  {\partial}_{22}{\mu}^2 + {\partial}_{12}{\mu}^1+x^2{\partial}_2{\mu}^1 + 2{\mu}^1  & = & {\nu}^1\\
{\partial}_{12}{\mu}^2 + {\partial}_{11}{\mu}^1+2x^2{\partial}_1{\mu}^1+x^2{\partial}_2{\mu}^2+(x^2)^2{\mu}^1-{\mu}^2 & = & {\nu}^2  
\end{array}  \right.  \fbox{ $ \begin{array}{ll}
1 & 2   \\
1 & \bullet 
\end{array} $ } \]
This system is involutive and the corresponding generating CC for the second member $({\nu}^1,{\nu}^2)$ is:  \\
\[  {\partial}_2{\nu}^2 - {\partial}_1{\nu}^1 - x^2 {\nu}^1=0  \]
Therefore ${\nu}^2$ is differentially dependent on ${\nu}^1$ but ${\nu}^1$ is also differentially dependent on ${\nu}^2$.\\
Multiplying the first equation by the test function ${\xi}^1$, the second equation by the test function ${\xi}^2$, adding and integrating by parts, we get the {\it canonical parametrization} ${\cal{D}}\xi = \eta $:  \\
\[   \left \{\begin{array}{rcll}
{\mu}^2 & \rightarrow & {\partial}_{22}{\xi}^1+{\partial}_{12}{\xi}^2-x^2{\partial}_2{\xi}^2-2{\xi}^2  & = {\eta}^2  \\
{\mu}^1&  \rightarrow & {\partial}_{12}{\xi}^1-x^2{\partial}_2{\xi}^1+{\xi}^1+{\partial}_{11}{\xi}^2  -2x^2{\partial}_1{\xi}^2+(x^2)^2{\xi}^2 & = {\eta}^1        
\end{array}  \right. \fbox{ $ \begin{array}{ll}
1 & 2   \\
1 & \bullet 
\end{array} $ } \]
of the initial system with zero second member. This system is involutive and the kernel of this parametrization has differential rank equal to $1$ because ${\xi}^1$ or ${\xi}^2$ can be given arbitrarily.  \\
Keeping now ${\xi}^1=\xi$ while setting ${\xi}^2=0$, we get the {\it first minimal parametrization} $\xi \rightarrow ({\eta}^1,{\eta}^2)$:  \\
\[  \left  \{  \begin{array}{ll}
{\partial}_{22}\xi & = {\eta}^2  \\
{\partial}_{12}\xi-x^2{\partial}_2\xi+\xi & ={\eta}^1
\end{array}  \right.  \fbox{ $ \begin{array}{ll}
1 & 2   \\
1 & \bullet 
\end{array} $ }  \]
This system is again involutive and the parametrization is minimal because the kernel of this parametrization has differential rank equal to $0$. With a similar comment, setting now ${\xi}^1=0$ while keeping ${\xi}^2={\xi}'$, we get the {\it second minimal parametrization} ${\xi}' \rightarrow ({\eta}^1,{\eta}^2)$:  \\
\[  \left  \{\begin{array}{ll}
{\partial}_{11}{\xi}'-2x^2{\partial}_1{\xi}'+(x^2)^2{\xi}' & = {\eta}^1  \\
{\partial}_{12}{\xi}'  - x^2{\partial}_2{\xi}' - 2{\xi}' & = {\eta}^2 
\end{array}   \right.  \]
which is again easily seen to be involutive by exchanging $x^1$ with $x^2$.  \\

Leaving now physics for mathematics, the content of the paper becomes clear enough:  \\
$\bullet$  In Section $2$ we provide a self-contained survey of the formal theory of systems of OD or PD equations, only caring about the results that will be {\it absolutely needed} for understanding the next Section.  \\
$\bullet$  In Section $3$ we provide in a similar way the main results of algebraic analysis and biduality, only caring about the results that will be {\it absolutely needed} for understanding the last Section.  \\
$\bullet$  In Section $4$ we combine these results in order to revisit the mathematical foundations of GR. \\

This paper is an extended written version of a series of lectures (10 h) given during a short intensive course organized by the Center of Theoretical Physics (CPT) in Marseille (may, 23-25, 2016) under the title "Algebraic Analysis, Double Duality and Applications in Mathematical Physics".  \\

Though the matters involved are difficult, we point out that we have only presented the {\it strict minimum} of mathematics needed in order to deal with the mathematical foundations of GR in the sense that {\it all results will be used} and advise the reader not to look at the last section without reading the two preceding ones in the order they are presented, even though they are deeply interacting between themselves. We 
do believe that all the results presented are new and cannot therefore even provide other modern references. \\

\newpage

\noindent
{\bf 2)  DIFFERENTIAL SYSTEMS} \\

If $E$ is a vector bundle over the base manifold $X$ with projection $\pi$ and local coordinates $(x,y)=(x^i,y^k)$ projecting onto $x=(x^i)$ for $i=1,...,n$ and $k=1,...,m$, identifying a map with its graph, a (local) section $f:U\subset X \rightarrow E$ is such that $\pi\circ f =id$ on $U$ and we write $y^k=f^k(x)$ or simply $y=f(x)$. For any change of local coordinates $(x,y)\rightarrow (\bar{x}=\varphi(x),\bar{y}=A(x)y)$ on $E$, the change of section is $y=f(x)\rightarrow \bar{y}=\bar{f}(\bar{x})$ such that ${\bar{f}}^l(\varphi(x)\equiv A^l_k(x)f^k(x)$. The new vector bundle $E^*$ obtained by changing the {\it transition matrix} $A$ to its inverse $A^{-1}$ is called the {\it dual vector bundle} of $E$. Differentiating with respect to $x^i$ and using new coordinates $y^k_i$ in place of ${\partial}_if^k(x)$, we obtain ${\bar{y}}^l_r{\partial}_i{\varphi}^r(x)=A^l_k(x)y^k_i+{\partial}_iA^l_k(x)y^k$. Introducing a multi-index $\mu=({\mu}_1,...,{\mu}_n)$ with length $\mid \mu \mid={\mu}_1+...+{\mu}_n$ and prolonging the procedure up to order $q$, we may construct in this way, by patching coordinates, a vector bundle $J_q(E)$ over $X$, called the {\it jet bundle of order} $q$ with local coordinates $(x,y_q)=(x^i,y^k_{\mu})$ with $0\leq \mid\mu\mid \leq q$ and $y^k_0=y^k$. Hence, we may use the notation $y_{xx}$ or $y_{(2)}$, $y_{12}$ or $y_{(1,1)}$ and so on. For a later use, we shall set $\mu+1_i=({\mu}_1,...,{\mu}_{i-1},{\mu}_i+1,{\mu}_{i+1},...,{\mu}_n)$ and define the operator $j_q:E \rightarrow J_q(E):f \rightarrow j_q(f)$ on sections by the local formula $j_q(f):(x)\rightarrow({\partial}_{\mu}f^k(x)\mid 0\leq \mid\mu\mid \leq q,k=1,...,m)$. Finally, a jet coordinate $y^k_{\mu}$ is said to be of {\it class} $i$ if ${\mu}_1=...={\mu}_{i-1}=0, {\mu}_i\neq 0$. As the background will always be clear enough, we shall use the same notation for a vector bundle and its set of sections [38-41].   \\

\noindent
{\bf DEFINITION 2.1}:  A {\it system} of PD equations of order $q$ on $E$ is a vector subbundle $R_q\subset J_q(E)$ locally defined by a constant rank system of linear equations for the jets of order $q$ of the form $ a^{\tau\mu}_k(x)y^k_{\mu}=0$. Its {\it first prolongation} $R_{q+1}\subset J_{q+1}(E)$ will be defined by the equations $ a^{\tau\mu}_k(x)y^k_{\mu}=0, a^{\tau\mu}_k(x)y^k_{\mu+1_i}+{\partial}_ia^{\tau\mu}_k(x)y^k_{\mu}=0$ which may not provide a system of constant rank as can easily be seen for $xy_x-y=0 \Rightarrow xy_{xx}=0$ where the rank drops at $x=0$.\\

The next definition will be crucial for our purpose.\\

\noindent
{\bf DEFINITION 2.2}: A system $R_q$ is said to be {\it formally integrable} if the $R_{q+r}$ are vector bundles $\forall r\geq 0$ (regularity condition) and no new equation of order $q+r$ can be obtained by prolonging the given PD equations more than $r$ times, $\forall r\geq 0$.\\

Finding an intrinsic test has been achieved by D.C. Spencer in 1965 [63] along coordinate dependent lines sketched by Janet as early as in 1920 [19, 38] and Gr\"{o}bner in 1940 [16], as we already said. The key ingredient, missing explicitly before the moderrn approach, is provided by the following definition.\\

\noindent
{\bf DEFINITION 2.3}: The family $g_{q+r}$ of vector spaces over $X$ defined by the purely linear equations $ a^{\tau\mu}_k(x)v^k_{\mu+\nu}=0$ for $ \mid\mu\mid= q, \mid\nu\mid =r $ is called the {\it symbol} at order $q+r$ and only depends on $g_q$.\\

The following procedure, {\it where one may have to change linearly the independent variables if necessary}, is the heart towards the next definition which is intrinsic even though it must be checked in a particular coordinate system called $\delta$-{\it regular} (See [13] and [14] for more details):\\

\noindent
$\bullet$ {\it Equations of class} $n$: Solve the maximum number $\beta={\beta}^n_q$ of equations with respect to the jets of order $q$ and class $n$. Then call $(x^1,...,x^n)$ {\it multiplicative variables}.\\
\[  - - - - - - - - - - - - - - - -  \]
$\bullet$ {\it Equations of class} $i$: Solve the maximum number of {\it remaining} equations with respect to the jets of order $q$ and class $i$. Then call $(x^1,...,x^i)$ {\it multiplicative variables} and $(x^{i+1},...,x^n)$ {\it non-multiplicative variables}.\\
\[ - - - - - - - - - - - - - - - - - \]
$\bullet$ {\it Remaining equations equations of order} $\leq q-1$: Call $(x^1,...,x^n)$ {\it non-multiplicative variables}.\\

\noindent
{\bf DEFINITION 2.4}: A system of PD equations is said to be {\it involutive} if its first prolongation can be achieved by prolonging its equations only with respect to the corresponding multiplicative variables. The numbers ${\alpha}^i_q=m(q+n-i-1)!/((q-1)!(n-i)!)-{\beta}^i_q$ will be called {\it characters} and ${\alpha}^1_q\geq ... \geq {\alpha}^n_q=\alpha=m-\beta $. For an involutive system, $(y^{{\beta}^n_q +1},...,y^m)$ can be given arbitrarily.  \\

Though the preceding description was known to Janet (he called it : "modules de formes en involution"), surprisingly he never used it explicitly. In any case, such a definition is far from being intrinsic and the hard step will be achieved from the Spencer cohomology that will also play an important part in the so-called {\it reduction to first order}, a result no so well known today as we shall see.\\

Let us consider $J_{q+1}(E)$ with jet coordinates $\{ y^l_{\lambda}\mid 0\leq \mid\lambda\mid\leq q+1\}$ and $J_1(J_q(E))$ with jet coordinates $\{z^k_{\mu},z^k_{\mu,i}\mid 0\leq \mid\mu\mid\leq q,i=1,...,n\}$. The canonical inclusion $J_{q+1}(E)\subset J_1(J_q(E))$ is described by the {\it two kinds} of equations:\\
\[   z^k_{\mu,i}-z^k_{\mu+1_i}=0 ,     \hspace{3cm}     0\leq \mid\mu\mid\leq q-1  \]
\[  z^k_{\mu+1_j,i}-z^k_{\mu+1_i,j}=0  , \hspace{3cm}   \mid\mu\mid=q-1  \]
or using the parametrization $z^k_{\mu,i}=y^k_{\mu+1_i}$ for $\mid\mu\mid=q$ with $z^k_{\mu}=y^k_{\mu}, \forall 0\leq \mid \mu \mid \leq q$. \\

Let $T$ be the tangent vector bundle of vector fields on $X$, $T^*$ be the cotangent vector bundle of 1-forms on $X$ and ${\wedge}^sT^*$ be the vector bundle of s-forms on $X$ with usual bases $\{dx^I=dx^{i_1}\wedge ... \wedge dx^{i_s}\}$ where we have set $I=(i_1< ... <i_s)$. Also, let $S_qT^*$ be the vector bundle of symmetric q-covariant tensors. Moreover, if  $\xi,\eta\in T$ are two vector fields on $X$, we may define their {\it bracket} $[\xi,\eta]\in T$ by the local formula $([\xi,\eta])^i(x)={\xi}^r(x){\partial}_r{\eta}^i(x)-{\eta}^s(x){\partial}_s{\xi}^i(x)$ leading to the {\it Jacobi identity} $[\xi,[\eta,\zeta]]+[\eta,[\zeta,\xi]]+[\zeta,[\xi,\eta]]=0, \forall \xi,\eta,\zeta \in T$. We have also the useful formula $[T(f)(\xi),T(f)(\eta)]=T(f)([\xi,\eta])$ where $T(f):T(X)\rightarrow T(Y)$ is the tangent mapping of a map $f:X\rightarrow Y$. Finally, we may introduce the {\it exterior derivative} $d:{\wedge}^rT^*\rightarrow {\wedge}^{r+1}T^*:\omega={\omega}_Idx^I \rightarrow d\omega={\partial}_i{\omega}_Idx^i\wedge dx^I$ with $I=\{i_1< ... <i_r\}$ and we have $d^2=d\circ d\equiv 0$ in the {\it Poincar\'{e} sequence}:\\
\[  {\wedge}^0T^* \stackrel{d}{\longrightarrow} {\wedge}^1T^* \stackrel{d}{\longrightarrow} {\wedge}^2T^* \stackrel{d}{\longrightarrow} ... \stackrel{d}{\longrightarrow} {\wedge}^nT^* \longrightarrow 0  \]

In a purely algebraic setting, one has [38,63]:  \\

\noindent
{\bf PROPOSITION 2.5}: There exists a map $\delta:{\wedge}^sT^*\otimes S_{q+1}T^*\otimes E\rightarrow {\wedge}^{s+1}T^*\otimes S_qT^*\otimes E$ which restricts to $\delta:{\wedge}^sT^*\otimes g_{q+1}\rightarrow {\wedge}^{s+1}T^*\otimes g_q$ and ${\delta}^2=\delta\circ\delta=0$.\\

{\it Proof}: Let us introduce the family of s-forms $\omega=\{ {\omega}^k_{\mu}=v^k_{\mu,I}dx^I \}$ and set $(\delta\omega)^k_{\mu}=dx^i\wedge{\omega}^k_{\mu+1_i}$. We obtain at once $({\delta}^2\omega)^k_{\mu}=dx^i\wedge dx^j\wedge{\omega}^k_{\mu+1_i+1_j}=0$ and 
$ a^{\tau \mu}_k(\delta \omega)^k_{\mu}=dx^i \wedge(  a^{\tau \mu}_k{\omega}^k_{\mu +1_i})=0$.  \\
\hspace*{12cm} Q.E.D.  \\

The kernel of each $\delta$ in the first case is equal to the image of the preceding $\delta$ but this may no longer be true in the restricted case we set:\\

\noindent
{\bf DEFINITION 2.6}: We denote by $B^s_{q+r}(g_q)\subseteq Z^s_{q+r}(g_q)$ and $H^s_{q+r}(g_q)=Z^s_{q+r}(g_q)/B^s_{q+r}(g_q)$ respectively the coboundary space $im(\delta)$, cocycle space $ker(\delta)$ and cohomology space at ${\wedge}^sT^*\otimes g_{q+r}$ of the restricted $\delta$-sequence which only depend on $g_q$ and may not be vector bundles. The symbol $g_q$ is said to be s-{\it acyclic} if $H^1_{q+r}=...=H^s_{q+r}=0, \forall r\geq 0$, {\it involutive} if it is n-acyclic and {\it finite type} if $g_{q+r}=0$ becomes trivially involutive for r large enough. Finally, $S_qT^*\otimes E$ is involutive $\forall q\geq 0$ if we set $S_0T^*\otimes E=E$. \\

\noindent
{\bf THEOREM 2.7}: ({\it Integrability/involutivity criterion}) A system $R_q \subset J_q(E)$ is formally integrable (involutive) if ${\pi}^{q+1}_q:R_{q+1} \rightarrow R_q$ is an epimorphism of vector bundles and $g_q$ is $2$-acyclic (involutif).  \\

From now on, we shall suppose that $R_q$ is involutive and that we are only dealing with vector bundles, in particular that $g_q$ is a vector bundle and that the projection $R_{q-1}$ of $R_q$ in $J_{q-1}(E)$ is thus also a vector bundle (See [38, 41, 44] for more details). The following technical result will prove to be quite useful later on for our purpose [38, 63]:   \\

\noindent
{\bf PROPOSITION 2.8}: We may define the {\it Spencer bundles} by the isomorphisms:  \\
\[  C_r={\wedge}^rT^*\otimes R_q/\delta ({\wedge}^{r-1}T^*\otimes g_{q+1}) \simeq \delta ({\wedge}^rT^*\otimes g_q)\oplus 
{\wedge}^rT^*\otimes R_{q-1}\]
In particular one has $C_0=R_q$ and $C_n={\wedge}^nT^*\otimes R_{q-1}$.\\

{\it Proof}: The first commutative ad exact diagram:\\
\[  \begin{array}{rcccccl}
      & 0 &  & 0 &  & 0 &    \\
      &\downarrow &  &\downarrow &  &  \downarrow &   \\
0\rightarrow &g_{q+1} &\rightarrow &T^*\otimes R_q &\rightarrow &C_1&\rightarrow 0  \\
      & \downarrow &  &\downarrow &  &\parallel &   \\
0\rightarrow &R_{q+1}&\rightarrow &J_1(R_q) & \rightarrow &C_1 & \rightarrow 0  \\
     & \downarrow &   &\downarrow  &  & \downarrow &  \\
0\rightarrow & R_q & = & R_q &\rightarrow & 0  \\
  & \downarrow &   & \downarrow &  &   &  \\
  &   0   &   &   0  &   &   &
  \end{array}   \]
 \noindent 
shows that $C_1\simeq T^*\otimes R_q/g_{q+1} $. The general case finally depends on the following commutative and exact diagram by using a (non-canonical) splitting of the right column:\\  
\[   \begin{array}{cccccl}
       0   &   &   0   &   &   0   &   \\
    \downarrow &  & \downarrow &  & \downarrow  &   \\
 {\wedge}^{r-1}T^*\otimes g_{q+1} & \stackrel{\delta}{ \rightarrow} & {\wedge}^rT^*\otimes g_q &
 \stackrel{\delta}{\rightarrow} & \delta ( {\wedge}^rT^*\otimes g_q) &  \rightarrow 0  \\
      \parallel  &  &  \downarrow  &  & \downarrow  &  \\
 {\wedge}^{r-1}T^*\otimes g_{q+1} &\rightarrow  &{\wedge}^rT^*\otimes R_q & \rightarrow &C_r &
 \rightarrow 0  \\
      \downarrow  &  & \downarrow &  & \downarrow &   \\
        0  &  \rightarrow &  {\wedge}^rT^*\otimes R_{q-1}&  = & {\wedge}^r T^*\otimes R_{q-1} & \rightarrow 0  \\
         &   &  \downarrow &  &  \downarrow  &  \\
         &   &  0  &  &  0  &  
       \end{array}     \]
When $r=n$, the equality $\delta ({\wedge}^{n-1}T^*\otimes g_{q+1})= {\wedge}^nT^*\otimes  g_q$ gives the last result. \\
\hspace*{12cm} Q.E.D.  \\
    
Accordingly, the inclusion $R_{q+1}\subset J_1(R_q)$ can be considered as a new first order system over $R_q$, called {\it first order reduction} or {\it Spencer form}. The same procedure is valid for the inclusion $J_{q+1} \subset J_1(J_q(E))$. One obtains [38-41, 44]:\\
 
 \noindent
{\bf PROPOSITION 2.9}: The first order reduction is formally integrable (involutive) whenever $R_q$ is formally integrable (involutive). In that case, the reduction has no longer any zero order equation.\\
  
 Having in mind control theory, {\it it just remains to modify the Spencer form} in order to generalize the Kalman form from OD equations to PD equations. Here is the procedure that must be followed in the case of a first order involutive system with no zero order equation.   \\

\noindent 
$\bullet$  Look at the equations of class n solved with respect to $y^1_n,...,y^{\beta}_n$.\\
$\bullet$  Use integrations by part like:\\
\[ y^1_n-a(x)y^{\beta +1}_n=d_n(y^1-a(x)y^{\beta +1})+{\partial}_na(x)y^{\beta +1}={\bar{y}}^1_n+{\partial}_na(x)y^{\beta +1}  \]
$\bullet$  Modify $y^1,...,y^{\beta} $ to ${\bar{y}}^1,...,{\bar{y}}^{\beta}$ in order to "{\it absorb}" the various $y^{\beta +1}_n,...,y^m_n$ {\it only appearing in the equations of class} n.  \\

\noindent
{\bf PROPOSITION 2.10}: The new equations of class $n$ only contain $y^{\beta +1}_i,...,y^m_i$ with $0\leq i\leq n-1$ while the equations of class $1,...,n-1$ no more contain $y^{\beta+1},...,y^m$ and their jets.\\

{\it Proof}: The first assertion comes from the absorption procedure. Now, if $y^m$ or $y^m_i$ should appear in an equation of class $\leq n-1$, prolonging this equation with respect to the non-multiplicative variable $x^n$ should bring $y^m_n$ or $y^m_{in}$ and (here involution is essential) we should get a linear combination of equations of various classes prolonged with respect to $x^1,...,x^{n-1}$ {\it only}, but this is impossible.\\
\hspace*{12cm} Q.E.D.  \\

A similar proof provides at once (See next Section for the definition):\\

\noindent
{\bf COROLLARY 2.11}: Any torsion element, if it exists, only depends on ${\bar{y}}^1,...,{\bar{y}}^{\beta}$.\\

For an involutive system of order $q$ in solved form, we shall use to denote by $y_{pri}$ the {\it principal jet coordinates}, namely the leading terms of the solved equations in the sense of involution. Accordingly, any formal derivative of a principal jet coordinate is again a principal jet coordinate. The remaining jet coordinates will be called {\it  parametric jet coordinates} and denoted by $y_{par}$. We shall use a "trick" in order to study the parametric jet coordinates. Indeed, the symbol of $j_q$ is the zero symbol and is thus trivially involutive at any order $q$. Accordingly, if we introduce the {\it multiplicative variables} $x^1,...,x^i$ for the parametric jets of order $q$ and class $i$, the formal derivative or a parametric jet of strict order $q$ and class $i$ by one of its multiplicative variables is uniquely obtained and cannot be a principal jet of order $q+1$ which is coming from a uniquely defined principal jet of order $q$ and class $i$. We have thus obtained the following technical Proposition which is very useful in actual practice: \\
   
\noindent
{\bf PROPOSITION 2.12}: The principal and parametric jets of strict order $q$ of an involutive system of order $q$ have the same Janet board if we extend it to all the classes that may exist for both sets, in particular the respective empty classes.   \\
 
\noindent
{\bf EXAMPLE 2.13}: With $n=3,m=1,q=2$, let us consider the linear second order system $R_2$ defined by the three PD equations\\
\[    {\Phi}^1\equiv Py=  y_{33}=0, \hspace{5mm}{\Phi}^2\equiv Qy=y_{23}-y_{11}=0,\hspace{5mm} {\Phi}^3\equiv Ry=y_{22}=0  \]
which is homogeneous and thus {\it automatically} formally integrable but $g_2$ is not involutive though finite type because $g_4=0$ (Exercise). Elementary computations of ranks of matrices shows that the $\delta$-map defined by a $3\times 3$ matrix:\\
\[    0\rightarrow  {\wedge}^2T^*\otimes g_3  \stackrel{\delta}{\longrightarrow} {\wedge}^3T^*\otimes g_2 \rightarrow 0  \]
is an isomorphism and $g_3$ is thus 2-acyclic, a crucial intrinsic property [38, 44] totally absent from any "old" work and that will be used in order to study the conformal group of space-time and Einstein equations. We have $y_{123} - y_{111}=0$ and thus $par =\{ y,y_1,y_2,y_3, y_{11},y_{12}, y_{13}, y_{111}\}$ with $ dim(R_3)=8=2^n$ according to Macaulay (See ([26, 48] for more details). Finally, comparing to the Poincar\'e sequence for ${\mathbb{R}}^3$, we notice the identities:  \\
\[   {\Psi}^1\equiv  Q{\Phi}^3-R{\Phi}^2=0, {\Psi}^2\equiv R{\Phi}^1-P{\Phi}^3=0, {\Psi}^3 \equiv P{\Phi}^2-Q{\Phi}^1= 0 \Rightarrow  
 P{\Psi}^1 + Q {\Psi}^2 + R{\Psi}^3 \equiv 0  \]
and obtain the strictly exact sequence made by second order operators:  \\
\[   0\rightarrow \Theta \rightarrow 1 \rightarrow 3 \rightarrow 3 \rightarrow 1 \rightarrow 0  \]
which is nevertheless far from being a Janet sequence because only $R_4$ is involutive.   \\

The main use of involution is to construct {\it canonical differential sequences} made up by successive {\it compatibility conditions} (CC). In particular, when $R_q$ is involutive, the linear differential operator ${\cal{D}}:E\stackrel{j_q}{\rightarrow} J_q(E)\stackrel{\Phi}{\rightarrow} J_q(E)/R_q=F=F_0$ of order $q$ with space of solutions $\Theta\subset E$ is said to be {\it involutive} and one has the canonical {\it linear Janet sequence} ([41], p 144):\\
\[  0 \longrightarrow  \Theta \longrightarrow E \stackrel{\cal{D}}{\longrightarrow} F_0 \stackrel{{\cal{D}}_1}{\longrightarrow}F_1 \stackrel{{\cal{D}}_2}{\longrightarrow} ... \stackrel{{\cal{D}}_n}{\longrightarrow} F_n \longrightarrow 0   \]
where each other operator is first order involutive and generates the CC of the preceding one while the {\it Janet bundles} are defined by:  \\
\[ F_r={\wedge}^rT^*\otimes J_q(E)/({\wedge}^rT^*\otimes R_q+\delta ({\wedge}^{r-1}T^*\otimes S_{q+1}T^*\otimes E))\]
For a later use in Section 4 and in the Conclusion, it is important to notice that the canonical Janet sequence, like the Poincar\'e sequence, can be "{\it cut at any place} ", that is can also be constructed anew from any intermediate operator. {\it The numbering of the Janet bundles has thus nothing to do with that of the Poincar\'{e} sequence for the exterior derivative}, contrary to what many physicists  still believe. Moreover, the fiber dimension of the Janet bundles can be computed at once inductively from the board of multiplicative and non-multiplicative variables that can be exhibited for $\cal{D}$ by working out the board for ${\cal{D}}_1$ and so on. For this, the number of rows of this new board is the number of dots appearing in the initial board while the number $nb(i)$ of dots in the column $i$ just indicates the number of CC of class $i$ for $i=1, ... ,n$ with $nb(i) < nb(j), \forall i<j$ and we have therefore:  \\

\noindent
{\bf THEOREM 2.14}: The successive first order operators ${\cal{D}}_1, ... , {\cal{D}}_n$ are {\it automatically} in reduced Spencer form. \\

\noindent
{\bf DEFINITION 2.15}: The Janet sequence is said to be {\it locally exact at} $F_r$ if any local section of $F_r$ killed by ${\cal{D}}_{r+1}$ is the image by ${\cal{D}}_r$ of a local section of $F_{r-1}$. It is called {\it locally exact} if it is locally exact at each $F_r$ for $0\leq r \leq n$. The Poincar\'{e} sequence is locally exact, that is a closed form is locally an exact form but counterexamples may exist ([38], p 373). More generally, a differential sequence is said to be {\it formally exact} if each operator involved generates the CC of the preceding one. It is said to be {\it strictly exact} ({\it involutive}) if all the operators are also formally integrable (involutive). It is said to be {\it canonical} if it is strictly exact {\it and} all the operators can be defined by a {\it single formula}, that is "{\it altogether} " and not only "{\it step by step} ".  \\

In actual practice, the following theorem will be of constant use, in particular for systems with constant coefficients that are not involutive [40,54]:   \\

\noindent
{\bf THEOREM 2.16}: If a differential operator ${\cal{D}}=\Phi\circ j_q:E \longrightarrow F_0$ is such that $R_q=ker(\Phi)$ is formally integrable and $s\geq 0$ is the smallest number of prolongations needed in such a way that the symbol $g_{q+s}={\rho}_s(g_q)$ becomes $2$-acyclic, then the order of the generating CC ${\cal{D}}_1:F_0 \longrightarrow F_1$ is equal to $s+1$.  \\

\noindent
{\bf EXAMPLE 2.17}: When studying the conformal Killing system ${\cal{D}}\xi={\cal{L}}(\xi) \omega=A(x)\omega$ for the Euclidean metric $\omega$, obtained by eliminating the function factor $A(x)$, we shall see in Section 4 that $F_0=\{ ({\Omega}_{ij})\in S_2T^*\mid tr(\Omega)\equiv {\omega}^{ij}{\Omega}_{ij}=0\}$ and this second order system is trivially formally integrable because it is an homogeneous system with constant coefficients.
We have the commutative diagram with exact rows and exact $\delta$-columns but the first:  \\

\[ \begin{array}{rcccccccl}
   & 0  &  & 0 & & 0 & &  & \\
   & \downarrow & & \downarrow & & \downarrow &  & & \\
 0\rightarrow &  g_4 & \rightarrow &S_4T^*\otimes T & \rightarrow & S_3T^*\otimes F_0 & \rightarrow & F_1 & \rightarrow 0 \\
&\downarrow &  &\downarrow & & \downarrow & &   \\
0\rightarrow & T^*\otimes g_3 & \rightarrow & T^*\otimes S_3T^*\otimes T & \rightarrow & T^*\otimes S_2T^*\otimes F_0 & \rightarrow & 0 &  \\
&\downarrow &  &\downarrow & & \downarrow & &   \\
0\rightarrow & {\wedge}^2T^*\otimes g_2 & \rightarrow & {\wedge}^2T^*\otimes S_2T^*\otimes T & \rightarrow & {\wedge}^2T^*\otimes T^*\otimes F_0 & \rightarrow & 0 &  \\
&\downarrow &  &\downarrow & & \downarrow & &  & \\
0\rightarrow & {\wedge}^3T^*\otimes g_1 & \rightarrow & {\wedge}^3T^*\otimes T^*\otimes T & \rightarrow & {\wedge}^3T^*\otimes F_0 & \rightarrow & 0 &  \\
&\downarrow &  &\downarrow & & \downarrow & &  & \\
 & 0  & & 0 & & 0 & & &
\end{array}  \]
leading to the short exact sequence $0 \rightarrow F_1 \rightarrow {\wedge}^2T^*\otimes g_2 \stackrel{\delta}{\longrightarrow} {\wedge}^3T^*\otimes g_1 \rightarrow 0$ with $F_1=H^2_2(g_1)\neq 0$. We have $dim(g_1)=4, dim(g_2)=3,g_3=0\Rightarrow g_4=0$ and respective fiber dimensions:  \\

\[ \begin{array}{rcccccccl}
   &   &  & 0 & & 0 & &  & \\
   &  & & \downarrow & & \downarrow &  & & \\
&  0 & \rightarrow &45 & \rightarrow & 50 & \rightarrow &5 & \rightarrow 0 \\
&&  &\downarrow & & \downarrow & &   \\
0\rightarrow & 0 & \rightarrow & 90 & \rightarrow & 90 & \rightarrow & 0 &  \\
&\downarrow &  &\downarrow & & \downarrow & &   \\
0\rightarrow & 9& \rightarrow & 54& \rightarrow & 45 & \rightarrow & 0 &  \\
&\downarrow &  &\downarrow & & \downarrow & &  & \\
0\rightarrow & 4 & \rightarrow & 9 & \rightarrow & 5& \rightarrow & 0 &  \\
&\downarrow &  &\downarrow & & \downarrow & &  & \\
 & 0  & & 0 & & 0 & & &
\end{array}  \]
It follows that $g_2$ and thus $g_1$ {\it cannot} be $2$-acyclic while $g_3={\rho}_2(g_1)=0$ is trivially involutive with $q=1,r=2$. Moreover, in order to convince the reader about the powerfulness of these new methods, we invite him to prove that $H^2_1(g_1)=0$ by exhibiting the short exact sequence $0 \rightarrow T^*\otimes g_2\stackrel{\delta}{\longrightarrow} {\wedge}^2T^*\otimes g_1\stackrel{\delta}{\longrightarrow} {\wedge}^3T^*\otimes T \rightarrow 0$. It does not seem that these Vessiot structure equations of order $3$ are known but this result has been recently checked by A. Quadrat with new computer algebra packages [54].  \\

\noindent
{\bf DEFINITION 2.18}: If $\chi={\chi}_idx^i \in T^*$ and we set ${\chi}_{\mu}=({\chi}_1)^{{\mu}_1}... ({\chi}_n)^{{\mu}_n}$, the map ${\sigma}_{\chi}({\cal{D}}): E \rightarrow F$ defined by the matrix $a^{\tau \mu}_k(x){\chi}_{\mu}$ is called the {\it symbol} of ${\cal{D}}$ at 
$\chi \in T^*$ and we have ([38], p 155-160):  \\

\noindent
{\bf   THEOREM 2.19}: If ${\cal{D}}$ is involutive, the {\it symbol sequence}:  \\
\[  0 \rightarrow ker({\sigma}_{\chi}({\cal{D}})) \rightarrow E \stackrel{{\sigma}_{\chi}({\cal{D}})}{\longrightarrow} F_0 \stackrel{{\sigma}_{\chi}({\cal{D}}_1)}{\longrightarrow}  ...  \stackrel{{\sigma}_{\chi}({\cal{D}}_n)}{\longrightarrow} F_n \rightarrow 0  \]
is exact if and only if $\chi \in T^*$ is such that the rank of ${\sigma}_{\chi}({\cal{D}})$ has its maximum value. \\
 
\noindent
{\bf COROLLARY 2.20}: We have:   \\
\[ {\sum}_{r=0}^n (-1)^r dim(F_r)=m-\alpha=\beta >0 \hspace{4mm}\Leftrightarrow \hspace{4mm} {\sum}_{r=0}^n(-1)^rdim(C_r)=\alpha  \]  \\

\noindent
{\bf EXAMPLE 2.21}: ([26],$\S 38$, p 40 where one can find the first intuition of formal integrability) With $n=3,m=1, q=2$, the system $y_{11}=0, y_{13}-y_2=0$ is neither formally integrable nor involutive. Indeed, we get $d_3y_{11}-d_1(y_{13}-y_2)=y_{12}$ and $d_3y_{12}-d_2(y_{13}-y_2)=y_{22}$, that is to say {\it each first and second} prolongation does bring a new second order PD equation. Considering the new system $y_{22}=0, y_{12}=0, y_{13}-y_2=0, y_{11}=0$, the question is to decide whether this system is involutive or not. One could use Janet/Gr\"{o}bner algorithm but with no insight towards involution. In such a simple situation, as there is no PD equation of class $3$, the evident permutation of coordinates $(1,2,3)\rightarrow (3,2,1)$ provides the following involutive second order system with one equation of class $3$, $2$ equations of class $2$ and $1$ equation of clas $1$:   \\
\[  \left\{  \begin{array}{lcl}
{\Phi}^4 \equiv y_{33}  & = & 0  \\
{\Phi}^3 \equiv y_{23} & = & 0  \\
{\Phi}^2 \equiv y_{22} & = &  0  \\
{\Phi}^1 \equiv y_{13}-y_2 & = &  0 
\end{array}
\right. \fbox{$\begin{array}{lll}
1 & 2 & 3 \\
1 & 2 & \bullet \\
1 & 2 & \bullet \\
1 & \bullet & \bullet
\end{array}$}  \]
We have ${\alpha}={\alpha}^3_2=0,{\alpha}^2_2=0,{\alpha}^1_2=2$ and the corresponding CC system is easily seen to be the following involutive first order system in reduced Spencer form:  \\
\[  \left\{  \begin{array}{lcl}
{\Psi}^4 \equiv  d_3{\Phi}^3-d_2{\Phi}^4  & = & 0  \\
{\Psi}^3 \equiv  d_3{\Phi}^2-d_2{\Phi}^3  & = & 0  \\
{\Psi}^2 \equiv  d_3{\Phi}^1-d_1{\Phi}^4+{\Phi}^3 & = &  0  \\
{\Psi}^1 \equiv  d_2{\Phi}^1-d_1{\Phi} ^3+ {\Phi}^2   & = &  0 
\end{array}
\right. \fbox{$\begin{array}{lll}
1 & 2 & 3  \\
1 & 2 & 3  \\
1 & 2 & 3  \\
1 & 2 & \bullet
\end{array}$}  \]
The final CC system is the involutive first order system in reduced Spencer form:  \\
\[  \left\{  \begin{array}{lcl}
\Omega \equiv  d_3{\Psi}^1-d_2{\Psi}^2+d_1{\Psi}^4-{\Psi}^3   & = & 0  
\end{array}
\right. \fbox{$\begin{array}{lll}
1 & 2 & 3 
\end{array}$}  \]
We get therefore the Janet sequence:    
\[    0 \longrightarrow  \Theta \longrightarrow 1 \stackrel{{\cal{D}}}{\longrightarrow} 4 \stackrel{{\cal{D}}_1}{\longrightarrow} 4 \stackrel{{\cal{D}}_2}{\longrightarrow} 1  \longrightarrow  0    \]
and check that the Euler-Poincar\'{e} characteristic, that is the alternate sum of dimensions of the Janet bundles, is $1-4+4-1=\alpha=0$ and thus $det({\sigma}_{\chi}({\cal{D}}_1))=0 $. Using the fact that $d_{33}$ commutes with $d_{13} - d_2$, we get the formally exact sequence $0 \rightarrow \Theta \rightarrow 1 \rightarrow 2 \rightarrow 1  \rightarrow 0$ with again $1-2+1=0$, which is formally exact but not strictly exact and thus far from being a Janet sequence.  \\

Equivalently, we have the involutive {\it first Spencer operator} $D_1:C_0=R_q\stackrel{j_1}{\rightarrow}J_1(R_q)\rightarrow J_1(R_q)/R_{q+1}\simeq T^*\otimes R_q/\delta (g_{q+1})=C_1$ of order one induced by the {\it Spencer operator} $D:R_{q+1}\rightarrow T^*\otimes R_q:{\xi}_{q+1} \rightarrow j_1({\xi}_q)-{\xi}_{q+1}=\{ {\partial}_i{\xi}^k_{\mu}-{\xi}^k_{\mu +1_i}\mid 0\leq \mid \mu \mid q \}$ which is well defined because both $J_{q+1}(E)$ and $T^*\otimes J_q(E)$ may be considered as sub-bundles of $J_1(J_q(E))$. Introducing the {\it Spencer bundles} $C_r={\wedge}^rT^*\otimes R_q/{\delta}({\wedge}^{r-1}T^*\otimes g_{q+1})$, the first order involutive ($r+1$)-{\it Spencer operator} $D_{r+1}:C_r\rightarrow C_{r+1}$ is induced by $D:{\wedge}^rT^*\otimes R_{q+1}\rightarrow {\wedge}^{r+1}T^*\otimes R_q:\alpha\otimes {\xi}_{q+1}\rightarrow d\alpha\otimes {\xi}_q+(-1)^r\alpha\wedge D{\xi}_{q+1}$. Indeed, differentiating the first equation below and substracting the second, we have:  \\
\[  a^{\tau \mu}_k(x){\xi}^k_{\mu}(x)\equiv 0, \hspace{3mm} a^{\tau \mu}_k(x){\xi}^k_{\mu +1_i}(x)+ {\partial}_ia^{\tau \mu}_k(x) {\xi}^k_{\mu}(x)\equiv 0 \Rightarrow  a^{\tau \mu}_k(x)({\partial}_i{\xi}^k_{\mu}(x) - {\xi}^k_{\mu + 1_i}(x))\equiv 0   \]
We obtain therefore the canonical {\it linear Spencer sequence} ([14], p 150):\\
\[    0 \longrightarrow \Theta \stackrel{j_q}{\longrightarrow} C_0 \stackrel{D_1}{\longrightarrow} C_1 \stackrel{D_2}{\longrightarrow} C_2 \stackrel{D_3}{\longrightarrow} ... \stackrel{D_n}{\longrightarrow} C_n\longrightarrow 0  \]
as the canonical Janet sequence for the first order involutive system $R_{q+1}\subset J_1(R_q)$.\\

The canonical Janet sequence and the canonical Spencer sequence are both induced by the Spencer operator along the following comutative diagrams ( See [41], p 391 for details):  \\

\[   \begin{array}{ccccccc}
{\wedge}^rT^*\otimes R_{q+1} &  \stackrel{D}{\longrightarrow}  &  {\wedge}^{r+1}T^*\otimes R_q   &  \hspace{1cm}  &
{\wedge}^rT^*\otimes J_{q+1}(E)  &   \stackrel{D}{\longrightarrow}  & {\wedge}^{r+1}T^*\otimes J_q(E)   \\
         \downarrow  &  &   \downarrow   &   &   \downarrow &   &  \downarrow      \\
 C_r &  \stackrel{D_{r+1}}{\longrightarrow}  &  C_{r+1}  &  \hspace{1cm}  &   F_r  &  \stackrel{{\cal{D}}_{r+1}}{\longrightarrow}&  F_{r+1}  \\
    \downarrow & & \downarrow &    &  \downarrow &   & \downarrow    \\
    0  &  &  0  &  &  0  &   &  0     

\end{array}   \]

They can be connected by a commutative diagram (See Section 3) where the Spencer sequence is induced by the locally exact central horizontal sequence which is at the same time the Janet sequence for $j_q$ and the Spencer sequence for $J_{q+1}(E)\subset J_1(J_q(E))$ ([41], p 153). Surprisingly, this result will become a key piece of machinery for the applications of Section 4 (See [50-55] for recent papers providing more details on applications of these results to engineering and mathematical physics, in particular continuum mechanics, gauge theory and general relativity).  \\

For a later use and in order to explain a result provided in the Introduction, we have:  \\

\noindent
{\bf PROPOSITION 2.22}: The Spencer sequence for the Lie operator describing the infinitesimal action of of a Lie group $G$ is 
(locally) isomorphic to the tensor product of the Poincar\'e sequence by the Lie algebra ${\cal{G}}=T_e(Þ)$ where $e\in G$ is the identity element.  \\

{\it Proof}: Using the notations of the Introduction, we may introduce a basis $\{ {\theta}_{\tau}={\theta}^i_{\tau}(x){\partial}_i \}$ of infinitesimal generators of the action with $\tau=1,...,dim(G)$ and the commutation relations $[{\theta}_{\rho},{\theta}_{\sigma}]=c^{\tau}_{\rho \sigma}{\theta}_{\tau} $ discovered by S. Lie giving the {\it structure constants} c of ${\cal{G}}$ (See [41] and [55] for more details). Hence any element $\lambda \in {\cal{G}}$ can be written $\lambda=\{{\lambda}^{\tau}=cst \}$. Gauging such an element, that is to say {\it replacing the constants by functions} or, equivalently, introducing a map $X\rightarrow {\wedge}^0T^*\otimes {\cal{G}}:(x) \rightarrow ({\lambda}^{\tau}(x))$, we may introduce locally a map ${\wedge}^0T^*\otimes {\cal{G}}\rightarrow T: {\lambda}^{\tau}(x) \rightarrow {\lambda}^{\tau}(x){\theta}^k_{\tau}(x)$ or, equivalently, vector fields $\xi=({\xi}^i(x){\partial}_i)\in T$ of the form ${\xi}^k(x)={\lambda}^{\tau}(x){\theta}^k_{\tau}(x)$, keeping the index $i$ for $1$-forms. More generally, we can introduce a map :  \\
\[  {\wedge}^rT^*\otimes {\cal{G}} \rightarrow {\wedge}^rT^*\otimes J_q(T)=\lambda \rightarrow \lambda \otimes  j_q(\theta)=X_q: {\lambda}^{\tau}(x) \rightarrow {\lambda}^{\tau}(x) {\partial}_{\mu}{\theta}^k_{\tau}(x) ={X}^k_{\mu,I}(x)dx^I  \]
that we can lift to the element $\lambda \otimes  j_{q+1}(\theta)=X_{q+1}\in {\wedge}^r T^*\otimes J_{q+1}(T)$. It follows from the definitions that $D_rX_q=DX_{q+1}$ by introducing any element of $C_r(T)$ through its representative $X_q\in {\wedge}^rT^*\otimes J_q(T)$. We obtain therefore the {\it crucial formula}:  \\
\[  D_rX_q=DX_{q+1}=D(\lambda \otimes j_{q+1}(\theta))=d\lambda\otimes j_q(\theta)+ (-1)^r\lambda \wedge Dj_{q+1}(\theta)=d\lambda \otimes j_q(\theta)   \]
allowing to identify, {\it at least locally}, the Spencer sequence for $j_q$ with the Poincar\'e sequence. We let the reader prove that the 
map ${\wedge}^0T^*\otimes {\cal{G}}\rightarrow J_q(T)$ is injective when the action is effective. We obtain therefore an isomorphism 
${\wedge}^0T^*\otimes {\cal{G}} \rightarrow R_q\subset J_q(T)$ when $q$ is large enough allowing to exhibit, {\it again at least locally}, an isomorphism between the canonical Spencer sequence and the tensor product of the Poincar\'e sequence by ${\cal{G}}$ when $q$ is large enough in such a way that $R_q$ is involutive, that is $g_q=0$. As shown in the Introduction, it is finally important to notice that such a property does not exist for the canonical Janet sequence.  \\
\hspace*{12cm}  Q.E.D.   \\

\noindent
{\bf REMARK 2.23}: We now provide the explicit form of the $n$ finite nonlinear {\it elations} of the conformal group of transformations and their infinitesimal counterpart with $ \forall 1\leq r,s,t \leq n$:\\
\[  y=\frac{x-x^2b}{1-2(bx)+b^2x^2} \hspace{5mm}  \Rightarrow \hspace{5mm} {\theta}_s= - \frac{1}{2} x^2 {\delta}^r_s{\partial}_r+{\omega}_{st}x^tx^r{\partial}_r  \hspace{5mm} \Rightarrow  \hspace{5mm}A_i(x)={\omega}_{si}{\lambda}^s(x),   \]
where the underlying metric is used for the scalar products $x^2,bx,b^2$ involved. The complexity of the corresponding formulas explains why the previous result showing the importance of second order jets have not been already known and used ([53], p 35).  \\

\noindent
{\bf EXAMPLE 2.24}: In order to emphasize the importance of dealing with vector bundles in the differential geometric setting of this section and with differential fields or projective modules in the differential algebraic setting of the next section, we provide a tricky example of a linear system with coefficients in a true differential field which is not just a field of rational functions in the independent variables. With $n=2, m=1, q=2$, let us consider the {\it non-linear second order system} 
${\cal{R}}_2$:  \\
\[\left\{  \begin{array}{lc}
 y_{22}-\frac{1}{3}(y_{11})^3&=0\\
  y_{12}-\frac{1}{2}(y_{11})^2&=0  
  \end{array}
  \right.  \fbox{$\begin{array}{cc}
  1 & 2 \\
  1& \bullet
  \end{array}$}     \]
obtained by equating to zero two differential polynomials. Doing crossed derivatives, it is easy to check that the system is involutive and allows to define a true differential extension $K$ of $k=\mathbb{Q}$ which is isomorphic to $k(y,y_1,y_2,y_{11}, y_{111},... )$ if we set for example $d_2y_1=y_{12}=\frac{1}{2}(y_{11})^2$ and so on. By linearization, we get the following {\it linear second order involutive system} $R_2$ defined over $K$:  \\
\[\left\{  \begin{array}{lc}
 Y_{22}-(y_{11})^2Y_{11}&=0\\
  Y_{12}-y_{11}Y_{11}&=0  
  \end{array}
  \right.  \fbox{$\begin{array}{cc}
  1 & 2 \\
  1& \bullet
  \end{array}$}     \]
The various symbols of the first system are vector bundles over ${\cal{R}}_2$ while the symbols of the second system are vector spaces over $K$. As an exercise in order to understand the problems that may arise in general, we invite the reader to study similarly the non-linear second order system $y_{22}-\frac{1}{2}(y_{11})^2=0, y_{12}-y_{11}=0$ and conclude. The interested reader may look at ([41],VI.B.3 ,p 273 and VI.B.7 p 275) for criteria providing differential fields and based on the Spencer $2$-acyclicity property of the symbol at order $q$ ([41], Prop. III.1.3, p 92 and Theorem III.C.1, p 95).  \\

\newpage

\noindent
{\bf 3) DIFFERENTIAL MODULES} \\

As a rough motivation for introducing modules and residues, let us recall that $\sqrt{3}=1,732...$ or $e^x=1+x+\frac{x^2}{2}+ ...$ cannot be stored on a computer. In order to avoid such a difficulty, we may introduce the field $K=\mathbb{Q}$ and a {\it polynomial} $P\equiv y^2-3\in K[y]$ or a (linear) {\it differential polynomial} $P\equiv dy-y\in K[d]y=Dy$ in order to consider the (prime) {\it ideal} $\mathfrak{p}\subset K[y]$ generated by $P$ or the {\it differential module of equations} $I\subset Dy$ generated similarly by $P$ in the corresponding short exact sequences of residues where $M=K[y]/\mathfrak{p}$ on one side and $M=Dy/DI$ on the other side:  \\
\[   0 \rightarrow \mathfrak{p} \rightarrow K[y] \stackrel{p}{\longrightarrow} M \rightarrow 0, \hspace{1cm}  
0 \rightarrow I \rightarrow Dy \stackrel{p}{\longrightarrow} M  \rightarrow 0  \]
while calling $\bar{y}$ the image of $y$ under the canonical projection $p$. Of course $\bar{y}$ can be denoted by {\it any other symbol} like $\eta$ or $\sqrt{3}$ or $e^x$ and the only problem will be to add the word "{\it differential} " in concepts coming from pure algebra. A similar approach has been use in " {\it differential algebra} " for dealing with nonlinear differential polynomials (See [21, 44] for more details).  \\

Let $A$ be a {\it unitary ring}, that is $1,a,b\in A \Rightarrow a+b,ab \in A, 1a=a1=a$ and even an {\it integral domain} ($ab=0\Rightarrow a=0$ or $b=0$) with {\it field of fractions} $K=Q(A)$. However, we shall not always assume that $A$ is commutative, that is $ab$ may be different from $ba$ in general for $a,b\in A$. We say that $M={}_AM$ is a {\it left module} over $A$ if $x,y\in M\Rightarrow ax,x+y\in M, \forall a\in A$ or a {\it right module} $M_B$ over $B$ if the operation of $B$ on $M$ is $(x,b)\rightarrow xb, \forall b\in B$. If $M$ is a left module over $A$ and a right module over $B$ with $(ax)b=a(xb), \forall a\in A,\forall  b\in B, \forall x\in M$, then we shall say that $M={ }_AM_B$ is a {\it bimodule}. Of course, $A={ }_AA_A$ is a bimodule over itself. The category of left modules over $A$ will be denoted by $mod(A)$ while the category of right modules over $A$ will be denoted by $mod(A^{\it op})$. We define the {\it torsion submodule} $t(M)=\{x\in M\mid \exists 0\neq a\in A, ax=0\}\subseteq M$ and $M$ is a {\it torsion module} if $t(M)=M$ or a {\it torsion-free module} if $t(M)=0$. We denote by $hom_A(M,N)$ the set of morphisms $f:M\rightarrow N$ such that $f(ax)=af(x)$. In particular $hom_A(A,M)\simeq M$ because $f(a)=af(1)$ and we recall that a sequence of modules and maps is exact if the kernel of any map is equal to the image of the map preceding it. \\

When $A$ is commutative, $hom(M,N)$ is again an $A$-module for the law $(bf)(x)=f(bx)$ as we have $(bf)(ax)=f(bax)=f(abx)=af(bx)=a(bf)(x)$. In the non-commutative case, things are more complicate and we have:\\

\noindent
{\bf LEMMA 3.1}: Given ${}_AM$ and ${}_AN_B$, then $hom_A(M,N)$ becomes a right module over $B$ for the law $(fb)(x)=f(x)b$. Similarly, given 
${Ê}_AM_B$ and ${ }_AN$, then $hom_A(M,N)$ becomes a left module over $B$ for the law $(bf)(x)=f(xb)$.\\

\noindent
{\it Proof}:  In order to prove the first result we just need to check the two relations:
\[ (fb)(ax)=f(ax)b=af(x)b=a(fb)(x),\]
\[ ((fb')b")(x)=(fb')(x)b"=f(x)b'b"=(fb'b")(x).\]
The proof of the second result could be achieved similarly.  \\
\hspace*{12cm}               Q.E.D. \\
 
\noindent
{\bf DEFINITION 3.2}: A module $F$ is said to be {\it free} if it is isomorphic to a (finite) power of $A$ called the {\it rank} of $F$ over $A$ and denoted by $rk_A(F)$ while the rank $rk_A(M)$ of a module $M$ is the rank of a maximum free submodule $F\subset M$. It follows from this definition that $M/F$ is a torsion module. In the sequel we shall only consider {\it finitely presented} modules, namely {\it finitely generated} modules defined by exact sequences of the type $F_1 \stackrel{d_1}{\longrightarrow} F_0 \longrightarrow M\longrightarrow 0$ where $F_0$ and $F_1$ are free modules of finite ranks $m_0$ and $m_1$ often denoted by $m$ and $p$ in examples. \\
A module $P$ is called {\it projective} if there exists a free module $F$ and another (projective) module $Q$ such that $P\oplus Q\simeq F$. Accordingly, a {\it projective (free) resolution} of $M$ is a long exact sequence $... \stackrel{d_3}{\longrightarrow} P_2 \stackrel{d_2}{\longrightarrow} P_1 \stackrel{d_1}{\longrightarrow} P_0 \stackrel{p}{\longrightarrow} M \longrightarrow 0 $ where $P_0, P_1, P_2, ... $ are projective (free) modules, $M=coker(d_1)=P_0/ im(d_1)$ and $p$ is the canonical projection. Such a situation may be shortly described by $X \stackrel{p}{\rightarrow } M \rightarrow 0$ where $X$ is a {\it complex} that may not be exact in general. \\
We have the useful proposition that we shall only prove in the commutative case [44]: \\

\noindent
{\bf PROPOSITION 3.3}: For any short exact sequence $0\rightarrow M' \stackrel{f}{\longrightarrow} M \stackrel{g}{\longrightarrow} M" \rightarrow 0$, we have the relation $rk_A(M)=rk_A(M')+rk_A(M")$.\\

\noindent
{\it Proof}: Whenever $x\in M$ and $0\neq s\in A$, the image of $\frac{1}{s}\otimes x\in K{\otimes}_AM$ induced by $g$ in $K{\otimes}_A$ is:  \\
\[  \frac{1}{s}\otimes g(x)=\frac{t}{st}\otimes g(x)=\frac{1}{st}\otimes tg(x)=\frac{1}{st}g(tx), \hspace{1cm}  \forall 0\neq t\in A  \]
Hence, the kernel of $g:K{\otimes}_AM \rightarrow K{\otimes}_AM"$ is made by all previous elements $x\in M$ such that $\exists 0\neq t\in A$ with $g(tx)=tg(x)=0$. As the initial sequence is exact, we may find $x' \in M'$ such that $tx=f(x')$, a result leading to:  \\
\[  \frac{1}{s}\otimes x=\frac{t}{st}\otimes x=\frac{1}{st}\otimes tx=\frac{1}{st}\otimes f(x')  \]
and the kernel of $g:K{\otimes}_AM \rightarrow K{\otimes }_AM"$ is thus equal to the image of $f:K{\otimes}_AM'\rightarrow 
K{\otimes}_AM$.  \\
Hence, if $F$ is a maximum free submodule of $M$, we have the quite useful short exact sequence $0 \rightarrow F \rightarrow M \rightarrow M/F \rightarrow 0$ where $M/F$ is a torsion module over $A$ and $K{\otimes}_AM/F=0 \Rightarrow K{\otimes}_AF\simeq K{\otimes}_AM \Rightarrow rk_A(M)=dim_K(K{\otimes}_AM)$ because, if we have $t\bar{x}=0$ with $t\neq 0$, we have thus:  \\
\[  \frac{1}{s}\otimes \bar{x}= \frac{t}{st}\otimes \bar{x}=\frac{1}{st}\otimes t\bar{x}=0 .\]   \\
\hspace*{12cm}   Q.E.D.   \\

Then, tensoring by $M$ over $A$ the short exact sequence $0 \rightarrow A \rightarrow K \rightarrow K/A \rightarrow 0$, we obtain the other useful long exact sequence:  \\
\[   0  \rightarrow t(M) \longrightarrow M \longrightarrow K{\otimes}_A M \rightarrow K/A {\otimes }_A M  \rightarrow 0  \]

The following proposition will be used many times in Section $4$, in particular for exhibiting the Weyl tensor from the Riemann 
tensor ([4],p 73)([61],p 33) :\\

\noindent
{\bf PROPOSITION 3.4}: If one has a short exact sequence:
\[0\longrightarrow
M' \stackrel{\stackrel{u}{\longleftarrow}}{\stackrel{f}{\longrightarrow}} M
     \stackrel{\stackrel{v}{\longleftarrow}}{\stackrel{g}{\longrightarrow}} M''   \longrightarrow 0  \]
then the three following conditions are equivalent:\\
$\bullet$ There exists a monomorphism $v:M''\rightarrow M$ called {\it lift} of $g$ and such that $g\circ v=id_{M''}$ .\\
$\bullet$ There exists an epimorphism $u:M\rightarrow M'$ called {\it lift} of $f$ and such that $u\circ f=id_{M'}$.\\
$\bullet$ There exist isomorphisms $\varphi=(u,g):M\rightarrow M'\oplus M''$ and $\psi=f+v:M'\oplus M''\rightarrow M$ that are inverse to each other and provide an isomorphism $M\simeq M'\oplus M''$ with $f\circ u+v\circ g=id_M$ and thus $ker(u)=im(v)$.  \\

\noindent
{\it Proof}: When $u$ is given with $u\circ f=id_{M'}$, the only tricky point is to induce $v$ by chasing in the following commutative and exact diagram:  \\

\[  \begin{array}{rcccccl}
  & 0 &  & 0  &  & 0 &  \\
  & \downarrow  &  &  \downarrow  &  &  \downarrow  &    \\
  0 \rightarrow & M' &   \stackrel{\stackrel{u}{\longleftarrow}}{\stackrel{f}{\longrightarrow}}  
 &  M & \stackrel{g}{\longrightarrow}  & M" & \rightarrow 0  \\
    &  \parallel &   & \parallel & & \hspace{2mm}{\downarrow} v &   \\
    0 \rightarrow & M' & \stackrel{f}{\longrightarrow} & M & \stackrel{id_M -f\circ u}{\longrightarrow} & M   \\
   &   \downarrow & & \downarrow  & &  & \\
   &  0 &  &  0 &  &  &
\end{array}   \]
Then, for any $x"\in M"$, we can find $x\in M$ such that $g(x)=x"$ and we obtain $g\circ v(x")=g\circ v\circ g(x)=g(x)=x"$ because $g\circ f=0$ and thus $g\circ v=id_{M"}$. It follows that $f\circ u\circ v+v\circ g\circ v=v\Rightarrow f\circ (u \circ v)=0 \Rightarrow u\circ v=0$ because $f$ is a monomorphism.  \\
\hspace*{12cm}  Q.E.D.   \\

\noindent

\noindent
{\bf DEFINITION 3.5}: In the above situation, we say that the short exact sequence {\it splits}. The short exact sequence $0 \rightarrow \mathbb{Z} \rightarrow  \mathbb{Q} \rightarrow \mathbb{Q}/\mathbb{Z} \rightarrow 0$ cannot split over $\mathbb{Z}$. \\

\noindent
{\bf DEFINITION 3.6}: A resolution of a short exact sequence $0 \rightarrow M' \stackrel{f}{\longrightarrow} M \stackrel{g}{\longrightarrow} M" \rightarrow 0 $ of $A$-modules is a short exact sequence $0 \rightarrow X' \stackrel{f}{\longrightarrow} X \stackrel{g}{\longrightarrow} X" \rightarrow 0 $ of exact complexes such that $X \stackrel{p}{\longrightarrow} M \rightarrow 0, X' \stackrel{p'}{\longrightarrow} M' \rightarrow 0, X" \stackrel{p"}{\longrightarrow} M" \rightarrow 0$ are resolutions and we shall say that the sequence of complexes is over the sequence of modules. Such a definition can also be used when the complexes are not exact and we have the long exact {\it connecting sequence} $... \rightarrow H_i(X) \rightarrow H_i(X") \rightarrow H_{i-1}(X') \rightarrow ... $ if we introduce the homology $H_i(X)$ of a decreasing complex $X_{i+1} \rightarrow X_i \rightarrow X_{i-1} $ with a similar result for the cohomology of increasing complexes. In particular, if any two are exact, the third is exact too ([44], Theorem II.1.15, p 196-203).  \\

Using the notation $M^*=hom_A(M,A)$, for any morphism $f:M\rightarrow N$, we shall denote by $f^*:N^*\rightarrow M^*$ the morphism which is defined by  $f^*(h)=h\circ f, \forall h\in hom_A(N,A)$ and satisfies $rk_A(f)=rk_A(im(f))=rk_A(f^*),\forall f\in hom_A(M,N)$(See [45], Corollary 5.3, p 179). We may take out $M$ in order to obtain the {\it deleted sequence} $... \stackrel{d_2}{\longrightarrow} P_1 \stackrel{d_1}{\longrightarrow} P_0 \longrightarrow 0$ and apply  $hom_A(\bullet,A)$ in order to get the sequence $... \stackrel{d^*_2}{\longleftarrow} P^*_1 \stackrel{d^*_1}{\longleftarrow} P^*_0 \longleftarrow 0$. \\

\noindent
{\bf PROPOSITION 3.7}: The {\it extension modules}  $ext^0_A(M)=ker(d^*_1)=hom_A(M,A)=M^*$ and $ext^i_A(M)=ker(d^*_{i+1})/im(d^*_i), \forall i\geq 1$ do not depend on the resolution chosen and are torsion modules for $i\geq 1$. Using $hom_A(\bullet,N)$, one can similarly define $ext^i_A(M,N)$ with $ext^0_A(,N)=hom_A(M,N)$ and the $ext^i_A(M,N)$ vanish $\forall i>0$ whenever $M$ is a projective module (See [8, 17, 32, 33, 44, 51, 61] for more details). \\

Let $ A$ be a {\it differential ring}, that is a commutative ring with $n$ commuting {\it derivations} $\{{\partial}_1,...,{\partial}_n\}$, that is ${\partial}_i{\partial}_j={\partial}_j{\partial}_i={\partial}_{ij}, \forall i,j=1,...,n$ while ${\partial}_i(a+b)={\partial}_ia+{\partial}_ib$ and ${\partial}_i(ab)=({\partial}_ia)b+a{\partial}_ib, \forall a,b\in A$. We shall use thereafter a differential integral domain $A$ with unit $1\in A$ whenever we shall need a {\it differential field} $\mathbb{Q}\subset K=Q(A)$ of coefficients, that is a field ($a\in K\Rightarrow 1/a\in K$) with ${\partial}_i(1/a)=-(1/a^2){\partial}_ia$, in order to exhibit solved forms for systems of partial differential equations as in the preceding section. Using an implicit summation on multi-indices, we may introduce the (noncommutative) {\it ring of differential operators} $D=A[d_1,...,d_n]=A[d]$ with elements $P=a^{\mu}d_{\mu}$ such that $\mid \mu\mid<\infty$ and $d_ia=ad_i+{\partial}_ia$. The highest value of ${\mid}\mu {\mid}$ with $a^{\mu}\neq 0$ is called the {\it order} of the {\it operator} $P$ and the ring $D$ with multiplication $(P,Q)\longrightarrow P\circ Q=PQ$ is filtred by the order $q$ of the operators. We have the {\it filtration} $0=D_{-1}\subset D_0\subset D_1\subset  ... \subset D_q \subset ... \subset D_{\infty}=D$. Moreover, it is clear that $D$, as an algebra, is generated by $A=D_0$ and $T=D_1/D_0$ with $D_1=A\oplus T$ if we identify an element $\xi={\xi}^id_i\in T$ with the vector field $\xi={\xi}^i(x){\partial}_i$ of differential geometry, but with ${\xi}^i\in A$ now. It follows that $D={ }_DD_D$ is a {\it bimodule} over itself, being at the same time a left $D$-module ${ }_DD$ by the composition $P \longrightarrow QP$ and a right $D$-module $D_D$ by the composition $P \longrightarrow PQ$ with $D_rD_s=D_{r+s}, \forall r,s \geq 0$ in any case. \\

If we introduce {\it differential indeterminates} $y=(y^1,...,y^m)$, we may extend $d_iy^k_{\mu}=y^k_{\mu+1_i}$ to ${\Phi}^{\tau}\equiv a^{\tau\mu}_ky^k_{\mu}\stackrel{d_i}{\longrightarrow} d_i{\Phi}^{\tau}\equiv a^{\tau\mu}_ky^k_{\mu+1_i}+{\partial}_ia^{\tau\mu}_ky^k_{\mu}$ for $\tau=1,...,p$. Therefore, setting $Dy^1+...+Dy^m=Dy\simeq D^m$ and calling $I=D\Phi\subset Dy$ the {\it differential module of equations}, we obtain by residue the {\it differential module} or $D$-{\it module} $M=Dy/D\Phi$, introducing the canonical projection $Dy \stackrel{p}{\longrightarrow} M \rightarrow 0$ and denoting the residue of $y^k_{\mu}$ by ${\bar{y}}^k_{\mu}$ when there can be a confusion. Introducing the two free differential modules $F_0\simeq D^{m_0}, F_1\simeq D^{m_1}$, we obtain equivalently the {\it free presentation} $F_1\stackrel{d_1}{\longrightarrow} F_0 \stackrel{p}{\longrightarrow} M \rightarrow 0$ of order $q$ when $d_1={\cal{D}}=\Phi \circ j_q$. We shall moreover assume that ${\cal{D}}$ provides a {\it strict morphism} (see below) or, equivalently, that the corresponding system $R_q$ is formally integrable ([15]). It follows that $M$ can be endowed with a {\it quotient filtration} obtained from that of $D^m$ which is defined by the order of the jet coordinates $y_q$ in $D_qy$. We have therefore the {\it inductive limit} $0=M_{-1} \subseteq M_0 \subseteq M_1 \subseteq ... \subseteq M_q \subseteq ... \subseteq M_{\infty}=M$ with $d_iM_q\subseteq M_{q+1}$ but it is important to notice that $D_rD_q=D_{q+r} \Rightarrow D_rM_q= M_{q+r}, \forall q,r\geq 0 \Rightarrow M=DM_q, \forall q\geq 0$ {\it in this particular case}. It also follows from noetherian arguments and involution that $D_ rI_q=I_{q+r}, \forall r\geq 0$ though we have in general only $D_rI_s\subseteq I_{r+s}, \forall r\geq 0, \forall s<q$. We shall set $G_q=M_q/M_{q-1}$ and introduce the {\it graded module} $G=gr(M)={\oplus}_qG_q$ which is a module over the polynomial ring $gr(D)\simeq K[\chi]$. As $A\subset D$, we may introduce the {\it forgetful functor} $for : mod(D) \rightarrow mod(A): { }_DM \rightarrow { }_AM$. In this paper, we shall go as far as possible with such an arbitrary differential ring $A$ though, in actual practice and thus in most of the examples considered, we shall use a differential field $K$ [21, 41]. We shall also assume that the ring $A$ is a noetherian ring (integral domain) in such a way that $D$ becomes a (both left and right) noetherian ring (integral domain).\\

More generally, introducing the successive CC as in the preceding section while changing slightly the numbering of the respective operators, we may finally obtain the {\it free resolution} of $M$, namely the exact sequence $\hspace{5mm} ... \stackrel{d_3}{\longrightarrow} F_2  \stackrel{d_2}{\longrightarrow} F_1 \stackrel{d_1}{\longrightarrow}F_0\stackrel{p}{\longrightarrow}M \longrightarrow 0 $ where $p$ is the canonical projection. Also, with a slight abuse of language, when ${\cal{D}}=\Phi \circ j_q$ is involutive as in section 2 and thus $R_q=ker( \Phi)$ is involutive, one should say that $M$ has an {\it involutive presentation} of order $q$ or that $M_q$ is {\it involutive} and $D_rM_q=M_{q+r}, \forall q,r \geq 0$ because $D_rD_q=D_{q+r}, \forall q,r \geq 0$.\\

\noindent
{\bf REMARK 3.8}:  In actual practice, one must never forget that ${\cal{D}}=\Phi \circ j_q$ {\it acts on the left on column vectors in the operator case and on the right on row vectors in the module case}. For this reason, when $E$ is a (finite dimensional) vector bundle over $X$/(finite dimensional) vector space over $K$, we may apply the correspondence $J_{\infty}(E) \leftrightarrow D{\otimes}_KE^* : J_q(E)\leftrightarrow D_q{\otimes}_K E^*$ with ${\pi}^{q+1}_q:J_{q+1}(E) \rightarrow J_q(E) \leftrightarrow D_q \subset D_{q+1}$ and $E^*=hom_K(E,K)$ between jet bundles and left differential modules in order to be able to use the {\it double dual isomorphism} $E\simeq E^{**}$ in both cases. We shall say that $D(E)=D\otimes _KE^*=ind(E^*)$ is the the left differential module {\it induced} by $E^*$. Hence, starting from a differential operator $E \stackrel{\cal{D}}{\longrightarrow}F$, we may obtain a finite presentation $D{\otimes}_KF^* \stackrel{{\cal{D}}^*}{\longrightarrow}D{\otimes}_KE^* \rightarrow M \rightarrow 0$ and conversely, while keeping the same operator matrix if we act on the right of row vectors. This is a rather subtle point in the litterature where sometimes a dot is used on the left of ${\cal{D}}$ in the module sense or on the right in the operator sense, depending whether we have an action on the right or on the left. We consider that this is a rather confusing notation because we have the composition ${\cal{D}}_1 \circ {\cal{D}}=0$ {\it along the arrows in the operator framework} while we have the composition ${{\cal{D}}^*}\circ {\cal{D}}^*_1=0 $ {\it along the arrows in the module framework}, like a transposion of matrices. In actual practice, it is much better to keep the same operator matrix {\it acting on the left of column vectors in the operator framework} but {\it acting by composition on the right of row vectors}, the main difference being the position of the indices for the implicit summations.  \\ 

 \noindent
{\bf EXAMPLE 3.9}: With $n=2, m=3, p=3 $ and $K=\mathbb{Q}$, let us consider the linear first order involutive system with only 
$1$ CC:  \\
\[ \left\{  \begin{array}{rccl}
y^2_2 & +y^3_2 & -y^3_1 & -y^2_1=0 \\
y^1_2 &-y^3_2  &-y^3_1  & -y^2_1=0 \\
y^1_1 &        &-2y^3_1 & -y^2_1=0 
\end{array} \right. 
  \fbox{$\begin{array}{cc}
  1 & 2 \\
  1 & 2 \\
  1& \bullet
  \end{array}$} \hspace{4mm} \Rightarrow \hspace{4mm} {\cal{D}}=  \left(\begin{array}{ccc}   
  0 & d_2-d_1 & d_2-d_1  \\
  d_2 & -d_1 & -d_2-d_1  \\
  d_1 & -d_1 & -2d_1
  \end{array}  \right)   \]
We have $det({\sigma}_{\chi}({\cal{D}}))=0$ but $max_{\chi}rk ({\sigma}_{\chi}({\cal{D}}))=2$, a result leading to $rk(M)=1$. Now, setting $z=y^1-y^2-2y^3$, we get $z_1=0, z_2=0$ and $z$ is a torsion element of $M$. However, setting $z'=y^2+y^3$, we have $z'_2-z'_1=0$ and $z'$ is also a torsion element of $M$ in such a way that $t(M)$ is generated by $z$ and $z'$. The torsion-free module $M/t(M)$ is defined by $y^3= - y^2=y^1$ and is a free module isomorphic to $D$ which is thus projective. Accordingly, the short exact sequence 
$0 \rightarrow t(M) \rightarrow M \rightarrow M/t(M) \rightarrow 0$ splits with $M\simeq t(M) \oplus M/t(M)$ leading to the inclusion $D\subset M$. We finally obtain the Janet sequence and corresponding resolution of $M$:  \\
\[ 0 \rightarrow \Theta \rightarrow 3 \stackrel{{\cal{D}}}{\longrightarrow } 3 \stackrel{{\cal{D}}_1}{\longrightarrow} 1 \rightarrow 0 \hspace{1cm}\Leftrightarrow \hspace{1cm} 0 \rightarrow D \stackrel{{\cal{D}}_1}{ \rightarrow} D^3 \stackrel{\cal{D}}{\rightarrow} D^3 \stackrel{p}{\rightarrow} M  \rightarrow 0  \]  \\
The reader may look at [52] for the {\it purity filtration} $ 0 \subset Dz \subset t(M) \subset M $ with strict inclusions and more details on this example.  \\

\noindent
{\bf EXAMPLE 3.10}: (See the Bose conjecture and Example 5.27 in [45], p 216) With $n=3,m=3$ and $K=\mathbb{Q}$, let us consider the differential module $M$ defined by the two differentially independent PD equations:  \\
\[   y^3_{12}- y^2_3 - y^3=0, \hspace{1cm}  y^3_{22} - y^1_3=0  \]                                            
We let the reader transform this systems into an involutive system as an exercise in order to find $rk_D(M)=1$ but we shall obtain the same result by pointing out that both $y^1$ and $y^2$ are differentially dependent on $y^3$ and thus $y^2$ is differentially dependent on $y^1$ [21, 41]. Elimination of $y^3$ provides the only CC $d_3z=0$ with $z= y^2_{22} - y^1_{12} +y^1$ and $t(M)$ is generated by $z$. One can also use the criterion with $5$ steps for testing the torsion-freeness. It follows that the torsion-free module $M'=M/t(M)$ can be defined by the $3$ PD equations:  \\
\[    y^3_{12}- y^2_3 - y^3=0, \hspace{1cm}  y^3_{22} - y^1_3=0 , \hspace{1cm}  y^2_{22}- y^1_{12} + y^1 =0 \]  
This system admits an injective parametrization: \\
\[  u_{22}=y^1, \hspace{1cm} u_{12}- u=   y^2, \hspace{1cm}        u_3=   y^3  \hspace{3mm} \Rightarrow \hspace{3mm}  
u=y^1_{12}-y^2_{22} - y^2 \]
a result  showing that $M' \simeq D$ is a free differential module which is therefore projective  and the short exact sequence $0 \rightarrow t(M) \rightarrow M \rightarrow M/t(M) \rightarrow 0$ splits according to Proposition $3.4$ with 
$M\simeq  t(M) \oplus M/t(M)$. It is difficult to find similar examples because, as we shall see in Section $4$ with Einstein equations, the existence of such a splitting is not always fulfilled (See [11, 12] and the corresponding criterion in [56-57] and [58], Theorem $4$).  \\
If we consider now the system ${\Phi}^2 \equiv y^2_3=y^3_{12}-y^3, {\Phi}^1\equiv y^1_3=y^3_{22}$. It is clear that the system without second member is trivially involutive with no CC and thus $y^3$ may be given arbitrary. The tricky question is to look for a {\it compatible 
 differential constraint} on $(y^1,y^2)$ in such a way that $y^3$ {\it could remain arbitrary} and, for example, $y^1_3=0 \Rightarrow 
y^3_{22}=0$ is not convenient. In order to find a possibility, let us consider the involutive system:  \\
\[ \left\{ \begin{array}{lcc}
y^2_{33} & = & d_3 {\Phi }^2  \\ 
y^1_{33} & = & d_3 {\Phi}^1  \\
y^2_{23} & = & d_2 {\Phi}^2  \\
y^1_{23} & = & d_2{\Phi}^1  \\
y^2_{22}- y^1_{12} + y^1 & = & 0  \\
y^2_{13} & = & d_1{\Phi}^2  \\
y^1_{13} & = & d_1 {\Phi}^1  \\
y^2_3     & = & {\Phi}^2  \\
y^1_3     & = & {\Phi}^1
\end{array} \right.     
 \fbox{$\begin{array}{ccc}
 1 & 2 & 3 \\
 1 & 2 & 3  \\
 1 & 2 & \bullet \\
 1 & 2  & \bullet  \\
 1 & 2  & \bullet  \\
 1 & \bullet & \bullet  \\
 1 & \bullet & \bullet  \\
 \bullet & \bullet & \bullet  \\
 \bullet & \bullet & \bullet 
 \end{array} $ }  \]                                                                                                                                                                                                                                                                                                                                                                                                                                                                                                                                                                                                                                                                                                                                                                                                                                                                                                                                                                                                                                                                                                                                                                                                                                                                                                                                                                                                                                                                                                                                                                                                                                                                                                                                                                                                                                                                                                                                                                                                                                                                                                                                                                                                                                                                 
We are left with the only CC $d_{22}{\Phi}^2 - d_{12}{\Phi}^1 + {\Phi}^1=0$ which is trivially satisfied because we have the 
{\it identity} $(y^3_{1222}-y^3_{22}) - y^3_{1222} + y^3_{22}\equiv 0 $. Though striking it may look like, we shall see in Section $4$ that this is just the situation considered for introducing gravitational waves !.  \\

\noindent
{\bf EXAMPLE 3.11}: The exterior derivative in the Janet sequence ${\wedge}^0T^* \stackrel{d}{\longrightarrow}T^* \stackrel{d}{\longrightarrow} {\wedge}^2T^*$ explains the above comments when $K=\mathbb{Q}$. In the operator sense, then ${\wedge}^0T^*$, $T^*$ {\it and} ${\wedge}^2T^*$ are both represented by column vectors though they are exterior forms. In the module framework we have the dual Janet sequence of left $D$-modules $D\otimes {\wedge}^2T \stackrel{d^*}{\longrightarrow} D\otimes T\stackrel{d^*}{\longrightarrow} D$ where:  \\
\[ P\otimes d_i \wedge d_j \rightarrow  Pd_i \otimes d_j -Pd_j \otimes d_i , \hspace{1cm}P\otimes d_i \rightarrow Pd_i \]
Accordingly, if $f\in {\wedge}^0T^*  \rightarrow \alpha=({\alpha}_i=d_if) \in T^*$ in the operator way, the implicit summation 
$P^i{\alpha}_i=P^i(d_if)=(P^id_i)f$ in the module way is explaining the Remark.  \\

\noindent
{\bf EXAMPLE 3.12}: In elasticity theory, we may rewrite the Beltrami parametrization of the Cauchy stress equations as follows, after exchanging the third row with the fourth row:  \\
\[      \left(  \begin{array}{cccccc}
d_1& d_2 & d_3 &0 & 0 & 0 \\
 0 & d_1 &  0 & d_2 & d_3 & 0 \\
 0 & 0 & d_1 & 0 & d_2 & d_3 
\end{array}  \right)  
 \left(  \begin{array}{cccccc}
 0 & 0 & 0 & d_{33} & - 2d_{23} & d_{22} \\
 0 & - d_{33} & d_{23} & 0 & d_{13} & - d_{12}  \\
 0 & d_{23} & - d_{22} & - d_{13} & d_{12} & 0 \\
 d_{33}& 0 & - 2 d_{13} & 0 & 0 & d_{11}  \\
 - d_{23} & d_{13} & d_{12}& 0 & - d_{11} & 0 \\
 d_{22} & - 2 d_{12} & 0 & d_{11}& 0 & 0 
 \end{array} \right)  \equiv   0    \]
 as an identity where $0$ on the right denotes the zero operator. However, the standard implicit summation used in continuum mechanics (See [49] for more details) is, when $n=3$:  \\
 \[   \begin{array}{rcl}
{\sigma}^{ij}{\Omega}_{ij} & = & {\sigma}^{11}{\Omega}_{11} + 2 {\sigma}^{12}{\Omega}_{12} + 2 {\sigma}^{13}{\Omega}_{13} + {\sigma}^{22} {\Omega}_{22} + 2{\sigma}^{23}{\Omega}_{23} + {\sigma}^{33}{\Omega}_{33}  \\
   &  =  & {\Omega}_{22}d_{33}{\Phi}_{11}+ {\Omega}_{33}d_{22}{\Phi}_{11}- 2 {\Omega}_{23}d_{23}{\Phi}_{11}+ ... \\
   &    & + {\Omega}_{23}d_{13}{\Phi}_{12}+{\Omega}_{13}d_{23}{\Phi}_{12}- {\Omega}_{12}d_{33}{\Phi}_{12}- {\Omega}_{33}d_{12}{\Phi}_{12} + ...
\end{array}  \]
because {\it the stress tensor density $\sigma$ is supposed to be symmetric} in continuum mechanics. \\
Integrating by parts in order to construct the adjoint operator as in the Introduction, we get:  \\
\[ \begin{array}{rcl}
 {\Phi}_{11} &  \longrightarrow  &  d_{33}{\Omega}_{22} + d_{22}{\Omega}_{33} - 2 d_{23}{\Omega}_{23} \\
 {\Phi}_{12} &  \longrightarrow   &  d_{13}{\Omega}_{23}+d_{23}{\Omega}_{13}-d_{33}{\Omega}_{12} - d_{12}{\Omega}_{33}
 \end{array}  \]
and so on, obtaining therefore the striking identification:  \\
\[       Riemann=ad(Beltrami)   \hspace{1cm} \Longleftrightarrow  \hspace{1cm}   Beltrami=ad(Riemann)  \]
between the (linearized ) Riemann tensor and the Beltrami parametrization. \\
As we already said, the brothers E. and F. Cosserat proved in 1909 that such an assumption may be too strong because it only takes into account density of forces and ignores density of couples, that is {\it must} be replaced by the so-called {\it Cosserat couple-stress equations}. We have proved in many books and papers that {\it these equations are just described by the formal adjoint of the Spencer operator $D_1$ for the Killing system}, a reason for using in physics the Spencer sequence rather than the Janet sequence [39, 40, 42, 68]. In any case, taking into account the factor $2$ involved by multiplying the second, third and fifth row by $2$, we get the new $6\times 6$ matrix with rank $3$:

\[   \left(  \begin{array}{cccccc}
 0 & 0 & 0 & d_{33} & - 2d_{23} & d_{22} \\
 0 & - 2d_{33} & 2d_{23} & 0 & 2d_{13} & - 2d_{12}  \\
 0 & 2d_{23} & - 2d_{22} & - 2d_{13} & 2d_{12} & 0 \\
 d_{33}& 0 & - 2 d_{13} & 0 & 0 & d_{11}  \\
 - 2d_{23} & 2d_{13} & 2d_{12}& 0 & - 2d_{11} & 0 \\
 d_{22} & - 2 d_{12} & 0 & d_{11}& 0 & 0 
 \end{array} \right)     \]

\hspace{2cm}    SYMMETRIC  MATRIX \hspace{5mm} $\Rightarrow$ \hspace{5mm}  SELF-ADJOINT  OPERATOR   \\ 

It is only in Section $4$ that we shall be able to explain the relation of this striking result  with Einstein equations but the reader must already understand that, if we need to revisit in such a deep way the mathematical foundations of elasticity theory, we also need to revisit in a similar way the mathematical foundations of EM and GR as in [50-53]. \\

In Section 2, the formal integrability of a system has been used in a crucial way in order to construct various differential sequences. Therefore, the algebraic counterpart provided by the next definition and proposition will also be used in a crucial way too in order to construct resolutions of a differential module [20, 44, 60, 62], though in a manner which is not so natural when dealing with applications to mathematical physics [7,8,17,32,33,61]. For this reason, we invite the reader to follow closely the arguments involved on the illustrating examples provided. To sart with, if $M$ and $N$ are two filtred differential modules and $f:M \rightarrow N$ is a differential morphism, that is a $D$-linear map with $f(Pm)=Pf(m), \forall P\in D$, then $f$ will be called an {\it homomorphism} of filtred modules if it induces $A$-linear maps $f_q=M_q \rightarrow N_q$. Chasing in the following commutative diagram:   \\
\[  \begin{array}{cccccl}
0  &   &  0  &  &  &     \\
\downarrow  &  & \downarrow  &  &  &  \\
M_q  & \stackrel{f_q}{\longrightarrow}  & N_q &  \rightarrow  &  coker(f_q) &  \rightarrow  0  \\
\downarrow &  \  & \downarrow  &  &  \downarrow &   \\
M  &  \stackrel{f}{\longrightarrow} & N & \rightarrow & coker(f) & \rightarrow  0
\end{array}   \]
while introducing $im(f)=I\subseteq N, im(f_q)=I_q\subseteq N_q$, we may state: \\

\noindent
{\bf DEFINITION 3.13}: A differential morphism $f$ is said to be a {\it strict homomorphism} if the two following equivalent properties hold:  \\
1) There is an induced monomorphism $0\rightarrow coker(f_q) \rightarrow coker(f), \forall q\geq 0$.  \\
2) $f_q(M_q)=f(M)\cap N_q $, that is $ I_q=I\cap N_q$.\\
A sequence made by strict morphisms will be called a {\it strict sequence}. In order to fulfill the conditions of the definition, it is most of the time necessary to "{\it shift} " the filtration of a differential module $M$ by setting $M(r)_q=M_{q+r}$ in such a way that $q$ could be negative and we shall therefore always assume that $M_q=0, \forall q\ll 0$. \\

\noindent
{\bf PROPOSITION 3.14}: If we have a strict short exact sequence $0 \rightarrow M' \stackrel{f}{\longrightarrow} M \stackrel{g}{\longrightarrow} M" \rightarrow 0$ in which $\exists q\gg 0$ such that $D_rM_q=M_{q+r}, \forall r\geq 0$, then $\exists q',q" \gg0$ such that $D_rM'_{q'}=M'_{q'+r}, D_rM"_{q"}=M"_{q"+r}, \forall r\geq 0$ and conversely. We may thus assume that $q=q'=q"$ in both cases by choosing $q \gg 0$. More generally, an exact sequence of filtred differential modules is strictly exact if and only if the associated sequence of graded modules is exact in a way dualizing the differential geometric framework, on the condition to shift conveniently the various filtrations involved.\\

\noindent
{\it Proof}: First of all, setting $G=gr(M), G'=gr(M'), G"=gr(M")$, we have the commutative and exact diagram:  \\
\[  \begin{array}{rcccccl}
  &  0  & &  0  & &   0  &   \\
  & \downarrow  &  & \downarrow  &  & \downarrow &    \\
0 \rightarrow & M'_{q-1} & \stackrel{f_{q-1}}{\longrightarrow}  & M_{q-1} & \stackrel{g_{q-1}}{\longrightarrow} & M"_{q-1} & \rightarrow 0 \\
  & \downarrow  &  & \downarrow  &  & \downarrow &    \\
0 \rightarrow & M'_q & \stackrel{f_q}{\longrightarrow}  & M_q & \stackrel{g_q}{\longrightarrow} & M"_q & \rightarrow 0 \\
& \downarrow  &  & \downarrow  &  & \downarrow &    \\
0 \rightarrow & G'_q & \stackrel{gr_q(f)}{\longrightarrow}  & G_q & \stackrel{gr_q(g)}{\longrightarrow} & G"_q & \rightarrow 0 \\
& \downarrow  &  & \downarrow  &  & \downarrow &    \\
 & 0 &  & 0  & &  0 & 
\end{array}  \]
Indeed, as $g$ is a strict epimorphism, it follows that $g_q$ is surjective $\forall q\geq 0$. Also, as $f$ is a monomorphism, then $f_q$ is also a monomorphism $\forall q \geq 0$ by restriction. Moreover, as $f$ is also strict, we obtain successively by chasing:  \\
\[     ker(g_q)=f(M')\cap M_q=f(M'_q)=f_q(M'_q)=im(f_q)  \]
It follows that the two upper rows are exact and the bottom row is thus exact too $\forall q\geq 0$ from the snake theorem in homological algebra [8, 17, 32, 44, 45, 52, 61, 62].  \\
This result provides the short exact sequence $0 \rightarrow G' \stackrel{gr(f)}{\longrightarrow} G \stackrel{gr(g)}{\longrightarrow} G" \rightarrow 0 $ of graded modules. \\
Let us now consider the following commutative diagram with maps such as $\xi \otimes m \rightarrow \xi m$ and where the upper row is exact because $D_1\simeq A \oplus T$ is free over $A$:\\
\[  \begin{array}{rcccccl}
0 \rightarrow & D_1{\otimes}_AM'_q & \stackrel{f}{\longrightarrow} & D_1{\otimes}_A M_q  & \stackrel{g}{\longrightarrow} & D_1{\otimes}_AM"_q & \rightarrow 0  \\
  & \downarrow & & \downarrow & & \downarrow &  \\
  0 \rightarrow & M'_{q+1} & \stackrel{f_{q+1}}{\longrightarrow} & M_{q+1} & \stackrel{g_{q+1}}{\longrightarrow} & M"_{q+1}  & \rightarrow 0  
  \end{array}  \]

If the central map is surjective, then the map on the right is also surjective, that is $D_1M_q=M_{q+1} \Rightarrow D_1M"_q=M"_{q+1}$ and thus $q=q"$. This is the typical situation met in a finite presentation of a system already considered. Moreover, $D_1M'_q\subseteq M'_{q+1} \Rightarrow TG'_q\subseteq G'_{q+1}$ like in the following commutative and exact diagrams where the left one is holding for a (formally integrable) system while the corresponding right one is holding for an arbitrary filtred module $M$ with $gr(M)=G$:  \\
\[ \begin{array}{rccclcrcccl}
 &   0  &   &  0  &                                               \hspace{1cm}       &                     &     &   0  &   &  0  &                                              \\                                         
    &  \downarrow  &  &  \downarrow  &                                           &     \hspace{1cm} & & \uparrow  &  &  \uparrow  &                    \\
0 \rightarrow & g_{q+1}  &  \rightarrow  &  T^* \otimes R_q &      &     \hspace{1cm} &     & T{\otimes}_AM_q  &  \rightarrow  &  G_{q+1} &  \rightarrow 0 \\
    &   \downarrow  &  &  \downarrow  &                                        &  \hspace{1cm}&   &  \uparrow  &  &  \uparrow  &                \\
 0 \rightarrow &R_{q+1} & \rightarrow & J_1(R_q) &               &   \hspace{1cm} &    &  D_1{\otimes}_AM_q & \rightarrow & M_{q+1} &    \\
  &  \downarrow  &  &  \downarrow  &                                             & \hspace{1cm}  &   &\uparrow  &  &  \uparrow  &               \\
0 \rightarrow &    R_q  &  =  & R_q  & \rightarrow 0                     &  \hspace{1cm}  &   0 \rightarrow &    M_q  &  =  & M_q  & \rightarrow 0    \\
   &  \downarrow  &  &  \downarrow  &                                           & \hspace{1cm} & &   \uparrow  &  &  \uparrow  &       \\
     &   0  &  &  0  &                                                                              &  \hspace{1cm} &  &   0  &  &  0  &    
      \end{array}       \]
 In these diagrams, the upper morphism is the composition $g_{q+1} \stackrel{\delta}{\longrightarrow} T^*\otimes g_q \rightarrow T^*\otimes R_q$ in the system diagram and the composition $T{\otimes}_AM_q \rightarrow T{\otimes}_AG_q \rightarrow G_{q+1}$ in the module diagram. Accordingly, a chase is showing that $D_1M_q=TM_q+M_q\subseteq M_{q+1}$ with equality if and only if  $TG_q=G_{q+1}$.\\
From noetherian arguments for polynomial rings in commutative algebra, it follows that $G'$ is finitely generated and we may choose for $q'$ the maximum order of a minimum set of generators.\\
 Conversely, if $D_rM'_{q'}=M'_{q'+r}, D_rM"_{q"}=M"_{q"+r}, \forall r\geq 0$, we may choose $q=sup(q',q")$ and we have thus $D_1M'_q=M'_{q+1}, D_1M"_q=M"_{q+1}\Rightarrow D_1M_q=M_{q+1}$, using again the snake theorem. \\
 As a byproduct, it is always possible to find $q\gg 0$ such that we could have {\it at the same time} $D_rM_q=M_{q+r}, D_rM'_q=M'_{q+r}, D_rM"_q=M"_{q+r}, \forall r\geq 0$ in the two situations considered.  \\
 We end this proof with a comment on the prolongation of symbols and graded modules which, in our opinion based on more than thirty years spent on computing and applying these dual concepts, is not easy to grasp. For this, let us consider the corresponding diagrams:  \\
 \[  \begin{array}{rccccccccl}
   & 0  &  &  0  &                                                                                     \hspace{1cm}                          &  0 &  & 0 &   \\
  & \downarrow &  & \downarrow &                                                               \hspace{1cm}       & \uparrow   &       & \uparrow  &    \\
0 \rightarrow & g_{q+1} & \rightarrow &  S_{q+1}T^* \otimes E  &\hspace{1cm}    &S_{q+1}T{\otimes}_AE^* & \rightarrow & G_{q+1} & \rightarrow 0  \\
   &\hspace{3mm}\downarrow\delta &  &\hspace{3mm}\downarrow\delta  &   \hspace{1cm} &\hspace{5mm} \uparrow{\delta}^*  &  &\hspace{5mm}  \uparrow {\delta}^* &   \\
0\rightarrow & T^*\otimes g_q & \rightarrow & T^*\otimes S_qT^*\otimes E &\hspace{1cm}     &T{\otimes}_AS_qT{\otimes}_A E^*  & \rightarrow & T{\otimes}_AG_q&\rightarrow 0 
\end{array}   \] 
Indeed, exactly as we have in general $R_{q+1}\subseteq {\rho}_1(R_q)\Rightarrow g_{q+1}\subseteq {\rho}_1(g_q)$, there is no corresponding concept in module theory without a reference to a presentation. In the differential geometric framework, ${\rho}_1(g_q)$ is the {\it reciprocal image} of $\delta$, that is the subset (not always a vector bundle !) of $S_{q+1}T^*\otimes E$ made by elements having an image in $T^*\otimes g_q$ under $\delta$.  \\
 \hspace*{12cm}       Q.E.D.    \\

\noindent
{\bf EXAMPLE 3.15}: Though this is not evident at first sight when $m=1, n=2, A=\mathbb{Q}[x^1,x^2]$, we invite the reader to prove that the third order linear system $y_{222}+x^2y_2=0, y_{111}+y_2-y=0$ has the same formal solutions as the third order system $y_{111}-y=0, y_2=0$ which is defined over $\mathbb{Q}$, a result leading to the generating involutive third order linear system $y_{222}=0, y_{122}=0, y_{112}=0, y_{111}-y=0, y_{22}=0, y_{12}=0,y_2=0$. We have $M_0=\{\bar{y}\},  M_1=\{\bar{y},{\bar{y}}_1\}, M_2=\{\bar{y},{\bar{y}}_1,{\bar{y}}_{11}\}=M$ while using only parametric jets because $d_1{\bar{y}}_{11}={\bar{y}}_{111}=\bar{y}$ and thus $D_1I_1=I_2, D_1I_2\subset I_3$ with a strict inclusion, $DI_3=I$. (See the similar Examples III.2.64 and III.3.11 in [44] for the details). \\

Roughly speaking, homological algebra has been created in order to find intrinsic properties of modules not depending on their presentations or even on their resolutions and we now exhibit another approach by defining the {\it formal adjoint} of an operator $P$ and an operator matrix 
$\cal{D}$:  \\

\noindent
{\bf DEFINITION 3.16}: Setting $P=a^{\mu}d_{\mu}\in D  \stackrel{ad}{\longleftrightarrow} ad(P)=(-1)^{\mid\mu\mid}d_{\mu}a^{\mu}   \in D $, we have $ad(ad(P))=P$ and $ad(PQ)=ad(Q)ad(P), \forall P,Q\in D$. Such a definition can be extended to any matrix of operators by using the transposed matrix of adjoint operators and we get:  
\[ <\lambda,{\cal{D}} \xi>=<ad({\cal{D}})\lambda,\xi>+\hspace{1mm} {div}\hspace{1mm} ( ... )  \]
from integration by part, where $\lambda$ is a row vector of test functions and $<  > $ the usual contraction. We quote the useful formulas $[ad(\xi),ad(\eta)]=ad(\xi)ad(\eta)-ad(\eta)ad(\xi)= - ad([\xi, \eta]), \forall \xi, \eta \in T$ ({\it care about the minus sign}) and $rk_D({\cal{D}})=rk_D(ad({\cal{D}}))$ as in ([41], p 339-341).\\

\noindent
{\bf LEMMA 3.17}: If $f\in aut(X)$ is a local diffeomorphisms on $X$, we may set $ x=f^{-1}(y)=g(y)$ and we have the {\it identity}:
\[   \frac{\partial}{\partial y^k}(\frac{1}{\Delta (g(y))} {\partial}_if^k(g(y))\equiv 0.   \]

\noindent
{\bf PROPOSITION 3.18}: If we have an operator $E\stackrel{\cal{D}}{\longrightarrow} F$, we may obtain by duality an operator ${\wedge}^nT^*\otimes E^*\stackrel{ad(\cal{D})}{\longleftarrow} {\wedge}^nT^*\otimes F^*$. \\

\noindent
{\bf EXAMPLE 3.19}: In order to understand how the Lemma is involved in the Proposition, let us revisit relativistic electromagnetism (EM) in the light of these results when $n=4$. First of all,  we have $dA=F \Rightarrow dF=0$ in the sequence ${\wedge}^1T^*\stackrel{d}{\longrightarrow} {\wedge}^2T^* \stackrel{d}{\longrightarrow} {\wedge}^3T^*$ and the {\it field equations} of EM (first set of Maxwell equations) are invariant under any local diffeomorphism $f\in aut(X)$. By duality, we get the sequence ${\wedge}^4T^*\otimes {\wedge}^1T \stackrel{ad(d)}{\longleftarrow} {\wedge}^4T^*\otimes {\wedge}^2T \stackrel{ad(d)}{\longleftarrow} {\wedge}^4T^*\otimes {\wedge}^3T$ which is locally isomorphic (up to sign) to ${\wedge}^3T^* \stackrel{d}{\longleftarrow} {\wedge}^2T^* \stackrel{d}{\longleftarrow} {\wedge}^1T^*$ and the {\it induction equations} ${\partial}_i{\cal{F}}^{ij}={\cal{J}}^j$ of EM (second set of Maxwell equations) are thus also invariant under any $f\in aut(X)$. Indeed, using the last lemma and the {\it identity} ${\partial}_{ij}f^l{\cal{F}}^{ij}\equiv 0$, we have: \\
\[\frac{\partial}{\partial y^k}(\frac{1}{\Delta}{\partial}_i f^k{\partial}_j f^l{\cal{F}}^{ij})=\frac{1}{\Delta} {\partial}_i f^k \frac{\partial}{\partial y^k}({\partial}_j f^l{\cal{F}}^{ij})=\frac{1}{\Delta}{\partial}_i({\partial}_j f^l{\cal{F}}^{ij})=\frac{1}{\Delta}{\partial}_j f^l{\partial}_i{\cal{F}}^{ij} \]
Accordingly, it is not correct to say that the conformal group is the biggest group of invariance of Maxwell equations in physics as it is only the biggest group of invariance of the Minkowski constitutive laws in vacuum [36]. Finally, according to Proposition $2.20$, both sets of equations can be parametrized {\it independently}, the first by the potential, the second by the so-called pseudopotential {\it in a totally independent way} (See the last section of this paper and [44], p 492 for more details). \\

Now, with operational notations, let us consider the two differential sequences:  \\
\[   \xi  \stackrel{{\cal{D}}}{\longrightarrow} \eta \stackrel{{\cal{D}}_1}{\longrightarrow} \zeta  \]
\[   \nu  \stackrel{ad({\cal{D}})}{\longleftarrow} \mu \stackrel{ad({\cal{D}}_1)}{\longleftarrow} \lambda   \]
where ${\cal{D}}_1$ generates all the CC of ${\cal{D}}$. Then ${\cal{D}}_1\circ {\cal{D}}\equiv 0 \Longleftrightarrow ad({\cal{D}})\circ ad({\cal{D}}_1)\equiv 0 $ but $ad({\cal{D}})$ may not generate all the CC of $ad({\cal{D}}_1)$. Passing to the module framework, we just recognize the definition of $ext^1_D(M)$. Now, exactly like we defined the differential module $M$ from $\cal{D}$, let us define the differential module $N$ from $ad(\cal{D})$. Then $ext^1_D(N)=t(M)$ does not depend on the presentation of $M$. In particular, if ${\cal{D}}$ is formally surjective (differentially independent PD equations), then $M$ is torsion-free if $d(N)\leq n-2$ and projective if $d(N)=-1$ that is if $ext^i(N)=0, \forall 0\leq i \leq n$. For example, the $div$ operator for $n=3$ is torsion-free but not projective because $ext^3(N)\neq 0$. More generally, changing the presentation of $M$ may change $N$ to $N'$ but we have [23], ([44],p 651), ([45],p 203): \\

\noindent
{\bf THEOREM 3.20}: The modules $N$ and $N'$ are {\it projectively equivalent}, that is one can find two projective modules $P$ and $P'$ such that $N\oplus P\simeq N' \oplus P'$ and we obtain therefore $ext^i_D(N)\simeq ext^i_D(N'), \forall i\geq 1$.  \\

Having in mind that $D$ is a $A$-algebra, that $A$ is a left $D$-module with the standard action $(D,A) \longrightarrow A:(P,a) \longrightarrow P(a):(d_i,a)\longrightarrow {\partial}_ia$ and that $D$ is a bimodule over itself, {\it we have only two possible constructions leading to the following two definitions}:  \\

\noindent
{\bf DEFINITION 3.21}: We may define the right $D$-module $M^*=hom_D(M,D)$ or the {\it inverse system} $R=hom_A(M,A)$ of $M$ and set $R_q=hom_A(M_q,A)$ as the {\it inverse system of order} $q$. \\

If $G=gr(M)$ is the graded module of $M$ with $G={\oplus}^{\infty}_{q=0}G_q$, we have the short exact sequences $0\rightarrow M_{q} \rightarrow M_{q+1} \rightarrow G_{q+1} \rightarrow 0$ of modules over $A$ and it is tempting to compare them to the dual short exact sequences $0\rightarrow g_{q+1}\rightarrow R_{q+1}\rightarrow R_q \rightarrow 0$ that were used in the previous section. However, applying $hom_A(\bullet,A)$ to the first sequence does not in general provide a short exact sequence (See the end of Example 3.30), unless the first sequence splits, that is if we replace vector bundles over $X$ used in Section 2 by finitely generated projective modules over $A$. We shall rather prefer to use the field of fractions $K=Q(A)$ in order to deal only with finite dimensional vector spaces over $K$ or use the fact that $K=Q(A)$ is an injective module over $A$ and deal with $hom_A(\bullet,K)$ in order to obtain exact sequences. From the injective limit of the filtration of $M$ we deduce the {\it projective limit} $R=R_{\infty} \longrightarrow ... \longrightarrow R_q \longrightarrow ... \longrightarrow R_1 \longrightarrow R_0$. It follows that $f_q\in R_q:y^k_{\mu} \longrightarrow f^k_{\mu}\in A$ with $a^{\tau\mu}_kf^k_{\mu}=0$ defines a {\it section at order} $q$ and we may set $f_{\infty}=f\in R$ for a {\it section} of $R$. For a ground field of constants $k$, this definition has of course to do with the concept of a formal power series solution. However, for an arbitrary differential ring $A$ or differential field $K$, {\it the main novelty of this new approach is that such a definition has nothing to do with the concept of a formal power series solution} as illustrated in the next examples. Nevertheless, in actual practice, it is always simpler to deal with a differential field $K$ in order to have finite dimensional vector spaces at each order $q$ for applications.\\

We shall now study with more details the module $M$ versus the system $R$ when $D=K[d]$. First of all, as $K$ is a field, we obtain in particular the Hilbert polynomial $dim_K(M_{q+r})=dim_K(R_{q+r})= \frac{{\alpha}^{n-r}_q}{d!}r^d+ ...$ where the intrinsic integer ${\alpha}^{n-r}_q$ is called the {\it multiplicity} of $M$ and is the smallest non-zero character, that is ${\alpha}^{n-r}_q\neq 0, {\alpha}^{n-r+1}= ... ={\alpha}^n_q=0$. We use to set $d_D(M)=d(M)=d \Rightarrow cd_D(M)=cd(M)=n-d=r, rk_D(M)=rk(M)=\alpha$. Accordingly, $M$ is a torsion module over $D$ if and only if 
$\alpha = 0$. Now, If $M$ is a module over $D$ and $m\in M$, then the cyclic differential submodule $Dm\subset M$ is defined by a system of OD or PD equations for one unknown and we may look for its codimension $cd(Dm)$. A similar comment can be done for any differential submodule $M'\subset M$. Sometimes, a single element $m\in M$, called {\it differentially primitive element}, may generate $M$ if $Dm=M$.   \\

As $D={ }_DD_D$ is a bimodule, then $M^*=hom_D(M,D)$ is a right $D$-module according to Lemma 3.1 and we may thus define a right module $N_D$ by the ker/coker long exact sequence $0\longleftarrow N_D \longleftarrow F_1^*\stackrel{{\cal{D}}^*}{ \longleftarrow} F^*_0 \longleftarrow M^* \longleftarrow 0$.\\

\noindent
{\bf THEOREM 3.22}: We have the {\it side changing} procedures $M={ }_DM \rightarrow M_D={\wedge}^nT^*{\otimes}_AM$ and $N_D \rightarrow N={}_DN=hom_A({\wedge}^nT^*,N_D)$ with ${}_D((M_D))=M$ and ${}_D(N_D)=N$. \\

\noindent
{\it Proof}: According to the above Theorem, we just need to prove that ${\wedge}^nT^*$ has a natural right module structure over $D$. For this, if $\alpha=adx^1\wedge ...\wedge dx^n\in {\wedge}^nT^*$ is a volume form with coefficient $a\in A$, we may set $\alpha.P=ad(P)(a)dx^1\wedge...\wedge dx^n$ when $P\in D$. As $D$ is generated by $A$ and $T$, we just need to check that the above formula has an intrinsic meaning for any $\xi={\xi}^id_i\in T$. In that case, we check at once:
\[  \alpha.\xi=-{\partial}_i(a{\xi}^i)dx^1\wedge...\wedge dx^n=-\cal{L}(\xi)\alpha \]
by introducing the Lie derivative of $\alpha$ with respect to $\xi$, along the intrinsic formula ${\cal{L}}(\xi)=i(\xi)d+di(\xi)$ where $i( )$ is the interior multiplication and $d$ is the exterior derivative of exterior forms. According to well known properties of the Lie derivative, we get :
\[\alpha.(a\xi)=(\alpha.\xi).a-\alpha.\xi(a), \hspace{5mm} \alpha.(\xi\eta-\eta\xi)=-[\cal{L}(\xi),\cal{L}(\eta)]\alpha=-\cal{L}([\xi,\eta])\alpha=\alpha.[\xi,\eta].  \]
Using the anti-isomorphism $ad:D \rightarrow D : P \rightarrow ad(P)$, we may also introduce the {\it adjoint functor} $ad : mod(D) \rightarrow mod (D^{op}): M \rightarrow ad(M)$ with $for(M)=for(ad(M))$ and $m.P=ad(P)m, \forall m\in M, \forall P\in D$. We obtain:  \\
\[    m.(PQ)=ad(PQ)m=(ad(Q)ad(P))m=ad(Q)(ad(P)m)=(m.P).Q, \forall P,Q\in D  \]
We have an $A$-linear isomorphism $ad(M)\simeq M_D:m \rightarrow m\otimes \alpha$ in $mod(D^{op})$. Indeed, with $\alpha=dx^1\wedge ... \wedge dx^n$ and any $d$ among the $d_i$ in place of $\xi$, we get $m.d=ad(d)m=-dm$ and $m.P=ad(P)m$ in $ad(M)$ while $(m\otimes \alpha)d=-dm\otimes \alpha, \forall m\in M$ because $div(d)=0$ and thus ${\cal{L}}(d)\alpha =0$. Accordingly, the previous isomorphism is right $D$-linear.  \\
In order to study the case of $D={ }_DD$, considered as a left $D$-module over $D$, we shall compare $ad(D), D_r$ and $D_D$. According to the last isomorphism obtained, we just need to study the isomorphim $ad(D)\simeq D_D: P \rightarrow ad(P)$. Indeed, we get $P \rightarrow P.Q=ad(Q)P\neq PQ$ and obtain therefore $P.Q \rightarrow ad(P.Q)=ad(ad(Q)P)=ad(P)ad(ad(Q))=ad(P)Q$, a result showing that this isomorphism is also right $D$-linear. \\
\hspace*{12cm}  Q.E.D.  \\
 
\noindent
{\bf REMARK 3.23}: The above results provide a new light on duality in physics. Indeed, as the Poincar\'{e} sequence is self-adjoint (up to sign) 
{\it as a whole} and the linear Spencer sequence for a system of finite type is locally isomorphic to copies of that sequence, it follows {\it in this case} from Proposition 3.7 that $ad(D_{r+1})$ parametrizes $ad(D_r)$ in the dual of the Spencer sequence while $ad({\cal{D}}_{r+1})$ parametrizes $ad({\cal{D}}_r)$ in the dual of the Janet sequence, {\it a result highly not evident at first sight because} ${\cal{D}}_r$ {\it and} $D_{r+1}$ {\it are totally different operators}. The reader may look at the next section or to [50, 51, 53, 55] for recent applications to mathematical physics, in particular to Gauge Theory and General Relativity. \\

Now, if $A,B$ are rings and ${ }_AM, { }_BL_A, { }_BN$ are modules, using the second part of Lemma 3.1 and the relation $l\otimes am=la\otimes m$ with the left action $b(l\otimes m)=bl\otimes m,\forall a\in A, \forall b\in B,\forall l\in L, \forall m\in M$, we may provide the so-called {\it adjoint isomorphism} as in ([61], Th 2.11, p. 37), saying that there is a one-to-one correspondence between maps of the form $L\otimes M \rightarrow N$ and maps of the form $M \rightarrow hom(L,N)$ or, fixing an element $m\in M$, providing a parametrization set of maps of the form $L \rightarrow N$ in both cases:  \\

\noindent
{\bf PROPOSITION 3.24}: \hspace{1cm}$   \varphi :hom_B(L{\otimes}_A M, N)\stackrel{\simeq}{\longrightarrow} hom_A(M, hom_B(L,N)) $  \\

\noindent
{\it Proof}: If $f:L{\otimes}_AM \rightarrow N$ is a $B$-morphism, one may define $\varphi(f):M \rightarrow hom_B(L,N)$ as a -A-morphism by the formula $(\varphi(f)(m))(l)=f(l\otimes m)$. It follows that $\varphi$ is a monomorphism because it is defined on the basis of simple tensors in $L{\otimes}_AM$ and it remains to check that it is an epimorphism by constructing an inverse $\psi$. For this, starting with a $A$-morphism $g:M\rightarrow hom_B(L,N)$, we just define $\psi(g)=f$ by $f(l\otimes m)=(g(m))(l)$. We have in particular: \\
\[  (\varphi(f)(am))(l)=f(l\otimes am)=f(la\otimes m)=(\varphi(f)(m))(la)=a(\varphi(f)(m))(l)  \]
and thus $\varphi(f)(am)=a(\varphi(f)(m))$ in a coherent way with $hom_A$ and Lemma 3.1. \\
\hspace*{12cm}   Q.E.D.   \\

With $M={ }_DM,L={ }_AD_D,N={ }_AA$ and $1\in A\subset D$, one obtains the isomorphism:  \\
\[  hom_A(M,A)=hom_A(D{\otimes}_  DM,A)\simeq hom_D(M,hom_A(D,A))\]
where $(\varphi (f))(1)=f(m)$ if we identify $m$ with $1\otimes m$. However, even though $M$ is a left $D$-module by assumption and $hom_A(D,A)$ is also a left $D$-modules with $(Qh)(P)=h(PQ), \forall h\in hom_A(D,A)$, for any $P,Q\in D$ because of Lemma 3.1, there is no similar reason "{\it a priori} " that $hom_A(M,A)$ could {\it also} be a left $D$-module. In particular, we have $(ah)(P)=h(Pa)\neq h(aP)=a(h(P))$, for any $a\in A$ and any $h \in hom_A(D,A)$, unless $A$ is a ring of constants.\\

\noindent
ACCORDINGLY, THIS APPROACH IS NOT CONVENIENT AND MUST BE MODIFIED WHEN $A$ IS A TRUE DIFFERENTIAL RING OR $K$ IS A TRUE DIFFERENTIAL FIELD, THAT IS WHEN $D$ IS NOT COMMUTATIVE.  \\  

The next crucial theorem will allow to provide the module counterpart of the differential geometric construction of the Spencer operator provided in Section 2 (Compare to [7] and [62]). For a more general approach, we shall consider a differential ring $A$ with unity $1$ and set $D=A[d]$. \\

\noindent
{\bf THEOREM 3.25}: When $M$ and $N$ are left $D$-modules, then $hom_A(M,N)$ and $M{\otimes}_AN$ are left $D$-modules. In particular $R=hom_A(M,A)$ is also a left $D$-module for the Spencer operator. Moreover, if $M$ and $N$ are right $D$-modules, then $hom_A(M,N)$ is a left $D$-module. Moreover, if $M$ is a left $D$-module and $N$ is a right $D$-module, then $M{\otimes}_AN$ is a right $D$-module. \\

\noindent
{\it Proof}:  Let us define for any $f\in hom_A(M,N)$:   \\
\[   (af)(m)=af(m)=f(am) \hspace{1cm} \forall a\in A, \forall m\in M\]
\[   (\xi f)(m)=\xi f(m)-f(\xi m)  \hspace{1cm}  \forall \xi ={\xi}^id_i\in T, \forall m\in M  \]
It is easy to check that $\xi a=a \xi+\xi (a)$ in the operator sense and that $\xi\eta -\eta\xi =[\xi,\eta]$ is the standard bracket of vector fields. We have in particular with $d$ in place of any $d_i$: \\
\[  \begin{array}{rcl}
((da)f)(m)=(d(af))(m)=d(af(m))-af(dm)&=&(\partial a)f(m)+ad(f(m))-af(dm)\\
       &=& (a(df))(m)+(\partial a)f(m)  \\
       &=& ((ad+\partial a)f)(m)
       \end{array}  \]
 We may then define for any $m\otimes n\in M{\otimes}_AN$ with arbitrary $m\in M$ and $n\in N$:   \\
 \[      a(m\otimes n)=am\otimes n=m\otimes an\in M{\otimes}_AN  \]
 \[  \xi (m\otimes n)=\xi m\otimes n + m\otimes \xi n \in M{\otimes}_AN   \]
 and conclude similarly with:   \\
 \[  \begin{array}{rcl}
  (da)(m\otimes n)=d(a(m\otimes n)) & = & d(am\otimes n)\\
                         & = &  d(am)\otimes n+am\otimes dn  \\
                             & = & (\partial a)m\otimes n + a(dm)\otimes n + am\otimes dn  \\
                                & = & (ad+\partial a)(m\otimes n)
                                \end{array}    \]
Using $A$ or $K=Q(A)$ in place of $N$, we finally get $(d_if)^k_{\mu}=(d_if)(y^k_{\mu})={\partial}_if^k_{\mu}-f^k_{\mu +1_i}$ that is {\it we recognize exactly the Spencer operator} that we have used in the second Section and thus:\\
\[  (d_i(d_jf))^k_{\mu}={\partial}_{ij}f^k_{\mu}-({\partial}_if^k_{\mu+1_j}+{\partial}_jf^k_{\mu+1_i})+f^k_{\mu+1_i+1_j} \Rightarrow d_i(d_jf)=d_j(d_if)=d_{ij}f \]
In fact, $R$ is the {\it projective limit} of ${\pi}^{q+r}_q:R_{q+r}\rightarrow R_q$ in a coherent way with jet theory [44]. In the more specific case of $hom_A(D,A)$, the upper index $k$ is not present and we have thus $(af)_{\mu}=af_{\mu}$ with $ (d_if)_{\mu}={\partial}_if_{\mu}-f_{\mu+1_i},\forall f\in D^*,\forall a\in A,\forall i=1,...,n$ (Compare to [26], chapter 4, where the Spencer operator is lacking). This left $D$-module structure on $hom_A(D,A)$ is quite different from the one provided by Lemma 3.1 but coincide with it up to sign when $A=k$.\\
Finally, if $M$ and $N$ are right $D$-modules, we just need to set $(\xi f)(m)=f(m\xi)-f(m)\xi, \forall \xi\in T, \forall m\in M$ and conclude as before. Similarly, if $M$ is a left $D$-module and $N$ is a right $D$-module, we just need to set $(m\otimes n)\xi=m\otimes n\xi - \xi m \otimes n$. \\
\hspace*{12cm}     Q.E.D.   \\

\noindent
{\bf REMARK 3.26}: The $A$-module $hom_A(M,N)$ cannot be endowed with any left or right differential structure when $M=M_D$ and 
$N={}_DN$. Similarly, the $A$-module $M{\otimes}_AN$ cannot be endowed with any differential structure when $M=M_D$ and 
$N=N_D$. Also, according to the previous results, when $M={}_DM$ is given one can construct:  \\
\hspace*{1cm} $\bullet$  The {\it right} $D$-module $M^*=hom_D(M,D)$ by using the bimodule structure of $D={}_DD_D$.  \\
\hspace*{1cm} $\bullet$  The {\it left } $D$-module $R=hom_A(M,A)$ for the Spencer operator.  \\
The second situation is the one studied by Macaulay in [26] and we provide a few examples.  \\

\noindent
{\bf REMARK 3.27}: A section $f\in R:y^k_{\mu} \rightarrow f^k_{\mu}\in A$ may not provide a formal power series solution. Accordingly, it may be useful to exhibit $f$ as a formal (in general infinite) summation $E\equiv f^k_{\mu}a^{\mu}_k=0$ called {\it modular equation} by Macaulay ([26], \S 59, p 67) and to set $d_iE\equiv ({\partial}_if^k_{\mu}-f^k_{\mu+1_i})a^{\mu}_k=0$. Equivalently, one can use ${\partial}_i$ on the coefficients of $E$ in $A$ and set $d_ia^{\mu}_k=0$ if ${\mu}_i=0$ or $d_ia^{\mu}_k=-a^{\mu-1_i}_k$ if ${\mu}_i>0$. When $A=K=k$ is a field of constants and $m=1$, we recover exactly the notation of Macaulay ({\it up to sign}) but the link with the Spencer operator has never been provided. \\

\noindent
{\bf EXAMPLE 3.28}: With $n=3, m=1, q=2, K=\mathbb{Q}(x^1,x^2,x^3)$, the tricky example $y_{33}-x^2 y_{11}=0, y_{22}=0$ provided by Janet (See [14] and [16] for more details) is such that $dim_K(R)=12 < \infty$ because $par=\{ y,y_1,y_2,y_3,y_{11}, y_{12},y_{13},y_{23},y_{111},y_{113}, y_{123},y_{1113}\}$. Also, $R$ can be generated by the unique modular equation $E\equiv a^{1113}+x^2a^{1333}+a^{12333}=0$ with $d_2E=0$, because $y_{12333}-y_{1113}=0, y_{1333}-x^2y_{1113}=0$ and all the jets of order $>5$ vanish (Exercise).\\

\noindent
{\bf EXAMPLE  3.29}: With $n=3,m=1,q+2$ and $K=\mathbb{Q}$, let us consider again the second order system $R_2$ defined by the $3$ PD equations:  \\
\[     y_{33}=0, \hspace{1cm}  y_{23} - y_{11}=0 , \hspace{1cm}  y_{22}=0  \]
Such a system is homogeneous and thus {\it automatically} formally integrable but $g_2$ with $dim(g_2)=3$ is not involutive though finite type because $dim(g_3)=1$ and $g_4=0$. An elementary computation of the rank of a $3 \times 3$ matrix shows that the $\delta$-map:   \\
\[      0 \rightarrow {\wedge}^2T^*\otimes g_3 \stackrel{\delta}{\longrightarrow} {\wedge}^3T^*\otimes g_2 \rightarrow 0   \]
is an isomorphism and thus $g_3$ is $2$-ayclic. This crucial intrinsic property, lacking from any "old " work, will be a key tool for studying the conformal group of space-time in the next section. We have the following commutative and exact diagram where $dim(E)=1, dim(F_0)=3, dim(F_1)=18-15=3$ (From the second row):  \\
\[  \begin{array}{rcccccccl}
   & 0 & & 0  & & 0  &  & 0  &   \\
   & \downarrow  &  & \downarrow  & & \downarrow  & & \downarrow   &  \\
0 \rightarrow  & g_5 & \rightarrow  & S_5 T^* & \rightarrow & S_3T^*\otimes F_0 & \rightarrow & T^*\otimes F_1  & \rightarrow 0  \\
  &\hspace{2mm} \downarrow \delta &  &\hspace{2mm} \downarrow \delta & &\hspace{2mm} \downarrow \delta &  & \parallel &   \\
 0 \rightarrow &  T^*\otimes g_4  & \rightarrow& T^*\otimes S_4T^*& \rightarrow  & T^* \otimes S_2T^*\otimes F_0& \rightarrow &T^* \otimes F_1&   \rightarrow 0  \\
   &\hspace{2mm} \downarrow  \delta &   & \hspace{2mm} \downarrow \delta &  &\hspace{2mm} \downarrow \delta  &  & \downarrow  &  \\
 0 \rightarrow & {\wedge}^2T^*\otimes  g_3  &\rightarrow  & {\wedge}^2 T^*\otimes S_3 T^* & \rightarrow  &{\wedge}^2T^* \otimes T^* \otimes F_0  &\rightarrow& 0  \\
   & \hspace{2mm} \downarrow \delta &  &\hspace{2mm}  \downarrow \delta &  &\hspace{2mm}  \downarrow\delta  &   \\
 0 \rightarrow & {\wedge}^3 T^*\otimes g_2 & \rightarrow & {\wedge}^3T ^* \otimes S_2 T^*  & \rightarrow &{\wedge}^3T^*\otimes F_0& \rightarrow &0  \\
   & \downarrow &  & \downarrow &  & \downarrow     \\
   &  0  &    &  0 &    & 0 &   &  & 
  \end{array}   \]
and the long exact sequence:  \\
\[   0 \rightarrow S_6T^* \rightarrow S_4T^* \otimes F_0 \rightarrow S_2T^* \otimes F_1 \rightarrow F_2 \rightarrow 0   \]
giving $dim(F_2)=28-45+18=1$ and providing the following free resolution with second order operators:  \\
\[   0 \rightarrow D  \rightarrow D^3 \rightarrow D^3 \rightarrow D   \rightarrow M  \rightarrow 0   \]
where the Euler-Poincar\'{e} characteristic is equal to $1-3+3-1=0$ as $M$ is defined by a finite type system. With a slight abuse of language while shifting the various filtrations, we may say that we have a strict resolution because all the operators involved, being homogeneous,  are formally integrable though not involutive. This is {\it not} a Janet sequence but we notice that the first and second Spencer sequences coincide because we have $dim(R)=dim(R_3)=1+3+3+1=8$ as $g_4=0$ and $par=\{y,y_1,y_2,y_3,y_{11}, y_{12},y_{13}, y_{111} \}$. We let the reader prove that $R\simeq R_3$ is generated by the {\it single} modular equation  
$E\equiv a^{111} + a^{123}=0 $.   \\

\noindent
{\bf EXAMPLE 3.30}: With $n=1, m=1, q=2, A=\mathbb{Q}[x]\Rightarrow K=\mathbb{Q}(x)$ an thus $k=\mathbb{Q}$, let us consider the second order system $y_{xx}-xy=0$. We successively obtain by prolongation $y_{xxx}-xy_x-y=0, y_{xxxx}-2y_x-x^2y=0, y_{xxxxx}-x^2y_x-4xy=0, y_{xxxxxx}-6xy_x-(x^3+4)y=0$ and so on. We obtain the corresponding board:Ê \\
\[  \begin{array}{r|c|c|c|c|c|c|c|l}
 order & y & y_x & y_{xx} & y_{xxx} & y_{xxxx} & y_{xxxxx} & y_{xxxxxx} &... \\
\hline
2 & -x & 0 & 1 & 0 & 0 & 0 & 0& ...   \\
3 & -1 & -x & 0 & 1 & 0 & 0& 0& ...   \\
4& -x^2 & -2  & 0 & 0 & 1 & 0& 0& ... \\
5& -4x & -x^2 & 0&0 & 0 & 1 & 0 & ... \\
6 & -(x^3+4) & -6x & 0 & 0 & 0 & 0 & 1 & ...
\end{array}  \]

Let us define the sections $f'$ and $f"$ by the following board where $d=d_x$: \\
\[  \begin{array}{r|c|c|c|c|c|c|c|l}
  section& y & y_x & y_{xx} & y_{xxx} & y_{xxxx} & y_{xxxxx} & y_{xxxxxx} &... \\
\hline
f' & 1 & 0 & x & 1 & x^2 & 4x & x^3+4& ...   \\
f" & 0 & 1 & 0 & x & 2 & x^2 & 6x& ...   \\
\hline
df' & 0 & -x  & 0 & -x^2 & -2x & - x^3 & -6x^2 & ... \\
df" & -1 & 0 & -x & -1 & -x^2 & -4x & -x^3-4 & ... \\
\end{array}  \]
in order to obtain $df'=-xf", df"=-f'$. Though this is not evident at first sight, {\it the two boards are orthogonal over} $K$ in the sense that each row of one board contracts to zero with each row of the other though only the rows of the first board do contain a finite number of nonzero elements. {\it It is absolutely essential to notice that the sections} $f'$ {\it and} $f"$ {\it have nothing to do with solutions}  because $df'\neq 0, df"\neq 0$ on one side and also because $d^2f'-xf'=-f"=\frac{1}{x}df'\neq 0$ even though $d^2f"-xf"=0$ on the other side. As a byproduct, $f'$ or $f"$ can be chosen separately as unique generating section of the inverse system over $K$ ({\it care}) and we may write for example $f' \rightarrow E'\equiv a^0+xa^{xx}+a^{xxx}+x^2 a^{xxxx}+ ... =0$ while $f" \rightarrow E"\equiv a^x+xa^{xxx}+2a^{xxxx}+ ... =0$. \\
Finally, setting $f=af'+bf"$, we have $ df=(\partial a)f'+(\partial b-xa)f"=0\Leftrightarrow {\partial}^2a-xa=0, b=\partial a$. If $a=P/Q$ with $P,Q\in \mathbb{Q}[x]$ and $Q\neq 0$, we obtain easily :Ê\\
\[  Q^2{\partial}^2P-2Q\partial P\partial Q-PQ {\partial}^2Q +  2P(\partial Q)^2-xPQ^2=0  \]
If $deg(P)=p, deg(Q)=q$, the four terms on the left have the same degree $p+2q-2$ while the last term has degree $p+2q+1$ and thus $Q\neq 0 \Rightarrow P=0 \Rightarrow a=0 \Rightarrow b=0$, a result showing that there is no solution in $K$. We invite the reader to treat similarly the case $xy_x-y=0$ as an exercise.\\

\noindent
{\bf COROLARY 3.31}: The structures of left $D$-modules existing therefore on $M{\otimes}_AN$ and $hom_A(N,L)$ are now coherent with the {\it adjoint isomorphism} for $mod(D)$:  \\
\[   \varphi :  hom_D(M{\otimes}_AN,L) \stackrel{\simeq}{\longrightarrow} hom_D(M,hom_A(N,L)) \hspace{5mm} ,\forall L,M,N\in mod(D)    \]
It follows that we have also $R=hom_A(M,A)\simeq hom_D(M,hom_A(D,A))$ but in a quite different framework.\\

\noindent
{\it Proof}: With $\varphi(f)=g$, the third result is entrelacing the two left structures that we have just provided through the formula $(g(m))(n)=f(m\otimes n)\in N$ defining the map $\varphi$ whenever $f\in hom_D(M{\otimes}_A N,L)$ is given. Using any $\xi\in T$, we get successively in $L$ (Compare to [7], Proposition 2.1.3, p 54):  \\
\[  \begin{array}{rcl}
(\xi(g(m)))(n)& = & \xi((g(m))(n))-(g(m))(\xi n)  \\
                                   & = & \xi(f(m\otimes n))-f(m\otimes \xi n)    \\
                                   & = & f(\xi(m\otimes n))-f(m\otimes \xi n)  \\
                                   & = & f(\xi m\otimes n+m\otimes \xi n)-f(m\otimes \xi n)  \\
                                   & = & f(\xi m\otimes n)  \\
                                   & = & (g(\xi m))(n)
\end{array}  \]
and thus $ \xi(g(m))=g(\xi m), \forall m\in M $ or simply $\xi\circ g=g\circ \xi$.   \\
For any $g\in hom_D(M,hom_K(N,L))$, we may define the inverse $\psi$ of $\varphi$ through the formula $\psi(g)(m\otimes n)=(g(m))(n)\in L$ by checking the bilinearity over $A$ of $(m,n) \rightarrow (g(m))(n)$ and studying as before the action of any $\xi\in T$.\\

The last result is more tricky and we provide two different proofs with $D^*=hom_A(D,A)$. \\
If $M$ is finitely presented, applying $hom_D(\bullet,D^*)$ to a free presentation $D^p\stackrel{{\cal{D}}}{\longrightarrow}D^m \rightarrow M \rightarrow 0 $, we obtain the exact sequence $0\rightarrow hom_D(M,D^*) \rightarrow D^{*m} \rightarrow D^{*p}$ because $hom_D(D,D^*)=D^*$. As any module over $D$ is a module over $A$, applying $hom_A(\bullet, A)$ to the same sequence, we get the exact sequence $0 \rightarrow hom_A(M,A) \rightarrow D^{*m} \rightarrow D^{* p}$ and thus an isomorphism $R=hom_A(M,A)\simeq hom_D(M, D^*)$.\\
More generally, because $A$ is a commutative ring, we have the isomorphism of left $D$-modues:  \\
\[  M{\otimes}_AN \simeq N{\otimes }_A M: m\otimes n \rightarrow n\otimes m  , \forall m\in M, \forall n\in N    \]
as we check at once for any $a\in A, \xi \in T, m\in M,n\in N$:  \\
\[ a(m\otimes n)=am\otimes n= m\otimes an \rightarrow n\otimes am= a(n\otimes m)  \]
\[  \xi (m\otimes n)=\xi m\otimes n + m \otimes \xi n \rightarrow n\otimes \xi m + \xi n\otimes m = \xi (n\otimes m) \]
ans we may therefore exchange $M$ and $N$. 
Acordingly, when $M={}_DM, N={}_DD, L={}_DA$ with $(P,a) \rightarrow P(a)\in A$, we obtain:   \\
\[ \begin{array}{rcl}
hom_D(M,hom_A(D,A))\simeq hom_D(M{\otimes}_AD,A)\simeq hom_D(D{\otimes}_AM,A) & \simeq & hom_D(D,hom_A(M,A))  \\ 
   & \simeq & hom_A(M,A)  
\end{array}  \]    
 \hspace*{12cm}   Q.E.D.    \\    

\noindent
{\bf REMARK 3.32}: We emphasize once more that the left $D$-structure on $hom_A(D,A)$ used in [7] is coming from the right action of $D$ on $D=D_D$ through the formula $(\xi f)(P)=f(P\xi), \forall \xi\in T, \forall f\in hom_A(D,A)$ and therefore does not provide in general the structure of differential module defined by the formula $(\xi f)(P)=\xi(f(P))-f(\xi P)$ as in the theorem.  \\

\noindent
{\bf COROLLARY 3.33}: Using the bimodule structure of $D={}_DD_D$, we get a right structure on $D_D {\otimes}_A {}_DM$ according to the last Theorem and a compatible left structure defined by $Q(P\otimes m)=QP\otimes m$. With a hat for omission, we may set :\\
\[  \begin{array}{rcl}
d^*(P\otimes m \otimes{\xi}_1\wedge ... \wedge {\xi}_r) & = & {\sum}_i (-1)^{i-1}(P\otimes m){\xi}_i\otimes {\xi}_1\wedge ...\wedge{\hat{\xi}}_i\wedge ...\wedge {\xi}_r   \\
   & & +{\sum}_{i<j}(-1)^{i+j}P\otimes m \otimes [{\xi}_i,{\xi}_j]\wedge {\xi}_1\wedge ... \wedge {\hat{\xi}}_i\wedge ... \wedge {\hat{\xi}}_j\wedge ... \wedge {\xi}_r  
\end{array}   \]
Comparing to the standard definition of the exterior derivative, it is easy to check that $d^*\circ d^*=0$ 

\hspace*{12cm}  Q.E.D.  \\

\noindent
{\bf REMARK 3.34}: When ${\cal{D}}=\Phi \circ j_q$ is an arbitrary but regular operator of order $q$, we may "{\it cut} " the Janet sequence at $F_0$ in two parts by introducing the systems $B_r= im({\rho}_r(\Phi))\subseteq J_r(F_0)$ with $B_0=F_0$ and $B_{r+1}\subseteq {\rho}_r(B_1)$ projecting onto $B_r, \forall r\geq 0$. When $\cal{D}$ is involutive, then $B_1\subseteq J_1(F_0)$ is also involutive with $B_{r+1}={\rho}_r(B_1), \forall r\geq 0$ and we have the commutative and formally exact "{\it fundamental diagram 1} " linking the second Spencer sequence and the Janet sequence:   \\
\[   \begin{array}{rccccccccccl}
  & &  & &0& &  0 &  &   &     &  0 &   \\
 & &  & &\downarrow & & \downarrow  & & &  & \downarrow &  \\
  &  0 \rightarrow & \Theta &\stackrel{j_q}{\longrightarrow}  &C_0  &\stackrel{D_1}{\longrightarrow} & C_1 & \stackrel{D_2}{\longrightarrow} & ...  & \stackrel{D_n}{\longrightarrow} & C_n & \rightarrow 0  \\
 &  &  &  &\downarrow & & \downarrow & & &  & \downarrow & \\
& 0 \rightarrow  &  E & \stackrel{j_q}{\longrightarrow} & C_0(E)&\stackrel{D_1}{\longrightarrow} & C_1(E) & \stackrel{D_2}{\longrightarrow} & ... & \stackrel{D_n}{\longrightarrow} & C_n(E) & \rightarrow 0 \\
 &  &  \parallel & &\hspace{5mm} \downarrow {\Phi}_0& & \hspace{5mm}\downarrow {\Phi}_1 & &  & & \hspace{5mm} \downarrow {\Phi}_n & \\
 0 \rightarrow &\Theta  \rightarrow & E  & \stackrel{{\cal{D}}}{\longrightarrow} &F_0  &\stackrel{{\cal{D}}_1}{\longrightarrow} & F_1 & \stackrel{{\cal{D}}_2}{\longrightarrow} &...  &  \stackrel{{\cal{D}}_n}{\longrightarrow} & F_n & \rightarrow 0 \\
 &  &  &  &\downarrow  &  & \downarrow & & & & \downarrow &    \\
 &  &  &  &0 &  & 0  & & && 0 &   
\end{array}   \]
where the epimorphisms ${\Phi}_1, ..., {\Phi}_n$ are successively induced by the epimorphism ${\Phi}_0=\Phi$, the canonical projection of $C_0(E)=J_q(E)$  onto $F_0=J_q(E)/R_q$ with $C_0=R_q$. It is known that the central sequence is locally exact. As we already pointed out that $g_q$ was a vector bundle, introducing the projection $R_{q-1}$ of $R_q$ into $J_{q-1}(E)$, we have $C_n\simeq {\wedge}^nT^*\otimes R_{q-1}$, $C_n(E)\simeq {\wedge}^nT^*\otimes J_{q-1}(E)$ and thus $F_n\simeq {\wedge}^nT^*\otimes (J_{q-1}(E)/R_{q-1})$. {\it It is not at all evident} that the dual of this diagram is nothing else but the resolution of the short exact sequence $0 \rightarrow I \rightarrow D^m \rightarrow M \rightarrow 0 $ considered in Definition 3.6. Indeed, dualizing the diagram of Proposition 2.8, we obtain at once the following commutative and exact diagram: \\
\[   \begin{array}{cccccl}
       0   &   &   0   &   &   0   &   \\
    \uparrow &  & \uparrow &  & \uparrow  &   \\
 {\wedge}^{r-1}T{\otimes}_A G_{q+1}  & \stackrel{{\delta}^*}{ \leftarrow} & {\wedge}^rT{\otimes}_A G_q &
  \leftarrow & Z ( {\wedge}^rT{\otimes}_A G_q) &  \leftarrow 0  \\
      \parallel  &  &  \uparrow  &  & \uparrow  &  \\
 {\wedge}^{r-1}T{\otimes}_A G_{q+1} &\leftarrow  &{\wedge}^rT{\otimes}_A M_q & \leftarrow &C_r^* &
 \leftarrow 0  \\
      \uparrow  &  & \uparrow &  & \uparrow &   \\
        0  &  \leftarrow &  {\wedge}^rT{\otimes}_A M_{q-1} &  = & {\wedge}^r T{\otimes} M_{q-1} & \leftarrow 0  \\
         &   &  \uparrow &  &  \uparrow  &  \\
        &   &  0  &  &  0  &  
       \end{array}     \]
Applying the dual Spencer operator ${\wedge}^rT{\otimes}_AM_q \rightarrow {\wedge}^{r-1}T{\otimes}_AM_{q+1}$, we obtain the strictly exact 
{\it second Spencer sequence} $SSP_q(M)$:   \\
\[  0 \rightarrow D{\otimes}_AC_n^* \rightarrow D{\otimes}_AC_{n-1}^* \rightarrow ... \rightarrow D{\otimes}_AC_1^* \rightarrow D{\otimes }_AM_q \rightarrow M \rightarrow 0    \]
which is a resolution of $M$ {\it stabilizing the filtration at order} $q$ {\it only} by means of induced differential modules. Accordingly, the last two differential morphisms, induced by the morphisms $P\otimes \xi \otimes m \rightarrow P\xi \otimes m - P\otimes \xi m$ and $P \otimes m \rightarrow Pm$ of the sequence $... \rightarrow D{\otimes}_AT{\otimes}_AM_q\rightarrow D{\otimes}_AM_{q+1} \rightarrow M \rightarrow 0$, dualize the exact sequence $0 \rightarrow R_{q+r+1} \rightarrow J_{r+1}(C_0) \rightarrow J_r(C_1)$ as in ([15], p 367-369). \\

In the opinion of the author based on thirty years of explicit applications to mathematical physics (general relativity, gauge theory, theoretical mechanics, control theory), the differential geometric framework is quite more natural than the differential algebraic framework. The simplest example being the fact that the so-called {\it Cosserat equations} of elasticity theory, discovered by the brothers Eug\`{e}ne and Fran\c{c}ois Cosserat as early as in 1909, are nothing else but the formal adjoint $ad(D_1)$ of the first Spencer operator $D_1$ for the {\it Killing equations} in Riemannian geometry [40, 47]. In particular, it must be noticed that the very specific properties of the Janet sequence, namely that it starts with an involutive operator of order $q\geq 1$ but the $n$ remaining involutive operators ${\cal{D}}_1, ..., {\cal{D}}_n$ are of order $1$ and in (reduced) Spencer form cannot be discovered from the differential module point of view. However, the importance of the torsion-free condition/test for differential modules is a novelty brought from the algebraic setting and known today to be a crucial tool for understanding control theory [44]. Finally, the situation in the present days arrived to a kind of "{\it vicious circle} " because the study of differential modules is based on filtration and thus formal integrability while computer algebra is based on Gr\"{o}bner bases as a way to sudy the same questions but by means of highly non-intrinsic procedures as we saw.  \\

   We may compare the differential algebraic framework with its differential geometric counterpart. Indeed, using notations coherent with the ones of the previous section, if now ${\cal{D}}=\Phi \circ j_q:E \rightarrow F$ is an operator of order $q$ with $dim(E)=m,dim(F)=p$, we may consider the exact sequences $0 \rightarrow R_{q+r} \rightarrow J_{q+r}(E) \stackrel{{\rho}_r(\Phi)}{\longrightarrow} J_r(F)$ by introducing the $r$-prolongation of $\Phi$, induce the Spencer operator $D:R_{q+r+1}\rightarrow T^*\otimes R_{q+r}$ when $r\geq 0$ and pass to the projective limit $R=R_{\infty}$. In actual practice, when $r=1$ we have $a^{\tau\mu}_kf^k_{\mu}=g^{\tau} \Rightarrow a^{\tau\mu}_kf^k_{\mu+1_i}+({\partial}_i a^{\tau\mu}_k)f^k_{\mu}=g^{\tau}_i $ and thus $a^{\tau\mu}_k({\partial}_if^k_{\mu}-f^k_{\mu+1_i})={\partial}_ig^{\tau}-g^{\tau}_i$, a procedure that can be easily extended to any value of $r>0$. As a byproduct, the link existing with infinite jets can be understood by means of the following commutative and exact diagram:  \\
\[   \begin{array}{rccccc}
0 \rightarrow & R & \rightarrow & J(E) & \stackrel{\rho(\Phi)}{\longrightarrow} & J(F)  \\
                     &\hspace{3mm} \downarrow d  &     &  \downarrow d  &    &   \downarrow d     \\
0 \rightarrow & T^* \otimes R & \rightarrow & T^*\otimes J(E) & \stackrel{\rho(\Phi)}{\longrightarrow} & T^*\otimes J(F) 
\end{array}   \]
where $df=dx^i\otimes d_if$. Hence, using the Spencer operator on sections, we may characterize $R$ by the following equivalent properties (See [41], Proposition 10, p 83 for a nonlinear version that can be used in Example 2.22): \\

\noindent
1) $f\in R$ is killed by ${\rho}_r(\Phi)$ ({\it no differentiation of} $f$ {\it is involved} ), $\forall r\geq 0$.  \\
2) $f\in R \Rightarrow d_if\in R$ ({\it a differentiation of} $f$ {\it is involved} ), $\forall i=1,...,n$.  \\ 

As an equivalent differential geometric counterpart of the above result, we may also define the {\it r-prolongations} ${\rho}_r(R_q)=J_r(R_q)\cap J_{q+r}(E)$ of a given system $R_q\subset J_q(E)$ of order $q$ by applying successively the following formula involving the Spencer operator of the previous section:  \\
\[    {\rho}_1(R_q)=J_1(R_q)\cap J_{q+1}(E)=\{ f_{q+1}\in J_{q+1}(E) \mid f_q\in R_q, Df_{q+1}\in T^*\otimes R_q\}   \]
Now, if we have another system $R_{q+1}\subseteq {\rho}_1(R_q) \subset J_{q+1}(E)$ of order $q+1$ {\it and projecting onto} $R_q$, we have the commutative and exact diagram:\\
\[  \begin{array}{rcccl}
  & 0  &  & 0  &   \\
  & \downarrow &  & \downarrow &   \\
  0 \rightarrow & g_{q+1}  & \rightarrow  & {\rho}_1(g_q) &     \\
  & \downarrow & &  \downarrow &  \\
  0 \rightarrow & R_{q+1} & \rightarrow & {\rho}_1(R_q) &   \\
  &  \downarrow &  & \downarrow &   \\
  0 \rightarrow & R_q & = & R_q & \rightarrow 0  \\
  &  \downarrow &  & \downarrow  &  \\
  & 0 &  & 0 &    
  \end{array}     \]
 Chasing in this diagram, it follows that $R_{q+1}={\rho}_1(R_q)$ if and only if $g_{q+1}={\rho}_1(g_q)$. Otherwise, we may start afresh with $R^{(1)}_q={\pi}^{q+1}_q(R_{q+1})$ (See Lemma III.2.46 in [44] for details).  \\

Dualizing the Spencer operator acting in the two diagrams presented at the end of the previous section, we get the two commutative diagrams of induced left $D$-modules:  \\
\[  \small  \begin{array}{ccccccc}
  0  &   &  0    &  &  0  &  &  0  \\
  \downarrow  &    &  \downarrow   &   &  \downarrow  &  &  \downarrow   \\
  D\otimes C^*_r &  \stackrel{D^*_r}{\longrightarrow}  &  D \otimes C^*_{r-1}  &  \hspace{2mm}  &  D \otimes F^*_r & \stackrel{{\cal{D}}_r^* }{\longrightarrow}  & D \otimes F^*_{r-1}   \\
  \downarrow  &    &  \downarrow  &  &  \downarrow  &    &  \downarrow   \\ 
  D \otimes {\wedge}^r T \otimes M_q   &   \stackrel{d^*}{\longrightarrow}   &  D \otimes {\wedge}^{r-1}T \otimes M_{q+1}      &   \hspace{2mm}  &  
D \otimes {\wedge}^rT \otimes D_q \otimes E^*  & \stackrel{d^*}{\longrightarrow}  &  D \otimes {\wedge}^{r-1}T \otimes D_{q+1}\otimes E^*
\end{array}  \]

\vspace{3mm}
\noindent
with the inclusions:   \\
\[  C^*_0 =M_q  \hspace{5mm} \Rightarrow   \hspace{5mm}  {\wedge}^r T\otimes M_{q-1}  \subset C^*_r  \subset {\wedge}^r T\otimes M_q  \subset  {\wedge}^rT \otimes M_{q+1}   \]

\noindent
{\bf THEOREM 3.35}: The operator $ d^* : D \otimes {\wedge}^r T \otimes M_q     \rightarrow   D \otimes {\wedge}^{r-1}T \otimes M_{q+1} $ is described by the formula:  \\
\[ \begin{array}{ccl}
 d^*(P\otimes m\otimes {\xi}_1\wedge ... \wedge {\xi}_r)  & = &  {\sum}_i (-1)^{i-1}(P\otimes m){\xi}_i \otimes {\xi}_1\wedge ... \wedge {\hat{\xi}}_i\wedge ... \wedge {\xi}_r  \\
   &  & +{\sum}_{i<j}(-1)^{i+j} (P\otimes m) \otimes  [{\xi}_i,{\xi}_j]\wedge ... \wedge {\hat{\xi}}_i \wedge ... \wedge {\hat{\xi}}_j \wedge ... \wedge {\xi}_r 
\end{array}   \]

\noindent
{\it Proof}: Let us introduce a left structure on $D\otimes M$ by using the left structure already exhibited in the last Theorem with ${}_DD$ and ${}_DM$ by setting $a(P\otimes m)=aP\otimes m=P\otimes am$ and $\xi(P\otimes m)=\xi P\otimes m+P\otimes \xi m$. Then let us introduce a right structure  with $D_D$ and ${}_DM$ by setting $(P\otimes m)Q=PQ\otimes m$. We check at once for any $P,Q\in D, \xi \in T, m\in M$:  \\
\[     (\xi(P\otimes m))Q=(\xi P\otimes m + P\otimes \xi m)Q=\xi P Q \otimes m+ PQ \otimes \xi m=\xi ((P\otimes m)Q)   \]
We may thus introduce a morphism $d^*: D\otimes {\wedge}^rT\otimes M \rightarrow D \otimes {\wedge}^{r-1}T\otimes M$ by the above formula:  \\
\[ \begin{array}{ccl}
 d^*(P\otimes m\otimes {\xi}_1\wedge ... \wedge {\xi}_r)  & = &  {\sum}_i (-1)^{i-1}(P\otimes m){\xi}_i \otimes {\xi}_1\wedge ... \wedge {\hat{\xi}}_i\wedge ... \wedge {\xi}_r  \\
   &  & +{\sum}_{i<j}(-1)^{i+j} (P\otimes m) \otimes  [{\xi}_i,{\xi}_j]\wedge ... \wedge {\hat{\xi}}_i \wedge ... \wedge {\hat{\xi}}_j \wedge ... \wedge {\xi}_r 
\end{array}   \]
where a "{\it hat} " is used for omission, or simply:  \\
\[  d^*(P\otimes m \otimes d_{i_1}\wedge ... \wedge d_{i_r}) = {\sum}_s (-1)^{s-1}(P\otimes m)d_{i_s}\otimes d_{i_1} \wedge ... \wedge {\hat{d}}_{i_s}\wedge ... \wedge d_{i_r}   \]
Having in mind a similar formula existing for the exterior derivative in the Poincar\'e sequence, it is easy to check that $d^* \circ d^*=0$. However, {\it it is not evident at all} to establish a link with the Spencer operator and we notice that there are almost no references to Spencer in the literature on $D$-modules [7, 20, 62]. For this, let us start with the simple example of the sequence:  \\
\[  0 \rightarrow  E \stackrel{j_2}{\longrightarrow}J_2(E) \stackrel{d}{\longrightarrow} T^*\otimes J_1(E)   \]
where we have used sections and the notation $d$ instead of $D$ in order to avoid any confusion. With $n=1,m=1, d=d_x$, we have the operators:  \\ 

\[   f \stackrel{j_2}{\longrightarrow} \left( \begin{array}{r} f  \\ df \\ d^2f  \end{array}  \right) , \hspace{1cm}  \left( \begin{array}{l} f  \\ f_x \\ f_{xx}  \end{array}  \right) \stackrel{d}{\longrightarrow}  \left(  \begin{array}{l}  df-f_x  \\ df_x-f_{xx}  \end{array} \right)   \]
and the operator matrix identity:  \\
\[   \left(  \begin{array}{ccc}  d & -1 & 0 \\ 0 & d & -1  \end{array} \right)  \left(  \begin{array}{c} 1 \\ d \\ d^2  \end{array}  \right) =                               
   \left( \begin{array}{c}  0  \\ 0 \end{array} \right)  \]
More genrally, we have in the operator sense:  \\
\[   P^{\mu,i}(d_if^k_{mu}-f^k_{\mu +1_i})=(P^{\mu,i} d_i)f^k_{\mu} - P^{\mu,i} f^k_{\mu + 1_i}   \]
that is a composition with $d_i$ on the right and a shift by one step to increasing order because $d_iy^k_{\mu}=y^k_{\mu + 1_i} $ and 
$\mid \mu + 1_i \mid = \mid \mu \mid + 1 $.  \\
\hspace*{12cm}   Q.E.D.  \\

\noindent
{\bf COROLLARY 3.36}: Dualizing the canonical Spencer sequence, we get the strictly exact canonical sequence of left $D$-modules and $D$-morphisms:  \\
\[  0 \rightarrow D\otimes C^*_n \stackrel{D^*_n}{\longrightarrow}  ... \stackrel{D^*_2}{\longrightarrow} D \otimes C^*_1 
\stackrel{D^*_1}{\longrightarrow}  D\otimes C^*_0 \rightarrow M \rightarrow 0  \] 
where $C^*_n={\wedge}^nT\otimes M_{q-1}, C^*_0= M_q$ and the last two morphisms are induced by:  \\
\[  D\otimes T \otimes M_q \rightarrow  D\otimes M_{q+1}: P\otimes \xi \otimes m \rightarrow P\xi \otimes m - P \otimes \xi m , \hspace{4mm} 
D\otimes M_q \rightarrow  M: P\otimes m \rightarrow Pm   \]

\noindent
{\bf COROLLARY 3.37}: Similarly, dualizing the canonical Janet sequence, we get the strictly exact canonical sequence of left $D$-modules and $D$-morphisms:  \\
\[  0 \rightarrow D \otimes F^*_n \stackrel{{\cal{D}}^*_n}{\longrightarrow}  ...  \stackrel{{\cal{D}}^*_1}{\longrightarrow}D\otimes F^*_0 \stackrel{{\cal{D}}^*}{\longrightarrow}  D\otimes E^*  \rightarrow M  \rightarrow 0  \]
where we have the short exact sequences:ÊÊÊ  \\
\[ 0 \rightarrow R_q \rightarrow J_q(E) \rightarrow F_0 \rightarrow 0 \hspace{5mm}  \Leftrightarrow \hspace{5mm}  0 \rightarrow F^*_0 \rightarrow D_q\otimes E^* \rightarrow M_q \rightarrow  0 \]                                                                                                                                                                                                                                                                                                                                                                                                                                                                                                                                                    
and the last morphism just provides the definition of $M$ in a more intrinsic way than the cokernel $D^p \rightarrow D^m \rightarrow M \rightarrow 0 $ already used. \\

 \newpage

\noindent
{\bf 4)  APPLICATIONS}  \\ 

   In this last section, we shall only deal wih linear or linearized differential operators. However, as explained with details in [38-41, 55], there is a nonlinear counterpart using the {\it nonlinear Janet sequence} coming from the {\it Vessiot structure equations} and a {\it nonlinear Spencer sequence}. However, the so-called {\it vertical machinery} involved, that is a systematic use of {\it fibered manifolds} and {\it vertical bundles}, is much more difficult though we have chosen the notations of this paper in such a way that the interested reader may easily adapt them. As for the quoted Vessiot structure equations, {\it they have been totally ignored during more than one century} for reasons that are not scientific at all (See [50] and the original letters presented in [40] for explanations).\\

Collecting all the results so far obtained, if a differential operator ${\cal{D}}$ is given in the framework of differential geometry, we may keep the same operator matrix in the framework of differential modules which are {\it left} modules over the ring $D$ of linear differential operators. We may also apply duality over $D$, that is apply $hom_D(\bullet,D)$, provided we deal now with {\it right} differential modules or use the operator matrix of $ad({\cal{D}})$ and deal again with {\it left} differential modules obtained through the $left \leftrightarrow right$ {\it conversion} procedure. In actual practice, it is essential to notice that {\it the new operator matrix may be quite different from the only transposed of the previous operator}, even if we are dealing with constant coefficients.\\

\noindent
{\bf DEFINITION 4.1}: If a differential operator $\xi \stackrel{\cal{D}}{\longrightarrow} \eta$ is given, a {\it direct problem} is to find (generating) {\it compatibility conditions} (CC) as an operator $\eta \stackrel{{\cal{D}}_1}{\longrightarrow} \zeta $ such that ${\cal{D}}\xi=\eta \Rightarrow {\cal{D}}_1\eta=0$. Conversely, given $\eta \stackrel{{\cal{D}}_1}{\longrightarrow} \zeta$, the {\it inverse problem} will be to look for $\xi \stackrel{\cal{D}}{\longrightarrow} \eta$ such that ${\cal{D}}_1$ generates the CC of ${\cal{D}}$ and we shall say that ${\cal{D}}_1$ {\it is parametrized by} ${\cal{D}}$ {\it if such an operator} ${\cal{D}}$ {\it is existing}.  \\

 \noindent
 {\bf REMARK 4.2}: Of course, solving the direct problem (Janet, Spencer) is {\it necessary} for solving the inverse problem. However, though the direct problem always has a solution, the inverse problem may not have a solution at all and the case of the Einstein operator is one of the best non-trivial PD counterexamples (Compare [44] to [67]). It is rather striking to discover that, in the case of OD operators, it took almost 50 years to understand that the possibility to solve the inverse problem was equivalent to the controllability of the corresponding control system (Compare [34] to [44]) and the situation will be probably similar in GR as the above result has been first found in 1994 according to the Introduction of this paper.  \\
 
As $ad(ad(P))=P, \forall P \in D$, any operator is the adjoint of a certain operator and we get:  \\

\noindent
{\bf FORMAL TEST  4.3}: The {\it double duality test} needed in order to check whether $t(M)=0$ or not and to find out a parametrization if $t(M)=0$ has 5 steps which are drawn in the following diagram where $ad({\cal{D}})$ generates the CC of $ad({\cal{D}}_1)$ and 
${\cal{D}}_1'$ generates the CC of ${\cal{D}}=ad(ad({\cal{D}}))$:  \\
\[  \begin{array}{rcccccl}
 & & & & &  {\zeta}' &\hspace{15mm} 5  \\
 & & & & \stackrel{{\cal{D}}'_1}{\nearrow} &  &  \\
4 \hspace{15mm}& \xi  & \stackrel{{\cal{D}}}{\longrightarrow} &  \eta & \stackrel{{\cal{D}}_1}{\longrightarrow} & \zeta &\hspace{15mm}   1  \\
 &  &  &  &  &  &  \\
 &  &  &  &  &  &  \\
 3 \hspace{15mm}& \nu & \stackrel{ad({\cal{D}})}{\longleftarrow} & \mu & \stackrel{ad({\cal{D}}_1)}{\longleftarrow} & \lambda &\hspace{15mm} 2
  \end{array}  \]
\vspace*{3mm}

\noindent
{\bf THEOREM 4.4}: We have ${\cal{D}}_1$ parametrized by ${\cal{D}} \Leftrightarrow {\cal{D}}_1={\cal{D}}'_1 \Leftrightarrow t(M)=0 \Leftrightarrow ext^1(N)=0 $ in the differential module framework. In particular, a necessary condition for an operator ${\cal{D}}_1$ to be parametrizable by an operator ${\cal{D}}$ is that 
$ max_{\chi}rk({\sigma}_{\chi}({\cal{D}}_1)) < dim(F_0)$ with a strict inequality.  \\
 
\noindent
{\bf COROLLARY 4.5}: In the differential module framework, if $F_1 \stackrel{{\cal{D}}_1}{\longrightarrow} F_0 \stackrel{p}{\longrightarrow} M \rightarrow 0$ is a finite free presentation of $M=coker({\cal{D}}_1)$ and we already know that $t(M)=0$ by using the preceding Theorem, then we may obtain an exact sequence $F_1 \stackrel{{\cal{D}}_1}{\longrightarrow} F_0 \stackrel{{\cal{D}}}{\longrightarrow} E $ of free differential modules where ${\cal{D}}$ is the parametrizing operator. However, there may exist other parametrizations $F_1 \stackrel{{\cal{D}}_1}{\longrightarrow} F_0 \stackrel{{\cal{D}}'}{\longrightarrow} E' $ called {\it minimal parametrizations} such that $coker({\cal{D}}')$ is a torsion module and we have thus $rk_D(M)=rk_D(E')$.  \\

\noindent
$\bullet$ $n=2$: The {\it Airy parametrization} of  the Cauchy stress equations when $n=2$ gives $rk(E)=1$ and we have thus only $1$ potential, namely the Airy function, that is the parametrization is trivially minimal. When constructing a dam as in the Introduction of [43], we may transform a problem of $3$-dimensional elasticity into a problem of $2$-dimensional elasticity by supposing that the axis $x^3$ is perpendicular to the river with ${\Omega}_{ij}(x^1,x^2), \forall i,j=1,2$ but ${\Omega}_{33}=0$ because of the rocky banks of the 
river and we may introduce the two {\it Lam\'{e} constants} $(\lambda,\mu)$ in order to describe the usual constitutive relations of an homogeneous isotopic medium as follows:  \\
\[ {\sigma}=\frac{1}{2}\lambda \, tr(\Omega)\, {\omega} + \mu \, {\Omega} \hspace{1cm}  \Leftrightarrow   \hspace{1cm}  
 \mu \, {\Omega}=    \sigma - \frac{\lambda}{2( \lambda + \mu)}\, tr(\sigma) \, \omega  \]
even though ${\sigma}^{33}=\frac{1}{2}\lambda ({\Omega}_{11}+{\Omega}_{22})=\frac{1}{2}\lambda tr(\Omega)\neq 0$ and thus 
${\sigma}^{33}=\frac{\lambda}{2(\lambda + \mu)}({\sigma}^{11} + {\sigma}^{22})$ where we have introduced the {\it Poisson coefficient} $\nu=\frac{\lambda}{2(\lambda + \mu)}$.  \\
Let us consider the right square of the diagram below with locally exact rows:  \\
\[   \begin{array}{ccccc}
   2  & \stackrel{Killing}{\longrightarrow} & 3 & \stackrel{Riemann}{\longrightarrow} & 1   \\
   \vdots &  & {\downarrow\uparrow} &  &  \vdots  \\
   2 & \stackrel{Cauchy}{\longleftarrow} & 3 & \stackrel{Airy}{\longleftarrow} & 1
\end{array}  \]
Taking into account the formula $5.1.4$ of [15] and substituting the Airy parametrization, we obtain:  \\
\[ R\equiv d_{11}{\Omega}_{22}+d_{22}{\Omega}_{11}-2d_{12}{\Omega}_{12}=0 \hspace{3mm} \Rightarrow 
\hspace{3mm} \mu \, R\equiv \frac{\lambda + 2 \mu}{2(\lambda +\mu)} \triangle \triangle \phi=0 \hspace{3mm} \Rightarrow 
\hspace{3mm} \triangle \triangle \phi=0  \]
Of course, the Airy potential $\phi$ has {\it nothing to do} with the perturbation $\Omega$ of the metric $\omega$. Also, as we shall see in the next paragraph, the origin of elastic waves is shifted by one step to the left square of the diagram.  \\

 \noindent
$\bullet$ $n=3$: The {\it Beltrami parametrization} of the Cauchy stress equations  when $n=3$ gives $rk(E)=6$ and we have thus $6$ potentials. However, Maxwell/Morera parametrizations of the stress equations when $n=3$ both give $rk(E')=3$ and we have thus $3$ potentials only. Paying a tribute to History, we shall set $Beltrami=ad(Riemann)$ and we have the following dual commutative and exact diagrams:  \\
\[  \begin{array}{rccccccl}
  &3 & \stackrel{Killing}{\longrightarrow } & 6 & \stackrel{Riemann}{\longrightarrow}& 6 & \stackrel{Bianchi}{\longrightarrow} & 3\rightarrow 0  \\
   &  &  &  &  &  &  &  \\
0 \leftarrow & 3 & \stackrel{Cauchy}{\longleftarrow} & 6 &\stackrel{Beltrami}{\longleftarrowÊ} & 6 &\longleftarrow &  3\\
                     &    &                                                             & \parallel &                                         &  \uparrow &                   &     \\
 &  &  & 6 & \stackrel{Maxwell}{\longleftarrow} & 3  &  &
\end{array}  \] 
Accordingly, the canonical parametrization has $6$ potentials while {\it any} minimal parametrization has $3$ potentials. We finally notice that the Cauchy operator is parametrized by the Beltrami operator which is {\it again} parametrized by the adjoint of the Bianchi operator obtained by linearizing the Bianchi identities existing for the Riemann tensor, {\it a property not held by any minimal parametrization} as we already noticed.\\
We now explain the origin and existence of elastic waves in this framework, pointing out first of all that any earthquake allows to verify the different types of waves propagating with different speeds. In addition to these types of waves, there also exists specific waves propagating on the surface of materials, like the {\it Rayleigh waves} discovered in $1885$, with an exponential decay of amplitude and a different speed 
$v_R$, inviting the reader to visit the so-called {\it whispering cupola} of St Paul's Cathedral in London. \\
For this, let us consider the left square of the diagram below with locally exact rows:  \\
\[   \begin{array}{ccccc}
   3  & \stackrel{Killing}{\longrightarrow} & 6 & \stackrel{Riemann}{\longrightarrow} & 6   \\
   \vdots &  & {\downarrow\uparrow} &  &  \vdots  \\
   3 & \stackrel{Cauchy}{\longleftarrow} & 6 & \stackrel{Beltrami}{\longleftarrow} & 6
\end{array}  \]
where the central vertical maps are described by the symmetric $6\times 6$ matrix of the well known constitutive laws for homogeneous isotropic media and its inverse:  \\
 \[  {\sigma}=\frac{1}{2}\lambda \, tr(\Omega)\, {\omega} + \mu \, {\Omega} \hspace{1cm}  \Leftrightarrow   \hspace{1cm}  
 \mu \, {\Omega}=    \sigma - \frac{\lambda}{3\, \lambda  + 2 {\mu}}\, tr(\sigma) \, \omega         \]
 with $tr(\sigma)={\omega}_{ij}{\sigma}^{ij}$ and $tr(\Omega)={\omega}^{ij}{\Omega}_{ij}$. Substituting in the Cauchy equations ${\partial}_i{\sigma}^{ij}=f^j$, we finally get:   \\
 \[   (\lambda + \mu)\,  \vec{\nabla}\, (\vec{\nabla}.\vec{\xi}\, )+ \mu \, \triangle \,  \vec{\xi} = \vec{f}   \] 
 Using the standard formula $\vec{\nabla} \wedge (\vec{\nabla}\wedge \vec{\xi}\,)= \vec{\nabla}(\vec{\nabla}. \vec{\xi}\, )- \triangle \vec{\xi}$, we have to consider two particular situations providing {\it longitudinal} and {\it transversal} waves with respective speeds 
 $v_R < v_T < v_L$ when $f^j=\rho\, {\partial}^2\vec{\xi}/\partial t^2$ with mass $\rho$ per unit volume [42]:  \\
 \noindent
 \[  \left\{   \begin{array}{rcrcl}
  \vec{\nabla}.\vec{\xi}=0  &  \hspace{1cm}\Rightarrow \hspace{1cm}&  \mu \triangle \vec{\xi}= \vec{f} &  \hspace{1cm} \Rightarrow \hspace{1cm} &  v_T=\sqrt{\frac{\mu}{\rho}}  \\
  \vec{\nabla}\wedge \vec{\xi}=0 & \hspace{1cm}\Rightarrow \hspace{1cm}& (\lambda + 2 \mu )\triangle \vec{\xi}= \vec{f} & \hspace{1cm}   \Rightarrow \hspace{1cm} & v_L=\sqrt{\frac{\lambda + 2\mu}{\rho}}  
\end{array} \right. \] 
Taking into account Proposition 3.18 and formula $(5.1.6)$ of [15] allowing to exhibit gravitational waves while dualizing in arbitrary dimension $n$, we may consider the change of stress functions with {\it inverse now depending on} $n$: \\
\[   {\bar{\Phi}}_{ij}={\Phi}_{ij}- \frac{1}{2}{\omega}_{ij}tr(\Phi) \hspace{6mm} \Leftrightarrow \hspace{6mm}
{\Phi}_{ij}= {\bar{\Phi}}_{ij} -  \frac{1}{(n-2)}{\omega}_{ij}tr(\bar{\Phi}) \]
It follows that, for $n=3$, we have ${\Phi}_{11}= - ({\bar{\Phi}}_{22}+{\bar{\Phi}}_{33})$ and ${\Phi}_{12}={\bar{\Phi}}_{12}$ that can be extended by circular permutation of $(1,2,3)$. We obtain therefore, whenever ${\partial}_i{\bar{\Phi}}^{ij}=0$: \\
\[ \begin{array}{rcl}
 {\sigma}^{11} & = & -d_{33}({\Phi}_{11}+{\Phi}_{33}) - d_{22}({\Phi}_{11}+{\Phi}_{22}) - 2 d_{23}{\Phi}_{23}   \\
   &  =  & - \triangle {\Phi}_{11} + d_{11}{\Phi}_{11}-(d_{33}{\Phi}_{33}+d_{23}{\Phi}_{23})-( d_{22}{\Phi}_{22}+ d_{23}{\Phi}_{23})  \\
   &  =  &  -\triangle {\Phi}_{11} +d_{11}{\Phi}_{11}+ d_{13}{\Phi}_{13} + d_{12}{\Phi}_{12}\\
   &  =  & -\triangle {\Phi}_{11}                                                                     
\end{array}   \]
\[  \begin{array}{rcl}
{\sigma}^{12} &  =  & d_{13}{\Phi}_{23}+d_{23}{\Phi}_{13} -d_{33}{\Phi}_{12}+ d_{12}({\Phi}_{11}+{\Phi}_{22})  \\
   & = & -\triangle {\Phi}_{12}+(d_{11}{\Phi}_{12}+d_{13}{\Phi}_{23})+(d_{22}{\Phi}_{12}+d_{23}{\Phi}_{13})+d_{12}({\Phi}_{11}+{\Phi}_{22})  \\
  & =  & -\triangle{\Phi}_{12}- d_{12}{\Phi}_{22}-d_{12}{\Phi}_{11}+d_{12}({\Phi}_{11}+{\Phi}_{22}) \\
  & =  & - \triangle {\Phi}_{12}
\end{array}  \]
where we have omitted the "bar\," and use the {\it formal} $d$ instead of the {\it partial} $\partial$ for simplicity. \\
Of course, the potential $\Phi$ has {\it nothing to do} with the perturbation $\Omega$ of the metric $\omega$.  \\

\noindent
$\bullet$ $n=4$ We shall prove below that the {\it Einstein parametrization} of the stress equations is neither canonical nor minimal in the following diagrams:  \\
\[   \begin{array}{rcccccccccl}
 &4 & \stackrel{Killing}{\longrightarrow} & 10 & \stackrel{Riemann}{\longrightarrow} & 20 & \stackrel{Bianchi}{\longrightarrow} & 20 & \longrightarrow & 6 & \rightarrow 0 \\
  &   &                                                            & \parallel &       & \downarrow  &  & \downarrow &  &   \\
 &    &                                                            &  10     & \stackrel{Einstein}{\longrightarrow}  & 10  & \stackrel{div}{\longrightarrow} & 4    & \rightarrow & 0                      &   \\  
  &   &  &  &  &  &  &  &  \\   
0 \leftarrow & 4 & \stackrel{Cauchy}{\longleftarrow} & 10 & \stackrel{Beltrami}{\longleftarrow} & 20 & \longleftarrow & 20 &  & & \\
                     &           &                                                         & \parallel &    & \uparrow  &  &  &    \\
  &    &      & 10 &  \stackrel{Einstein}{\longleftarrow} & 10 &  &  &  & & 
\end{array}   \]
obtained by using the fact that {\it the Einstein operator is self-adjoint}, where by Einstein operator we mean the linearization of the Einstein equations at the Minkowski metric, the $6$ terms being exchanged between themselves [45, 50]. Indeed, setting $E_{ij}=R_{ij}- \frac{1}{2} {\omega}_{ij}tr(R)$ with $tr(R)={\omega}^{ij}R_{ij}$, it is essential to notice that the {\it Ricci operator is not self-adjoint} because we have for example:  \\
\[  {\lambda}^{ij} ({\omega}^{rs}d_{ij}{\Omega}_{rs}) \stackrel{ad}{\longrightarrow} ({\omega}^{rs}d_{ij}{\lambda}^{ij})
{\Omega}_{rs}  \]
and $ad$ provides a term appearing in $- {\omega}_{ij}tr(R)$ but {\it not} in $2R_{ij}$ because we have, as in $(5.1.4)$ of [15]:  \\
\[  tr(\Omega)={\omega}^{rs}{\Omega}_{rs}  \hspace{1cm} \Rightarrow \hspace{1cm}
 tr(R)={\omega}^{rs}d_{rs} tr(\Omega)- d_{rs}{\Omega}^{rs}  \]

The upper $div$ induced by $Bianchi$ has {\it nothing to do} with the lower $Cauchy$ stress equations, contrary to what is still believed today while the $10$ {\it on the right} of the lower diagram has {\it nothing to do} with the perturbation of a metric which is the $10$ {\it on the left} in the upper diagram. It also follows that the Einstein equations in vacuum cannot be parametrized as we have the following diagram of operators recapitulating the five steps of the parametrizability criterion (See [44, 45] for more details or [55, 67] for a computer algebra exhibition of this result): \\
\[  \begin{array}{rcccl}
  &  &  &\stackrel{Riemann}{ }  & 20   \\
  & &  & \nearrow &    \\
 4 &  \stackrel{Killing}{\longrightarrow} & 10 & \stackrel{Einstein}{\longrightarrow} & 10  \\
  & & & &  \\
 4 & \stackrel{Cauchy}{\longleftarrow} & 10 & \stackrel{Einstein}{\longleftarrow} & 10 
\end{array}  \]

As a byproduct, we are facing {\it only two} possibilities, both leading to a contradiction:  \\
1) If we use the operator $S_2T^* \stackrel{Einstein}{\longrightarrow} S_2T^*$ in the geometrical setting, the $S_2T^*$ on the left has indeed {\it someting to do} with the perturbation of the metric but the $S_2T^*$ on the right has {\it nothing to do} with the stress. \\
2) If we use the adjoint operator ${\wedge}^nT^*\otimes S_2T\stackrel{ad(Einstein)}{\longleftarrow} {\wedge}^nT^*\otimes S_2T$ in the physical setting, then ${\wedge}^nT^*\otimes S_2T$ on the left has of course {\it something to do} with the stress but the ${\wedge}^nT^*\otimes S_2T$ on the right has {\it nothing to do} with the perturbation of a metric. \\

These purely mathematical results question the origin and existence of gravitational waves.  \\

\noindent
$\bullet$ It remains therefore to compute all the dimensions and ranks for an arbitrary dimension $n\geq 3$. For this, we notice that the successive prolongations ${\rho}_r(\Phi):J_{q+r}E \rightarrow J_r(F_0)$ defined by $d_{\nu}{\Phi}^{\tau}=z^{\tau}_{\nu}$ for $0\leq \mid \nu \mid \leq r$ have kernel $R_{q+r}$. The {\it symbol morphism} ${\sigma}_r(\Phi): S_{q+r}T^*\otimes E \rightarrow S_rT^*\otimes F_0$ with kernel $g_{q+r}$ is induced by the projection of ${\rho}_r(\Phi)$ onto ${\rho}_{r-1}(\Phi)$ (See [40], p 163 or [41], p 253] for details). If we use such a procedure for a first order system with no zero or first order CC, we have $q=1, E=T,F_0=J_1(T)/R_1$. The Killing 
system  $R_1$ is formally integrable ($R_2$ involutive) if and only if $\omega$ has {\it constant Riemannian curvature}:  \\
\[   {\rho}^k_{l,ij}=c({\delta}^k_i{\omega}_{lj} - {\delta}^k_j{\omega}_{li} )  \]
with $c=0$ when $\omega$ is the flat Minkowski metric [14, 41, 50]. In general, we may apply the Spencer 
$\delta$-map to the top row obtained with $r=2$ in order to get the first commutative diagram allowing to determine $F_1$:  \\    
\[  \begin{array}{rcccccccl}
   &  0 & & 0 & & 0 &  &  &   \\
   & \downarrow & & \downarrow & & \downarrow & & &  \\
0\rightarrow & g_3 & \rightarrow &  S_3T^*\otimes T & \rightarrow & S_2T^*\otimes F_0& \rightarrow & F_1 & \rightarrow 0  \\
   & \hspace{2mm}\downarrow  \delta  & & \hspace{2mm}\downarrow \delta & &\hspace{2mm} \downarrow \delta & & &  \\
0\rightarrow& T^*\otimes g_2&\rightarrow &T^*\otimes S_2T^*\otimes T & \rightarrow &T^*\otimes T^*\otimes F_0 &\rightarrow & 0 &  \\
   &\hspace{2mm} \downarrow \delta &  &\hspace{2mm} \downarrow \delta & &\hspace{2mm}\downarrow \delta &  &  &   \\
0\rightarrow & {\wedge}^2T^*\otimes g_1 & \rightarrow & \underline{{\wedge}^2T^*\otimes T^*\otimes T} & \rightarrow & {\wedge}^2T^*\otimes F_0 & \rightarrow & 0 &  \\
   &\hspace{2mm}\downarrow \delta  &  & \hspace{2mm} \downarrow \delta  &  & \downarrow  & &  &  \\
0\rightarrow & {\wedge}^3T^*\otimes T & =  & {\wedge}^3T^*\otimes T  &\rightarrow   & 0  &  &  &   \\
    &  \downarrow  &  &  \downarrow  &  &  &  &  &  \\
    &  0  &   & 0  & &  &  &  &
\end{array}  \]
with exact rows and exact columns but the first that may not be exact at ${\wedge}^2T^*\otimes g_1$. We shall denote by $B^2(g_1)$ the {\it coboundary} as the image of the central $\delta$, by $Z^2(g_1)$ the {\it cocycle} as the kernel of the lower $\delta$ and by $H^2(g_1)=Z^2(g_1)/B^2(g_1)$ the {\it Spencer} $\delta$-{\it  cohomology} at ${\wedge}^2T^*\otimes g_1$.  \\
Going one step further on in the differential sequence and using the fact that the Riemann operator and the Weyl operator are both second order operators when $n\geq 4$, we may define the vector bundle $F_2$ by the top row of the following second commutative diagram in order to look for the corresponding first order {\it Bianchi operator} $F_1 \rightarrow F_2$:    
  
 \[  \begin{array}{rcccccccccl}
   & 0  &  & 0  & &  0  &  & 0  &  &     \\
   & \downarrow  &  &  \downarrow & & \downarrow & & \downarrow & &  & \\
0 \rightarrow & g_4 & \rightarrow &S_4T^*\otimes T& \rightarrow &S_3T^ *\otimes F_0 &\rightarrow & T^*\otimes F_1&\rightarrow & F_2 & \rightarrow 0 \\
 & \downarrow  &  &  \downarrow & & \downarrow & & \parallel & &  \\
0 \rightarrow & T^*\otimes g_3 & \rightarrow &T^*\otimes S_3T^*\otimes T& \rightarrow &T^*\otimes S_2T^ *\otimes F_0 &\rightarrow & T^*\otimes F_1& \rightarrow & 0 &  \\
 & \downarrow  &  &  \downarrow & & \downarrow & & \downarrow & & &  \\
0 \rightarrow &{\wedge}^2 T^*\otimes g_2 & \rightarrow &{\wedge}^2T^*\otimes S_2T^*\otimes T& \rightarrow &{\wedge}^2T^*\otimes T^ *\otimes F_0 &\rightarrow &
 0&&& \\
& \downarrow  &  &  \downarrow & & \downarrow & &  & & & \\
0 \rightarrow &{\wedge}^3 T^*\otimes g_1 & \rightarrow &\underline{{\wedge}^3T^*\otimes T^*\otimes T}& \rightarrow &{\wedge}^3T^*\otimes F_0 &\rightarrow & 0 & &&  \\
  & \downarrow  &  &  \downarrow & & \downarrow & & & & &  \\
0 \rightarrow &{\wedge}^4 T^*\otimes T & = &{\wedge}^4T^*\otimes  T& \rightarrow &0 & & & &&  \\
 & \downarrow  &  &  \downarrow & &  & & & & & \\
 & 0  &  & 0  & &    &  &   &   &  & 
\end{array}  \]

In the classical Killing system, $g_1\subset T^*\otimes T$ is defined by ${\omega}_{rj}(x){\xi}^r_i+{\omega}_{ir}(x){\xi}^r_j=0 \Rightarrow {\xi}^r_r=0, g_2=0,g_3=0$. Applying the previous diagram, we discover that the {\it Riemann tensor} $({\rho}^k_{l,ij})\subset {\wedge}^2T^*\otimes T^*\otimes T$ is a section of the vector bundle $ F_1=H^2(g_1)=Z^2(g_1)$ with:  \\
\[ \begin{array}{ccl}dim(F_1) & = & (n^2(n+1)^2/4)-(n^2(n+1)(n+2)/6)  \\
     &  =  &  (n^2(n-1)^2/4)-(n^2(n-1)(n-2)/6)  \\
     & =  & n^2(n^2-1)/12
     \end{array}  \]
by using either the top row or the left column and call (linearized) {\it Riemann operator} the second order operator $F_0 \rightarrow F_1$. We obtain at once the well known properties of the (linearized) Riemann tensor through the chase involved, namely $({\rho}^k_{l,ij})\in {\wedge}^2T^*\otimes T^*\otimes T$ is killed by both $\delta$ and ${\sigma}_0(\Phi)$. However, we have no indices for $F_1$ and cannot therefore exhibit the {\it Ricci tensor} or the {\it Einstein tensor} of GR by means of the usual {\it contraction} or {\it trace}. We recall briefly their standard definitions by stating ${\rho}_{ij}={\rho}_{ji}={\rho}^r_{i,rj} \Rightarrow tr(\rho)={\omega}^{ij}{\rho}_{ij}\Rightarrow {\epsilon}_{ij}={\rho}_{ij}-\frac{1}{2}{\omega}_{ij}tr(\rho)$. Similarly, going one step further, the (linearized) {\it Bianchi operator} is the first order operator $F_1 \rightarrow F_2$ with $F_2=H^3(g_1)=Z^3(g_1)\Rightarrow dim(F_2)=dim({\wedge}^4T^*\otimes T)-dim({\wedge}^3T^*\otimes g_1)=n^2(n^2-1)(n-2)/24$ as in ([40], p 168-171). This approach is relating for the first time the concept of {\it Riemann tensor candidate}, introduced by Lanczos and others, to the Spencer $\delta$-cohomology of the Killing symbols.  \\

Counting the differential ranks is now easy because $R_1$ is formally integrable with finite type symbol and thus $R_2$ is involutive with symbol $g_2=0$. We get:  \\
\[ rk(Killing)=rk(Cauchy)=n \Rightarrow rk(Riemann)=dim(S_2T^*)-n=(n(n+1)/2) -n=n(n-1)/2\]
\[ rk(Bianchi)=(n^2(n^2-1)/12)-(n(n-1)/2)=n(n-1)(n-2)(n+3)/12 \]
that is $rk(Bianchi)=3$ when $n=3$ and $rk(Bianchi)=14=20-6$ when $n=4$. Collecting all the results, we obtain that the canonical parametrization needs $n^2(n^2-1)/12$ potentials while any minimal parametrization only needs $n(n-1)/2$ potentials [54]. The Einstein parametrization is thus " {\it in between} " because $n(n-1)/2< n(n+1)/2 < n^2(n^2-1)/12, \forall n\geq 4$. We may summarize the previous results by means of the following initial part of a differential sequence which is not a Janet sequence because the classical Killing operator is not involutive:  \\
\[   0 \rightarrow \Theta \rightarrow T  \overset{Killing}{\underset{1}{\longrightarrow}} S_2T^* \overset{Riemann}{\underset{2}{\longrightarrow}} F_1 \overset{Bianchi}{\underset{1}{\longrightarrow}} F_2 \underset{1}{\longrightarrow} F_3\rightarrow ...  \]
   
 The {\it conformal Killing system} ${\hat{R}}_1\subset J_1(T) $ is defined by eliminating the function $A(x)$ in the system ${\cal{L}}(\xi)\omega=A(x)\omega$. It is also a {\it Lie operator} $\hat{\cal{D}}$ with solutions $\hat{\Theta}\subset T$ satisfying $[\hat{\Theta},\hat{\Theta}]\subset \hat{\Theta}$. Its symbol ${\hat{g}}_1$ is defined by the linear equations ${\omega}_{rj}{\xi}^r_i+{\omega}_{ir}{\xi}^r_j - \frac{2}{n}{\omega}_{ij}{\xi}^r_r=0$ which do not depend on any conformal factor and is finite type when $n\geq 3$ because $g_3=0$ but ${\hat{g}}_2$ is {\it now} $2$-acyclic {\it only when} $n\geq 4$ and $3$-acyclic {\it only when} $n\geq 5$ 
 [38-41, 53, 55]. It is known that ${\hat{R}}_2$ and thus ${\hat{R}}_1$ too (by a chase) are formally integrable if and only if $\omega$ has zero {\it Weyl tensor}:  \\
 \[  {\sigma}^k_{l,ij}\equiv {\rho}^k_{l,ij} - \frac{1}{(n-2)}({\delta}^k_i{\rho}_{lj} - {\delta}^k_j{\rho}_{li} +{\omega}^{ks}({\omega}_{lj}{\rho}_{si} - {\omega}_{li}{\rho}_{sj})) + \frac{1}{(n-1)(n-2)}({\delta}^k_i{\omega}_{lj} - {\delta}^k_j{\omega}_{li})tr(\rho)=0  \]
We may use later on the formula $id_M-f\circ u=v\circ g$ of Proposition 3.4 in order to split the short exact sequence induced by the inclusions $R_1 \subset {\hat{R}}_1 \Rightarrow g_1 \subset {\hat{g}}_1$:  \\
\[ 0 \longrightarrow S_2T^*\longrightarrow F_1  \longrightarrow {\hat{F}}_1 \longrightarrow  0  \]
according to the Vessiot structure equations, in particular if $\omega$ has constant Riemannian curvature and thus ${\rho}_{ij}={\rho}^r_{i,rj}=c(n-1){\omega}_{ij} \Rightarrow tr(\rho)={\omega}^{ij}{\rho}_{ij}=cn(n-1)$. Using the same diagrams as before, we discover that the Weyl tensor is a section of the vector bundle ${\hat{F}}_1=H^2({\hat{g}}_1)\neq Z^2({\hat{g}}_1)$. As a byproduct, the (linearized) {\it Weyl operator} ${\hat{F}}_0 \rightarrow {\hat{F}}_1$ is of order $2$ with a symbol ${\hat{h}}_2\subset S_2T^*\otimes {\hat{F}}_0$ which is {\it not} $2$-acyclic by applying the $\delta$-map to the short exact sequence:  \\
 \[  0 \rightarrow {\hat{g}}_{3+r} \longrightarrow S_{3+r}T^*\otimes T \stackrel{{\sigma}_{2+r}(\Phi)}{\longrightarrow} {\hat{h}}_{2+r} \rightarrow  0  \]
and chasing through the commutative diagram thus obtained with $r=0,1,2$. As ${\hat{h}}_3$ becomes $2$-acyclic after one prolongation of ${\hat{h}}_2$ {\it only}, it follows that {\it the generating CC for the Weyl operator are of order} $2$ {\it when} $n=4$ {\it and order} $1$ {\it only when} $n\geq 5$, a result that can be checked by computer algebra [55, 59]. Accordingly, the so-called Bianchi identities for the Weyl tensor that can be found in the literature {\it are not} CC {\it at all} in the strict sense of the definition as they do not involve only the Weyl tensor. These results could not have been discovered by Lanczos and followers because the formal theory of Lie pseudogroups and the Vessiot structure equations are still not known today.  \\ 

With more details when $n=4$, we have the short exact sequence:  \\
\[ 0 \rightarrow   S_4T^*\otimes T \longrightarrow S_3T^ *\otimes {\hat{F}}_0   \longrightarrow  T^*\otimes {\hat{F}}_1  \rightarrow  0 \]
 because we have $dim ({\hat{F}}_2)= -4\times 35  + 20 \times 9  -  4 \times 10 = -140 + 180 - 40 = 0  $   \\
 
 We may also use the snake lemma in order to exhibit the two short exact sequences:  \\ 
 \[  0 \rightarrow Z^3({\hat{g}}_1) \rightarrow {\wedge}^3T^*\otimes {\hat{g}}_1 \stackrel{\delta}{\longrightarrow}{\wedge}^4T^* \otimes T \rightarrow 0  \hspace{4mm} \Rightarrow  \hspace{4mm}0 \rightarrow {\wedge}^2 T^*\otimes {\hat{g}}_2 \stackrel{\delta}{\longrightarrow }ÊZ^3({\hat{g}}_1) \rightarrow {\hat{F}}_2 \rightarrow 0  \]
 \[   dim({\hat{F}}_2)=   (dim({\wedge}^3T^*\otimes {\hat{g}}_1) - dim ({\wedge}^4T^*\otimes T))  )         - dim({\wedge}^2 T^* \otimes {\hat{g}}_2)   = ((4 \times 7) - 4)  - (6\times 4)=0   \]                                                                                                                                                                                                                                                                                                                                                                                                                                                                                                                                                                                                                                                                                                                                                                                                                                                                                                                                                                                                                                                                                                                                                                                                                                                                                                                                                                                                                                                                                                                                                                                                                                                                                                                                                                                                                                                                                                                                                                                                                               
Hence the generating CC for the Weyl operator are of order $2$ when $n=4$ and this result can be checked by computer algebra 
[11, 12, 55, 59] in a coherent way with the following long exact sequence:  \\
\[   0 \rightarrow S_5T^*\otimes T \rightarrow S_4T^*\otimes {\hat{F}}_0 \rightarrow S_2T^*\otimes {\hat{F}}_1 \rightarrow {\hat{F}}_2 \rightarrow 0  \]
providing $ dim({\hat{F}}_2)=(10\times 10) - (35\times 9)+(56\times 4)=9$ and the strictly exact differential sequence:  \\
\[ 0\rightarrow \hat{\Theta} \rightarrow 4 \overset{CKilling}{\underset{1}{\longrightarrow}} 9 
\overset{Weyl}{\underset{2}{\longrightarrow}} 10 \underset{2}{\longrightarrow} 9 \underset{1}{\longrightarrow} 4 \rightarrow 0 \]
with respective orders under the arrows, because the Euler-Poincar\'e characteristic must vanish.  \\

\newpage  

We may summarize these results, which do not seem to be known, by the following differential sequences where the order of an operator is written under its arrow:  \\

\noindent 
 $\bullet$ $n=3$: $ \hspace{2cm} 3 \underset{1}{\longrightarrow} 5 \underset{3}{\longrightarrow }5 \underset{1}{ \longrightarrow }3 \rightarrow 0  $  \\
 $\bullet$ $n=4$: $ \hspace{2cm} 4 \underset{1}{\longrightarrow} 9 \underset{2}{\longrightarrow} 10 \underset{2}{\longrightarrow} 9 \underset{1}{\longrightarrow} 4  \rightarrow 0  $ \\                                           
 $\bullet$ $n=5$: $ \hspace{2cm}  5 \underset{1}{\longrightarrow} 14 \underset{2}{\longrightarrow} 35 \underset{1}{\longrightarrow} 35 \underset{2}{\longrightarrow} 14 \underset{1}{\longrightarrow} 5 \rightarrow 0  $\\
 
 We shall revisit the previous results by showing that, {\it in fact}, all the maps and splittings existing for the Killing operator are coming from maps and splittings existing for the conformal Killing operator, {\it though surprising it may look like}. As these results are based on a systematic use of the Spencer operator, they are neither known nor acknowledged.  \\ 
 
\noindent
{\bf PROPOSITION 4.6}: Recalling that $F_1=H^2(g_1)=Z^2(g_1)$ in the Killing case, we have the commutative diagram:  \\ 

 \[   \begin{array}{ccccc}
 0  &   &  0  &  &  0   \\
 \downarrow  &   &  \downarrow  &  &  \downarrow  \\
 Z^2(g_1) & \subset &  Z^2(T^*\otimes T) &  \longrightarrow  & S_2T^*  \\
 \downarrow  &   &  \downarrow  &  & \hspace{3mm} \downarrow  \delta  \\
 {\wedge}^2T^*\otimes g_1  &  \subset  &  {\wedge}^2T^*\otimes T^*\otimes T  &  \longrightarrow &  T^*\otimes T  \\
 \hspace{3mm} \downarrow  \delta  &   & \hspace{3mm}  \downarrow  \delta  &  &  \hspace{3mm}  \downarrow  \delta   \\
 {\wedge}^3T^*\otimes T  & =  &  {\wedge}^3T^*\otimes T    & \longrightarrow  &  {\wedge}^2T^*  \\
  \downarrow  &   &  \downarrow  &  &  \downarrow  \\
  0  &   &  0  &  &  0  
  \end{array}  \]
 
 \noindent
{\it Proof}: First of all, we must point out that the surjectivity of the bottom $\delta$ in the central column is well known from the exactness of the $\delta$-sequence for $S_3T^*$ and thus also after tensoring by $T$. However, the surjectivity of the bottom $\delta$ in the left clumn is {\it not evident at all} as it comes from a delicate circular chase in the preceding diagram, using the fact that the Riemann and Weyl operators are second order operators. Then, setting ${\varphi}_{ij}={\rho}^r_{r,ij}= - {\varphi}_{ji}$ and ${\rho}_{ij}={\rho}^r_{i,rj}\neq {\rho}_{ji}$, we may define the central map by ${\rho}^k_{l,ij} \rightarrow {\rho}_{ij} - \frac{1}{2}{\varphi}_{ij}$ and the bottom map by $\omega\otimes \xi \rightarrow  i(\xi)\omega$ by introducing the interior product $i( )$. We obtain at once 
$-({\rho}^r_{r,ij}+{\rho}^r_{i,jr}+{\rho}^r_{j,ri})=({\rho}_{ij} - \frac{1}{2}{\varphi}_{ij}) - ({\rho}_{ji} - \frac{1}{2}{\varphi}_{ji})$ and the bottom diagram is commutative, clearly inducing the upper map. If we restrict to the Killing symbol, then ${\varphi}_{ij}=0$ and we obtain ${\rho}_{ij} - {\rho}_{ji}=0 \Rightarrow ({\rho}_{ij}={\rho}_{ji})\in S_2T^*$, that is the classical contraction allowing to obtain the Ricci tensor from the Riemann tensor but {\it there is no way to go backwards with a canonical lift}. A similar comment may be done for the conformal Killing symbol and the $\frac{1}{2}$ coefficient. \\
\hspace*{12cm}    Q.E.D.   \\

Using the previous diagram allowing to define both $F_1=H^2(g_1)$ and ${\hat{F}}_1=H^2({\hat{g}}_1)$, we obtain the commutative and exact diagram:  \\
\[  \begin{array} {ccccccc}
  &  0  &  &  0  &  &  0  &   \\ 
  & \downarrow  &  &  \downarrow &  &  \downarrow  &  \\
0 \rightarrow &  Z^2(g_1) &  \longrightarrow &  {\wedge}^2T^*\otimes g_1 & \stackrel{\delta}{\longrightarrow}  &  {\wedge}^3T^*\otimes T &  \rightarrow 0  \\ 
 & \downarrow  &  &  \downarrow &  &  \parallel  &  \\
 0 \rightarrow &  Z^2({\hat{g}}_1) &  \longrightarrow & {\wedge}^2T^*\otimes {\hat{g}}_1  &  \stackrel{\delta}{\longrightarrow}  &  {\wedge}^3T^*\otimes T &  \rightarrow 0  \\ 
   & \downarrow  &  &  \downarrow &  &  \downarrow  &  \\
 0 \rightarrow &  {\wedge}^2T^*  & = & {\wedge}^2T^* & \longrightarrow & 0 &     \\
 & \downarrow  &  &  \downarrow &  &  \ &  \\
  &  0  &  &  0  & &  &
 \end{array}  \]

\noindent
{\bf THEOREM 4.7}: We have the following commutative and exact "{\it fundamental diagram 2} ":  \\
 \[ \begin{array}{rcccccccll}
 & & & & & & & 0 & &\\
 & & & & & & & \downarrow && \\
  & & & & & 0& & S_2T^* & & \\
  & & & & & \downarrow & & \downarrow  & & \\
   & & & 0 &\longrightarrow & Z^2(g_1) & \longrightarrow & H^2(g_1)  & \longrightarrow 0 & \\
   & & & \downarrow & & \downarrow & &  \downarrow  &  &\hspace{5mm} JANET\\
   & 0 &\longrightarrow & T^*\otimes {\hat{g}}_2 & \stackrel{\delta}{\longrightarrow} & Z^2({\hat{g}}_1) & \longrightarrow & H^2({\hat{g}}_1) & \longrightarrow 0 & \\
    & & & \downarrow & & \downarrow & & \downarrow     &  & \\
 0 \longrightarrow & S_2T^* & \stackrel{\delta}{\longrightarrow}& T^*\otimes T^* &\stackrel{\delta}{\longrightarrow} & {\wedge}^2T^* & \longrightarrow & 0 &  & \\
   & & & \downarrow &  & \downarrow & & & & \\
   & & & 0 & & 0 & & &  &\\
  & & & & &  & & & &  \\ 
   &&&& SPENCER &&&&&
   \end{array}  \]
\[   \vspace{2mm}\]
The following theorem will provide {\it all} the classical formulas of both Riemannian and conformal geometry in one piece but in a totally unusual framework {\it not depending on any conformal factor}:    \\
 
\noindent
{\bf THEOREM 4.8}: All the short exact sequences of the preceding diagram split in a canonical way, that is in a way compatible with the underlying tensorial properties of the vector bundles involved. With more details:   \\
\[  \begin{array}{ccl}
T^*\otimes T^* \simeq S_2T^* \oplus {\wedge}^2 T^* &  \Rightarrow & Z^2({\hat{g}}_1)= Z^2(g_1) + \delta (T^*\otimes  {\hat{g}}_2) \simeq Z^2(g_1) \oplus  {\wedge}^2T^*  \\
 & \Rightarrow  & H^2(g_1)\simeq H^2({\hat{g}}_1) \oplus S_2T^* 
 \end{array}  \]
 
\noindent
{\it Proof}: First of all, we recall that:   \\
\[  g_1 =\{{\xi}^k_i \in T^*\otimes T\mid {\omega}_{rj}{\xi}^r_i+{\omega}_{ir}{\xi}^r_j=0 \}\subset {\hat{g}}_1=\{{\xi}^k_i\in T^*\otimes T \mid {\omega}_{rj}{\xi}^r_i+{\omega}_{ir}{\xi}^r_j - \frac{2}{n}{\omega}_{ij}{\xi}^r_r=0\}  \]
\[  \Rightarrow   \hspace{1cm} 0=g_2 \subset {\hat{g}}_2= \{ {\xi}^k_{ij}\in S_2T^*\otimes T \mid  n{\xi}^k_{ij}={\delta}^k_i{\xi}^r_{rj} +{\delta}^k_j{\xi}^r_{ri} - {\omega}_{ij}{\omega}^{ks}{\xi}^r_{rs} \}     \]
Now, if $({\tau}^k_{li,j})\in T^*\otimes {\hat{g}}_2$, then we have:  \\
\[  n {\tau}^k_{li,j}={\delta}^k_l{\tau}^r_{ri,j} + {\delta}^k_i {\tau}^r_{rl,j}-{\omega}_{li}{\omega}^{ks}{\tau}^r_{rs,j}  \] 
 and we may set ${\tau}^r_{ri,j}={\tau}_{i,j}\neq {\tau}_{j,i}$ with $({\tau}_{i,j})\in T^*\otimes T$ and such a formula does not depend on any conformal factor. We have:  \\
\[  \delta ({\tau}^k_{li,j})=({\tau}^k_{li,j} - {\tau}^k_{lj,i})=({\rho}^k_{l,ij}) \in B^2( {\hat{g}}_1)\subset Z^2({\hat{g}}_1)  \]
 with:   \\
 \[  Z^2({\hat{g}}_1)= \{ ({\rho}^k_{l,ij})\in {\wedge}^2T^*\otimes {\hat{g}}_1)\mid \delta ({\rho}^k_{l,ij})=0 \}\Rightarrow {\varphi}_{ij}={\rho}^r_{r,ij}\neq 0 \]
 \[   \delta ({\rho}^k_{l,ji}) = ({\sum}_{(l,i,j)}{\rho}^k_{l,ij}={\rho}^k_{l,ij} + {\rho}^k_{i,jl} + {\rho}^k_{j,li}) \in {\wedge}^3T^*\otimes T  \]
 \noindent
$\bullet$ The splitting of the lower row is obtained by setting $({\tau}_{i,j})\in T^*\otimes T^* \rightarrow (\frac{1}{2}({\tau}_{i,j} + {\tau}_{j,i}))\in S_2T^*$ in such a way that $({\tau}_{i,j}={\tau}_{j,i}={\tau}_{ij}\in S_2T^*) \Rightarrow \frac{1}{2}({\tau}_{ij}+{\tau}_{ji})={\tau}_{ij}$. \\
Similarly, $({\varphi}_{ij}= - {\varphi}_{ji})\in {\wedge}^2T^*  \rightarrow (\frac{1}{2}{\varphi}_{ij})\in T^*\otimes T^*$ and 
$ (\frac{1}{2}{\varphi}_{ij} - \frac{1}{2}{\varphi}_{ji})=({\varphi}_{ij}) \in {\wedge}^2T^*$.  \\ 

\noindent
$\bullet$ The splitting of the central vertical column is obtained by using Proposition $3.4 $ from a lift of the epimorphism $Z^2({\hat{g}}_1) \rightarrow {\wedge}^2T^* \rightarrow 0$ which is obtained by lifting $({\varphi}_{ij})\in {\wedge}^2T^*$ to $(\frac{1}{2}{\varphi}_{ij})\in T^*\otimes T^*$, setting ${\tau}^r_{ri,j}=\frac{1}{2} {\varphi}_{ij}$ and applying $\delta$ to obtain $({\tau}^r_{ri,j}-{\tau}^r_{rj,i}=\frac{1}{2} {\varphi}_{ij} - \frac{1}{2}{\varphi}_{ji}={\varphi}_{ij})\in B^2({\hat{g}}_1)\subset Z^2({\hat{g}}_1)$.\\

\noindent
$\bullet$ Now, let us define $({\rho}_{i,j}={\rho}^r_{i,rj}\neq {\rho}_{j,i})\in T^*\otimes T^*$. Hence, elements of $Z^2(g_1)$ are such that:  \\
\[  {\varphi}_{ij}={\rho}^r_{r,ij}=0, {\varphi}_{ij}-{\rho}_{i,j}+ {\rho}_{j,i}=0 \Rightarrow ({\rho}_{ij}={\rho}_{i,j}={\rho}_{j,i}={\rho}_{ji}) \in S_2T^*\]
while elements of $Z^2({\hat{g}}_1)$ are such that:  \\
\[ ({\rho}^r_{r,ij}={\varphi}_{ij}={\rho}_{i,j}- {\rho}_{j,i}={\tau}_{i,j}-{\tau}_{j,i}\neq 0 ) \in {\wedge}^2T^*  \]
Accordingly, $({\rho}_{i,j}- \frac{1}{2}{\varphi}_{ij}={\rho}_{j,i}- \frac{1}{2}{\varphi}_{ji}) \in S_2T^*$. More generally, we may consider ${\rho}^k_{l,ij}- ({\tau}^k_{li,j}-{\tau}^k_{lj,i})$ with ${\tau}^r_{ri,j}=\frac{1}{2}{\varphi}_{ij}$. Such an element is killed by $\delta$ and thus belongs to $Z^2({\hat{g}}_1)$ because each member of the difference is killed by $\delta$. However, we have 
${\rho}^r_{r,ij}-({\tau}^r_{ri,j}-{\tau}^r_{rj,i})={\varphi}_{ij} - {\varphi}_{ij}=0$ and the element does belong indeed to $Z^2(g_1)$, providing a lift $Z^2({\hat{g}}_1) \rightarrow Z^2(g_1) \rightarrow 0$.  \\

\noindent
$\bullet$ Of course, the most important result is to split the right column.For this, using again Proposition 3.4, we may take into account the fact that $(id_M - f \circ u)\circ f=f-f\circ id_{M'}=f-f=0$ in order to obtain a lift of $H^2(g_1) \rightarrow H^2({\hat{g}}_1) \rightarrow 0$ if we know a lift $H^2(g_1)\rightarrow S_2T^* \rightarrow 0 $. As this will be the hard step, we first need to describe the monomorphism $0 \rightarrow S_2T^* \rightarrow H^2(g_1)$ which is in fact produced by a diagonal north-east snake type chase. Let us choose $({\tau}_{ij}={\tau}_{i,j}={\tau}_{j,i}={\tau}_{ji})\in S_2T^* \subset T^* \otimes T^*$. Then, we may find $({\tau}^k_{li,j})\in T^* \otimes {\hat{g}}_2$ by deciding that ${\tau}^r_{ri,j}={\tau}_{i,j}={\tau}_{j,i}={\tau}^r_{rj,i}$ in $Z^2({\hat{g}}_1)$ and apply 
$\delta$ in order to get ${\rho}^k_{l,ij}={\tau}^k_{li,j} - {\tau}^k_{k,lj,i}$ such that ${\rho}^r_{r,ij}={\varphi}_{ij}=0$ and thus $({\rho}^k_{l,ij})  \in Z^2(g_1)=H^2(g_1)$. We obtain:  \\
\[ \begin{array}{rcl}
n{\rho}^k_{l,ij} & = &  {\delta}^k_l{\tau}^r_{ri,j}-{\delta}^k_l{\tau}^r_{rj,i}+{\delta}^k_i {\tau}^r_{rl,j}-{\delta}^k_j{\tau}^r_{rli}-{\omega }^{ks}({\omega}_{li}{\tau}^r_{rs,j} - {\omega}_{lj}{\tau}^r_{rs,i })   \\
  & =  & ({\delta}^k_i{\tau}_{lj} - {\delta}^k_j{\tau}_{li}) -{\omega}^{ks}({\omega}_{li}{\tau}_{sj} - {\omega}_{lj}{\tau}_{si})                          \\
\end{array}  \]
Contracting in $k$ and $i$ while setting simply $tr(\tau) ={\omega}^{ij}{\tau}_{ij}, tr(\rho)={\omega}^{ij}{\rho}_{ij}$, we get:  \\
\[  n {\rho}_{ij}=n{\tau}_{ij}-{\tau}_{ij}-{\tau}_{ij}+{\omega}_{ij} tr(\tau)=(n-2){\tau}_{ij}+{\omega}_{ij}tr(\tau)=n{\rho}_{ji} \Rightarrow ntr(\rho)=2(n-1)tr(\tau) \]              
Substituting, we finally obtain ${\tau}_{ij}=\frac{n}{(n-2)}{\rho}_{ij} - \frac{n}{2(n-1)(n-2)}{\omega}_{ij}tr(\rho)$ and thus the tricky formula:  \\
\[  {\rho}^k_{l,ij}=\frac{1}{(n-2)}({\delta}^k_i{\rho}_{lj} - {\delta}^k_j{\rho}_{li}) - {\omega}^{ks}({\omega}_{li}{\rho}_{sj}-{\omega}_{lj}{\rho}_{si})) - \frac{1}{(n-1)(n-2)}({\delta}^k_i{\omega}_{lj}-{\delta}^k_j{\omega}_{li})tr(\rho)  \]
Contracting in $k$ and $i$, we check that ${\rho}_{ij}={\rho}_{ij}$ indeed, obtaining therefore the desired canonical lift $H^2(g_1) \rightarrow S_2T^* \rightarrow 0: {\rho}^k_{i,lj} \rightarrow  {\rho}^r_{i,rj}={\rho}_{ij}$. Finally, using again Proposition 3.4, the epimorphism $H^2(g_1) \rightarrow H^2({\hat{g}}_1) \rightarrow 0$ is just described by the formula:  \\
\[  {\sigma}^k_{l,ij}={\rho}^k_{l,ij}-\frac{1}{(n-2)}({\delta}^k_i{\rho}_{lj} - {\delta}^k_j{\rho}_{li}-{\omega}^{ks}({\omega}_{li}{\rho}_{sj}-{\omega}_{lj}{\rho}_{si})) + \frac{1}{(n-1)(n-2)}({\delta}^k_i{\omega}_{lj}-{\delta}^k_j{\omega}_{li})tr(\rho)  \]
which is just the way to define the Weyl tensor. We notice that ${\sigma}^r_{r,ij}={\rho}^r_{r,ij}=0$ and ${\sigma}^r_{i,rj}=0$ by using indices or a circular chase showing that $Z^2({\hat{g}}_1)=Z^2(g_1) + \delta (T^*\otimes {\hat{g}}_2)$. This purely algebraic result only depends on the metric $\omega$ and does not depend on any conformal factor. In actual practice, the lift $H^2(g_1) \rightarrow S_2T^*$ is described by ${\rho}^k_{l,ij}\rightarrow {\rho}^r_{i,rj}={\rho}_{ij}={\rho}_{ji}$ but it is not evident at all that the lift $H^2({\hat{g}}_1) \rightarrow H^2(g_1)$ is described by the strict inclusion ${\sigma}^k_{l,ij} \rightarrow {\rho}^k_{l,ij}={\sigma}^k_{l,ij}$ providing a short exact sequence as in Proposition $3.4$ because ${\rho}_{ij}={\rho}^r_{i,rj}={\sigma}^r_{i,rj}=0$ by composition.\\

\noindent
$\bullet$ The splitting of the central row could be obtained similarly by using the fact that the diagram is symmetric with respect to the north-west axis.  \\
 \hspace*{12cm}   Q.E.D.   \\
 
 \noindent
 {\bf COROLLARY 4.9}: When $n\geq 4$, each component of the Weyl tensor is a torsion element killed by the Dalembertian whenever the Einstein equations in vacuum are satisfied by the metric. Hence, there exists a second order  operator ${\cal{Q}}$ such that we have an identity:  \\
 \[   \Box \circ Weyl = {\cal{Q}} \circ Ricci    \]
 
 \noindent
 {\it Proof}: According to Proposition 3.7, each extension module $ext^i(M)$ is a torsion module, $\forall i\geq 1$. It follows that each additional CC in ${\cal{D}}'_1$ which is not already in ${\cal{D}}_1$ is a torsion element as it belongs to this module. One may also notice that:  \\
 \[  rk_D(Einstein)=\frac{n(n+1)}{2}- n=\frac{n(n-1) }{2}  \hspace{3mm}, \hspace{3mm}rk_D(Riemann)=\frac{n(n+1)}{2}- n=\frac{n(n-1)}{2}   \] 
The differential ranks of the Einstein and Riemann operators are thus equal, but {\it this is a pure coincidence} because $rk_D(Einstein)$ has only to do with the $div$ operator induced by contracting the Bianchi identities, while $rk_D(Riemann)$ has only to do with the classical Killing operator and the fact that the corresponding differential module is a torsion module because we have a Lie group of transformations having $n + \frac{n(n-1)}{2}=\frac{n(n+1}{2}$ parameters (translations + rotations). Hence, as the Riemann operator is a direct sum of the Weyl operator and the Einstein or Ricci operator according to the previous theorem, each component of the Weyl operator must be killed by a certain operator whenever the Einstein or Ricci equations in vacuum are satisfied. Also, as no coordinate may have a particular importance, there is a good deal of chance that there is a unique operator for each component. It has been a great surprise for my former PhD student A. Quadrat (INRIA) when, after using very recent computer algebra packages that he has developped for studying extension modules, he discovered that ... it was just the Dalembertian operator and the computer produced {\it automatically} the identity of the Theorem [59, 60].  \\
It is at that time that the author of this paper, who has bee a student of A. Lichnerowicz ([18], exercise 7.7]) and Y. Choquet-Bruhat ([10], p 206]), just remembered a technical result that has been, many times but in vain, compared with the EM wave equations $\Box F=0$ easily obtained when the second set of Maxwell equations in vacuum is satisfied, {\it avoiding therefore the Lorenz (no "t" !) gauge condition for the EM potential}. Indeed, let us start with the Minkowski constitutive law with electric constant ${\epsilon}_0$ and magnetic constant ${\mu}_0$ such that ${\epsilon}_0{\mu}_0 c^2=1$ in vacuum:  \\
\[ {\cal{F}}^{rs}=\frac{1}{{\mu}_0}{\hat{\omega}}^{ri}{\hat{\omega}}^{sj}F_{ij}\sim {\omega}^{ri}{\omega}^{sj} F_{ij}\]
where ${\hat{\omega}}_{ij}={\mid det(\omega)\mid}^{-1/n}{\omega}_{ij}\Rightarrow  \mid det(\hat{\omega})\mid=1$, $F\in {\wedge}^2T^*$ is the EM field and the induction ${\cal{F}}$ is thus a contravariant skewsymmetric $2$-tensor density. From the Maxwell equations we have:
\[  {\partial}_rF_{ij} + {\partial}_iF_{jr} + {\partial}_jF_{ri}=0, \hspace{3mm}{\nabla}^r{\cal{F}}_{ri}=0 \hspace{3mm} \Rightarrow \hspace{3mm}{\nabla}^rF_{ri}=0 \]
\[   \Rightarrow  \hspace{3mm}  \Box F_{ij}={\nabla}^r{\nabla}_rF_{ij}={\nabla}^r ({\nabla}_iF_{rj}- {\nabla}_jF_{ri})=0   \]

We reproduce now this classical but tricky computation using essentially the Bianchi identities:  \\
\[   {\sum}_{(r,i,j)}{\nabla}_r{\rho}^k_{l,ij}\equiv {\nabla}_r {\rho}^k_{l,ij}+ {\nabla}_i{\rho}^k_{l,jr} +{\nabla}_j{\rho}^k_{l,ri}=0 \hspace{3mm}\Rightarrow \hspace{3mm} {\nabla}^r{\rho}_{rl,ij} - {\nabla}_i{\rho}_{lj}+{\nabla}_j{\rho}_{li}=0  \] 
\[{\nabla}^r({\sum}_{(r,i,j)}{\nabla}_r{\rho}_{kl,ij})    \equiv  {\nabla}^r{\nabla}_r{\rho}_{kl,ij} + {\nabla}^r{\nabla}_i{\rho}_{kl,jr} + {\nabla}^r{\nabla}_j{\rho}_{kl, ri}=0  \] 
\[\Rightarrow  \hspace{3mm}  {\nabla}^r{\nabla}_r{\rho}_{kl,ij} + {\nabla}_i{\nabla}^r{\rho}_{kl,jr} + {\nabla}_j{\nabla}^r{\rho}_{kl, ri}+[ {\nabla}^r,{\nabla}_i] {\rho}_{kl,jr} +[{\nabla}^r,{\nabla}_j]{\rho}_{kl,ri}=0 \]
\[\Rightarrow  \hspace{3mm}  {\nabla}^r{\nabla}_r{\rho}_{kl,ij} + {\nabla}_i{\nabla}^r{\rho}_{kl,jr} + {\nabla}_j{\nabla}^r{\rho}_{kl, ri}+ (\sum quadratic)=0 \]

\[{\rho}_{kl,ij}=-{\rho}_{ki,jl}-{\rho}_{kj,li}={\rho}_{ik,jl}+{\rho}_{jk,li}=({\rho}_{ij,kl}+{\rho}_{il,jk}) + ({\rho}_{jl,ki} + {\rho}_{ij,kl})=2{\rho}_{ij,kl} - {\rho}_{kl,ij}\]
\[  \Rightarrow  \hspace{3mm}{\rho}_{kl,ij}={\rho}_{ij,kl } \hspace{3mm} \Rightarrow \hspace{3mm}  {\nabla}^r{\rho}_{ij,rl}={\nabla}_i{\rho}_{lj} -{\nabla}_j{\rho}_{li} \hspace{3mm}  \Rightarrow \hspace{3mm}  {\nabla}^r{\rho}_{jr} - \frac{1}{2}{\nabla}_jtr(\rho)=0  \]
\[ \Box {\rho}_{kl,ij}= ({\nabla}_i  ({\nabla}_k{\rho}_{lj} - {\nabla}_l{\rho}_{kj})) - (i \leftrightarrow  j)  + (\sum quadratic)\]
Of course, we have:  \\
\[ \Box {\rho}_{ij}={\nabla}^r{\nabla}^s{\rho}_{si,rj}+{\nabla}^r{\nabla}_j{\rho}_{ir}= {\nabla}^r{\nabla}^s{\rho}_{rj,si}+{\nabla}^r{\nabla}_j{\rho}_{ir} = {\nabla}^r({\nabla}_r{\rho}_{ij}-{\nabla}_j{\rho}_{ir}) + {\nabla}^r{\nabla}_j{\rho}_{ir}=\Box {\rho}_{ij}\]
because the Ricci tensor only satisfies ${\nabla}^r{\rho}_{jr} -\frac{1}{2} {\nabla}_j tr(\rho)=0$.  \\
Linearizing at the Euclidean metric for $n=2,3$ or at the Minkowski metric for $n=4$, we get:  \\
\[ \Box R_{kl,ij}= d_i(d_kR_{lj} - d_lR_{kj}) -  d_j(d_kR_{li} - d_lR_{ki}) \]
The Corollary follows at once by using the splitting formula:  \\
\[ {\sigma}^k_{l,ij}={\rho}^k_{l,ij} - (  \sum {\rho}_{rs}) \hspace{3mm}\Rightarrow \hspace{3mm}  {\Sigma}^k_{l,ij}={R}^k_{l,ij} -(  \sum {R}_{rs})  \]

Finally, using the $div$-type relation satisfied by the Einstein tensor:  \\
\[ {\epsilon}_{ij}={\rho}_{ij} - \frac{1}{2}{\omega}_{ij}tr(\rho) \hspace{3mm}  \Rightarrow\hspace{3mm} E_{ij}=R_{ij} - \frac{1}{2}{\omega}_{ij} tr(R) \hspace{3mm} \Rightarrow \hspace{3mm} d_rE^r_j={\omega}^{ri}d_rE_{ij}=d_r R^r_j-\frac{1}{2}d_jtr(R)=0  \]  
we get:  \\
\[ \begin{array}{ccl}
\Box R_{ij} & =  &\Box R_{ij} - d_r(d_iR^r_j) - d_j(d_rR^r_i) + d_j(d_itr(R))  \\
  &  =  & \Box R_{ij} -  d_i(d_rR^r_j) - d_j(d_rR^r_i )+ d_{ij}tr(R)  \\
  &  =  & \Box R_{ij} - \frac{1}{2}d_i (d_jtr(R)) - \frac{1}{2}d_j(d_itr(R))+d_{ij}tr(R)  \\
  &  =  & \Box R_{ij} 
 \end{array}   \]
in a coherent manner with the corresponding non-linear result already obtained.  \\
\hspace*{12cm} Q.E.D.   \\

More generally, we have:  \\

\noindent
{\bf COROLLARY 4.10}: Constitutive relations ${C}_1\leftrightarrow {\wedge}^nT^*\otimes C_1^*$ provide wave equations for the field.  \\

\noindent
{\it Proof}: Using Proposition 2.20 and the last Theorem, we may identify the field as a section $(A^{\tau}_i(x)dx")\in T^*\otimes {\cal{G}}$ killed by $d$, that is such that ${\partial}_iA^{\tau}_j - {\partial}_jA^{\tau}_i=0$. By duality and a result first found by H. Poincar\'e in $1901$ [37, 46], we may introduce the parametrization ${\wedge}^0T^*\otimes {\cal{G}} \stackrel{d}{\longrightarrow} T^*\otimes {\cal{G}}: {\lambda}^{\tau}(x) \rightarrow {\partial}_i{\lambda}^{\tau}(x)=A^{\tau}_i(x)$ and obtain the induction equations from the following variational procedure:  \\
\[  F={\int}_V\varphi(A)dx\Rightarrow \delta F= {\int}_V\frac{\partial \varphi}{\partial A}\delta Adx={\int}_V{\cal{A}}^i_{\tau}
{\partial}_i\delta{\lambda}^{\tau}dx= - {\int}_V({\partial}_i{\cal{A}}^i_{\tau})\delta  {\lambda}^{\tau}dx + div (...)   \]
Hence, induction equations in vacuum can be written as ${\partial}_i{\cal{A}}^i_{\tau}=0$ and constitutive relations establish a (self-adjoint) isomorphism $A \leftrightarrow {\cal{A}}$ (See [43] for more details and examples). If we have only one constitutive coefficient like in Corollary $4.8$ for EM, that is ${\cal{A}}\sim A$ locally, then we obtain ${\omega}^{ij}{\partial}_{ij}A=0$. \\
\hspace*{12cm}   Q.E.D.   \\

The reader may understand that if this new approach brings the need to revisit the mathematical foundations of GR, it also brings the need to revisit the mathematical foundations of {\it Gauge Theory} (GT) as well, because we have seen in the Introduction and will justify in the Conclusion that the EM field is a section of $T^*\otimes {\hat{g}}_2\subset T^*\otimes {\hat{R}}_2={\hat{C}}_1$.  \\

 The next Example will prove at once that {\it Algebraic Analysis may provide results that cannot be obtained or even imagined in a classical framework}.  \\

\noindent
{\bf EXAMPLE 4.11}: We shall study for simplicity the case $n=4$ with $K=\mathbb{Q}$ but the generalization to an arbitrary dimension $n\geq 4$ is immediate. First of all we have the long exact sequence:  \\
\[  0 \rightarrow D^6 \longrightarrow D^{20} \stackrel{Bianchi}{\longrightarrow} D^{20} \stackrel{Riemann}{\longrightarrow} D^{10} \stackrel{Killing}{\longrightarrow}D^4 \rightarrow M \rightarrow 0  \]
which is a resolution of the differential module $M=coker(Killing)$ and we check that we have indeed $6 - 20 + 20 - 10 + 4 =0$.  Accordingly, we have $N'=coker(Riemann)\simeq im(Killing) \subset D^4 $ and thus $N'$ is torsion-free with $rk(N')=4-0=4=n$ because $rk(M)=0$. \\
It follows that $N'$ just describes the so-called " {\it gauge transformations}" used in the study of gravitational waves because it is isomorphic to the submodule of $D^4$ generated by the classical Killing equations ([15], (5.1.9), p 135]).\\
Now, using the lift exhibited in the last Theorem, the fact that $div$ is induced by $Bianchi$ and the short exact sequence:  \\
\[   0\rightarrow F_2 \rightarrow {\wedge}^3T^*\otimes g_1 \stackrel{\delta}{\longrightarrow} {\wedge}^4T^*\otimes T \rightarrow 0  \hspace{3mm}\Rightarrow \hspace{3mm} F_2 \subset {\wedge}^3T^*\otimes T^*\otimes T  \]
we have the following commutative and exact diagram where $N=coker(Einstein)$:  \\
\[  \begin{array}{rcccccccccl}
  & & &  &  &  &  &  &  &  0 &  \\
  & & &  &  &  &  &  &  &  \downarrow & \\
  & &  & 0 &  & 0  & & 0 & & t(N) &    \\
  &  &  &  \downarrow &  & \downarrow  &  & \downarrow & & \downarrow &  \\
  &  0  & \longrightarrow &  D^4 &  \stackrel{div}{\longrightarrow} & D^{10} & \stackrel{Einstein}{\longrightarrow} & D^{10} & \longrightarrow & N & \rightarrow 0  \\
 & \downarrow & & \downarrow &  & \downarrow &  & \parallel & & \downarrow &  \\
    0 \rightarrow &D^6 &\longrightarrow & D^{20} & \stackrel{Bianchi}{\longrightarrow} & D^{20} & \stackrel{Riemann}{\longrightarrow} & D^{10} & \longrightarrow &  N' &  \rightarrow 0  \\
   & \parallel & & \downarrow  &  &\downarrow &  & \downarrow &  & \downarrow &   \\
   0\rightarrow & D^6 & \longrightarrow & D^{16} & \longrightarrow & D^{10}  &  & 0 &  & 0 &  \\
   & \downarrow &  & \downarrow &  & \downarrow &  &  &  &  &  \\
   &  0 &  & 0  &  &  0  &  & &  &  &
    \end{array}  \]
 The monomorphism $0\rightarrow D^4 \rightarrow D^{10}$ dualizes the composition of epimorphisms:  \\
 \[     {\wedge}^3T^* \otimes T^*\otimes T \rightarrow {\wedge}^2T^*\otimes T^* \stackrel{\omega}{\longrightarrow} {\wedge}^2T^*\otimes T \rightarrow T^*  \]
 describing the two successive contractions of indices needed in the last Corollary and the use of the metric $\omega$ for raising an index.
We obtain from Proposition 3.3 that $rk(N)=10 - 10 + 4=4=rk(N')=n$. It also follows from Proposition 3.3 that the kernel of the canonical induced epimorphism $N \rightarrow N' \rightarrow 0$ is the torsion module $t(N)$ because its rank is $rk(N)-rk(N')=4-4=0$ and thus 
 $N'\simeq N/t(N)$ is {\it effectively} a torsion-free module. With more details, if $L$ is the kernel of the epimorphism $N \rightarrow N'$ is a torsion module, we have thus $L\subseteq t(N)$ in the following commutative and exact diagram:  \\
\[ \begin{array}{rcccl}
  & 0 & & 0 & \\
    &  \downarrow &  & \downarrow &     \\
    0 \rightarrow & L & \longrightarrow & t(N) &   \\
     & \downarrow  &  & \downarrow &  \\
    0 \rightarrow & N &  = & N & \rightarrow 0  \\
    &  \downarrow &  & \downarrow &   \\
     & N' & \longrightarrow & N/t(N) & \rightarrow 0  \\
     & \downarrow & & \downarrow &  \\
     & 0  &  & 0 &   
     \end{array}  \]
where $N/t(N)$ is a torsion-free module by definition. A snake chase allows to prove that the cokernel of the monomorphism $L \rightarrow t(N)$ is isomorphic to the kernel of the induced epimorphism $N' \rightarrow N/t(N)$ and must be therefore, at the same time, a torsion module and a torsion-free module, a result leading to a contradiction unless it is zero and thus $L=t(N)$.   \\ 
Moreover, we may introduce the cokernels of the canonical monomorphisms on the left side of the last big diagram, in particular that of $0\rightarrow D^{10} \rightarrow D^{20}$ which is isomorphic to $D^{10}$ and a snake/diagonal chase in the previous diagram allows to exhibit the long exact connecting sequence:  \\
 \[  0 \rightarrow D^6 \longrightarrow D^{16} \longrightarrow D^{10} \longrightarrow N \longrightarrow N' \rightarrow 0  \]
 providing the long exact sequence:  \\
 \[  0 \rightarrow D^6 \longrightarrow D^{16} \longrightarrow D^{10} \longrightarrow t(N) \rightarrow 0  \]
 This is an additional reason to bring the Weyl tensor and the Weyl operator in order to describe the $10$ generators of $t(N)$ in a way similar to the one used in Example 3.10. However, it must be noticed that one cannot find canonical morphisms between the classical and conformal resolutions constructed similarly because we recall that, for $n=4$ ({\it only}), the CC of the Weyl operator are of order $2$ and {\it not} $1$ like the Bianchi CC for the Riemann operator. \\
 
However, it follows from the last Theorem that the short exact sequence $0 \rightarrow D^{10} \longrightarrow D^{20} \longrightarrow D^{10} \rightarrow 0$ splits with $D^{20}\simeq D^{10} \oplus D^{10}$ but the existence of a canonical lift $D^{20} \rightarrow D^{10} \rightarrow 0$ in the above diagram does not allow to split the right column and thus $N\neq N' \oplus t(N)$.  \\
Hence, one can only say that the space of solutions of Einstein equations in vacuum contains the generic solutions of the Riemann operator which are parametrized by arbitrary vector fields. As for the torsion elements, we have $t(N)=coker(D^{16} \rightarrow D^{10})$ and we may thus represent them by the components of the Weyl tensor, killed by the Dalembertian. This module interpretation of the so-called gauge transformations and torsion elements may thus question the proper origin and existence of gravitational waves because $coker(div)$ on the left part of the diagram has {\it strictly nothing to do} with the generalized Cauchy stress tensor which cannot appear {\it anywhere} in this diagram as we already said.  \\
 
 Of course, nonlinear versions using the corresponding {\it nonliner Spencer sequences} exist but are much more difficult and out of the scope of this paper [39-41, 49].
 
 \newpage

\noindent
{\bf 5) CONCLUSION}\\

When constructing inductively the Janet sequences for two involutive systems $R_q \subset {\hat{R}}_q \subset J_q(E)$, {\it the Janet sequence for} $R_q$ {\it projects onto the Janet sequence for} ${\hat{R}}_q$, that is we may define inductively canonical epimorphisms 
$F_r \rightarrow {\hat{F}}_r \rightarrow 0$ for $r=0, 1,...,n$. This result can also be obtained from the general formulas allowing to define the Janet bundles globally by chasing in the following commutative and exact diagram:  \\
\[  \begin{array}{rcccccl} 
  &  0  &  &  &  &  &  \\
   & \downarrow &  &  &  & &  \\
   0 \rightarrow & {\wedge}^rT^*\otimes R_q + \delta ({\wedge}^{r-1}T^*\otimes S_{q+1}T^*\otimes E) & \rightarrow & {\wedge}^rT^*\otimes J_q(E) & \rightarrow & F_r & \rightarrow 0  \\
  & \downarrow &  & \parallel &  & \downarrow &  \\
   0 \rightarrow & {\wedge}^rT^*\otimes {\hat{R}}_q + \delta ({\wedge}^{r-1}T^*\otimes S_{q+1}T^*\otimes E) & \rightarrow & {\wedge}^rT^*\otimes J_q(E) & \rightarrow & {\hat{F}}_r & \rightarrow 0  \\
 &  &  &  &  &\downarrow  \\
 &  &  &  &  &   0
\end{array}  \] 
 It follows fom the short exact sequences $0 \rightarrow C_r \rightarrow C_r(E) \stackrel{{\Phi}_r}{\longrightarrow} F_r \rightarrow 0$ allowing to define the Spencer bundles inductively that the {\it kernels} of the canonical epimorphisms $F_r \rightarrow {\hat{F}}_r \rightarrow 0$ are isomorphic to the {\it cokernels} of the canonical monomorphisms $0 \rightarrow C_r \rightarrow {\hat{C}}_r \subset C_r(E)$ and we may say that {\it Janet and Spencer play at see-saw}. This result can also be obtained from the formulas allowing to define the Spencer bundles globally by chasing in the following commutative and exact diagram:  \\
 \[  \begin{array}{rcccccl}
   &  0  &  &  0 &  &  0  &  \\
   & \downarrow &  &  \downarrow  & & \downarrow  \\
   0 \rightarrow & \delta ({\wedge}^{r-1}T^*\otimes g_{q+1} ) & \rightarrow & {\wedge}^rT^*\otimes R_q & \rightarrow & C_r & \rightarrow 0   \\
              & \downarrow & & \downarrow & & \downarrow  &  \\
  0 \rightarrow & \delta({\wedge}^{r-1}T^*\otimes {\hat{g}}_{q+1} ) &                                                                                                         \rightarrow & {\wedge}^rT^*\otimes {\hat{R}}_q & \rightarrow & {\hat{C}}_r & \rightarrow 0  \\
 \end{array}   \]
 showing that {\it the Spencer sequence for} $R_q$ {\it is contained into the Spencer sequence for} ${\hat{R}}_q$.\\
When dealing with applications, we have set $E=T$ and considered systems of finite type Lie equations determined by Lie groups of transformations. Accordingly, we have obtained in particular $C_r={\wedge}^rT^*\otimes R_q \subset {\wedge}^rT^*\otimes {\hat{R}}_q ={\hat{C}}_r \subset C_r(T)$ when comparing the classical and conformal Killing systems, but {\it these bundles have never been used in physics}. Therefore, instead of the classical Killing system $R_2\subset J_2(T)$ defined by $\Omega \equiv {\cal{L}}(\xi)\omega=0$ {\it and} $\Gamma\equiv {\cal{L}}(\xi)\gamma=0$ or the conformal Killing system ${\hat{R}}_2\subset J_2(T)$ defined by $\Omega\equiv {\cal{L}}(\xi)\omega=A(x)\omega$ and ${\Gamma} \equiv {\cal{L}}(\xi)\gamma= ({\delta}^k_iA_j(x) +{\delta} ^k_j A_i(x) -{\omega}_{ij}{\omega}^{ks}A_s(x)) \in S_2T^*\otimes T$, we may introduce the {\it intermediate differential system} ${\tilde{R}}_2 \subset J_2(T)$ defined by ${\cal{L}}(\xi)\omega=A\omega$ with $A=cst$ and $\Gamma \equiv {\cal{L}}(\xi)\gamma=0 $, for the 
{\it Weyl group} obtained by adding the only dilatation with infinitesimal generator $x^i{\partial}_i$ to the Poincar\'e group. We have $R_1\subset {\tilde{R}}_1={\hat{R}}_1$ but the strict inclusions $R_2 \subset {\tilde{R}}_2 \subset {\hat{R}}_2$ and we discover {\it exactly} the group scheme already considered in the Introduction, both with the need to {\it shift by one step to the left} the physical interpretation of the various differential sequences used. Indeed, as ${\hat{g}}_2\simeq T^*$, the first Spencer operator ${\hat{R}}_2\stackrel{D_1}{\longrightarrow} T^*\otimes {\hat{R}}_2$ is induced by the usual Spencer operator ${\hat{R}}_3 \stackrel{D}{\longrightarrow} T^*\otimes {\hat{R}}_2:(0,0,{\xi}^r_{rj},{\xi}^r_{rij}=0) \rightarrow (0,{\partial}_i0-{\xi}^r_{ri}, {\partial}_i{\xi}^r_{rj}- 0)$ and thus projects by cokernel onto the induced operator $T^* \rightarrow T^*\otimes T^*$. Composing with $\delta$, it projects therefore onto $T^*\stackrel{d}{\rightarrow} {\wedge}^2T^*:A \rightarrow dA=F$ as in EM and so on by using he fact that $D_1$ 
{\it and} $d$ {\it are both involutive} or the composite epimorphisms ${\hat{C}}_r \rightarrow {\hat{C}}_r/{\tilde{C}}_r\simeq {\wedge}^rT^*\otimes ({\hat{R}}_2/{\tilde{R}}_2) \simeq {\wedge}^rT^*\otimes {\hat{g}}_2\simeq {\wedge}^rT^*\otimes T^*\stackrel{\delta}{\longrightarrow}{\wedge}^{r+1}T^*$. The main result we have obtained is thus to be able to increase the order and dimension of the underlying jet bundles and groups as we have [53-55]:  \\
\hspace*{2cm}  {\it POINCARE GROUP} $\subset$ {\it WEYL GROUP} $\subset$ {\it CONFORMAL GROUP} \\
that is  $10 < 11 < 15$ when $n=4$ like in the introduction, proving therefore that:  \\   
{\it The mathematical structures of electromagnetism and gravitation only depend on second order jets}.\\

As all these results are of a purely mathematical nature and not known, it will not be possible to ignore them any longer in a near future.  \\

 \newpage

\noindent
{\bf REFERENCES}  \\

\noindent
[1] Adler, F.W.: \"{U}ber die Mach-Lippmannsche Analogie zum zweiten Hauptsatz, Anna. Phys. Chemie, 22, 578-594 (1907).  \\
\noindent
[2] Airy, G.B.:  On the Strains in the Interior of Beams, Phil. Trans. Roy. Soc.London, 153, 1863, 49-80 (1863).  \\  
\noindent
[3] Arnold, V.: M\'{e}thodes Math\'{e}matiques de la M\'{e}canique Classique, Appendice 2 (G\'{e}od\'{e}siques des m\'{e}triques invariantes \`{a} gauche sur des groupes de Lie et hydrodynamique des fluides parfaits), MIR, Moscow (1974,1976). \\
\noindent
[4] Assem, I.: Alg\`ebres et Modules, Masson, Paris (1997).  \\
\noindent
[5] Beltrami, E.: Osservazioni sulla Nota Precedente, Atti Reale Accad. Naz. Lincei Rend., 5, 141-142 (1892).  \\
\noindent
[6] Birkhoff, G.: Hydrodynamics, Princeton University Press (1954).  \\
\noindent
[7] Bjork, J.E. (1993) Analytic D-Modules and Applications, Kluwer (1993).  \\ 
\noindent
[8] Bourbaki, N.: Alg\`{e}bre, Ch. 10, Alg\`{e}bre Homologique, Masson, Paris (1980). \\
\noindent
[9] de Broglie, L.: Thermodynamique de la Particule isol\'{e}e, Gauthiers-Villars, Pris 1964).  \\
\noindent
[10] Choquet-Bruhat, Y.: Introduction to General Relativity, Black Holes and Cosmology, Oxford University Press (2015).  \\
\noindent
[11] Chyzak, F.,Quadrat, A., Robertz, D.:  Effective algorithms for parametrizing linear control systems over Ore algebras,
Appl. Algebra Engrg. Comm. Comput., 16, 319-376, 2005. \\
\noindent
[12] Chyzak, F., Quadrat, A., Robertz, D.: {\sc OreModules}: A symbolic package for the study of multidimensional linear systems,
Springer, Lecture Notes in Control and Inform. Sci., 352, 233-264, 2007.\\
http://wwwb.math.rwth-aachen.de/OreModules  \\
\noindent
[13] Cosserat, E., \& Cosserat, F.: Th\'{e}orie des Corps D\'{e}formables, Hermann, Paris, 1909.\\
\noindent
[14] Eisenhart, L.P.: Riemannian Geometry, Princeton University Press, Princeton (1926).  \\
\noindent
[15] Foster, J., Nightingale, J.D.: A Short Course in General relativity, Longman (1979).  \\
\noindent
[16] Gr\"{o}bner, W.: \"{U}ber die Algebraischen Eigenschaften der Integrale von Linearen Differentialgleichungen mit Konstanten Koeffizienten, Monatsh. der Math., 47, 247-284 (1939).\\
\noindent
[17] Hu,S.-T.: Introduction to Homological Algebra, Holden-Day (1968).  \\
\noindent
[18] Hughston, L.P., Tod, K.P.: An Introduction to General Relativity, London Math. Soc. Students Texts 5, Cambridge University Press 
(1990). \\
\noindent
[19] Janet, M.: Sur les Syst\`{e}mes aux D\'{e}riv\'{e}es Partielles, Journal de Math., 8, 65-151 (1920). \\
\noindent 
[20] Kashiwara, M.: Algebraic Study of Systems of Partial Differential Equations, M\'{e}moires de la Soci\'{e}t\'{e} 
Math\'{e}matique de France, 63 (1995) (Transl. from Japanese of his 1970 MasterÕs Thesis).  \\
\noindent
[21] Kolchin, E.R.: Differential Algebra and Algebraic groups, Academic Press, New York (1973).  \\
\noindent
[22] Kumpera, A., \& Spencer, D.C.: Lie Equations, Ann. Math. Studies 73, Princeton University Press, Princeton (1972).\\
\noindent
[23] Kunz, E.: Introduction to Commutative Algebra and Algebraic Geometry, BirkhaŸser (1985).  \\
\noindent
[24] Lippmann, G.: Extension du Principe de S. Carnot \`{a} la Th\'{e}orie des P\'{e}nom\`{e}nes \'{e}lectriques, C. R. Acad/ Sc. Paris, 82, 1425-1428 (1876).  \\
\noindent
[25] Lippmann, G.: \"{U}ber die Analogie zwischen Absoluter Temperatur un Elektrischem Potential, Ann. Phys. Chem., 23, 
994-996 (1907).  \\
\noindent
[26]  Macaulay, F.S.: The Algebraic Theory of Modular Systems, Cambridge (1916).  \\
\noindent
[27] Mach, E.: Die Geschichte und die Wurzel des Satzes von der Erhaltung der Arbeit, p 54, Prag: Calve (1872).  \\
\noindent
[28]ÊMach, E.: Prinzipien der W\"{a}rmelehre, 2, Aufl., p 330, Leipzig: J.A. Barth (1900).  \\
\noindent
[29] Maxwell, J.C.: On Reciprocal Figures, Frames and Diagrams of Forces, Trans. Roy. Soc. Ediinburgh, 26, 1-40 (1870).  \\
\noindent
[30] Morera, G.: Soluzione Generale della Equazioni Indefinite dellÕEquilibrio di un Corpo Continuo, Atti. Reale. Accad. dei Lincei, 1, 137-141+233(1892).  \\
\noindent
[31] Nordstr\"{o}m, G.: Einstein's Theory of Gravitation and Herglotz's Mechanics of Continua, Proc. Kon. Ned. Akad. Wet., 19, 884-891 (1917). \\
\noindent
[32] Northcott, D.G.: An Introduction to Homological Algebra, Cambridge university Press (1966).  \\
\noindent
[33] Northcott, D.G.: Lessons on Rings Modules and Multiplicities, Cambridge University Press (1968).  \\
\noindent
[34] Oberst, U.: Multidimensional Constant Linear Systems, Acta Appl. Math., 20, 1-175 (1990).  \\ 
\noindent
[35] Oberst, U.: The Computation of Purity Filtrations over Commutative Noetherian Rings of Operators and their Applications to Behaviours, Multidim. Syst. Sign. Process. (MSSP) 26, 389-404 (2013).  \\
http://dx.doi.org/10.1007/s11045-013-0253-4   \\
\noindent
[36] Ougarov, V.: Th\'{e}orie de la Relativit\'{e} Restreinte, MIR, Moscow, 1969, (french translation, 1979).\\
\noindent
[37] Poincar\'{e}, H.: Sur une Forme Nouvelle des Equations de la M\'{e}canique, C. R. Acad\'{e}mie des Sciences Paris, 132 (7) (1901) 369-371.  \\
\noindent
[38] Pommaret, J.-F.: Systems of Partial Differential Equations and Lie Pseudogroups, Gordon and Breach, New York, 1978; Russian translation: MIR, Moscow, 1983.\\
\noindent
[39] Pommaret, J.-F.: Differential Galois Theory, Gordon and Breach, New York, 1983.\\
\noindent
[40] Pommaret, J.-F.: Lie Pseudogroups and Mechanics, Gordon and Breach, New York, 1988.\\
\noindent
[41] Pommaret, J.-F.: Partial Differential Equations and Group Theory, Kluwer, 1994.\\
http://dx.doi.org/10.1007/978-94-017-2539-2    \\
\noindent
[42] Pommaret, J.-F.: Fran\c{c}ois Cosserat and the Secret of the Mathematical Theory of Elasticity, Annales des Ponts et Chauss\'ees, 82, 59-66 (1997) (Translation by D.H. Delphenich).  \\
\noindent
[43] Pommaret, J.-F.: Group Interpretation of Coupling Phenomena, Acta Mechanica, 149 (2001) 23-39.\\
http://dx.doi.org/10.1007/BF01261661  \\
\noindent
[44] Pommaret, J.-F.: Partial Differential Control Theory, Kluwer, Dordrecht, 2001.\\
\noindent
[45] POMMARET, J.-F.: Algebraic Analysis of Control Systems Defined by Partial Differential Equations, in "Advanced Topics in Control Systems Theory", Springer, Lecture Notes in Control and Information Sciences 311 (2005) Chapter 5, pp. 155-223.\\
\noindent
[46] Pommaret, J.-F.: Arnold's Hydrodynamics Revisited, AJSE-mathŽmatiques, 1, 1, 2009, pp. 157-174.  \\
\noindent
[47] Pommaret, J.-F.: Parametrization of Cosserat Equations, Acta Mechanica, 215 (2010) 43-55.\\
http://dx.doi.org/10.1007/s00707-010-0292-y  \\
\noindent
[48] Pommaret, J.-F.: Macaulay Inverse Systems revisited, Journal of Symbolic Computation, 46, 1049-1069 (2011). \\
\noindent
[49] Pommaret, J.-F.: Spencer Operator and Applications: From Continuum Mechanics to Mathematical Physics, in "Continuum Mechanics-Progress in Fundamentals and Engineering Applications", Dr. Yong Gan (Ed.), ISBN: 978-953-51-0447--6, InTech, 2012, Available from: \\
http://dx.doi.org/10.5772/35607   \\
\noindent
[50] Pommaret, J.-F.: The Mathematical Foundations of General Relativity Revisited, Journal of Modern Physics, 4 (2013) 223-239. \\
 http://dx.doi.org/10.4236/jmp.2013.48A022   \\
  \noindent
[51] Pommaret, J.-F.: The Mathematical Foundations of Gauge Theory Revisited, Journal of Modern Physics, 5 (2014) 157-170.  \\
http://dx.doi.org/10.4236/jmp.2014.55026  \\
 \noindent
[52] Pommaret, J.-F.: Relative Parametrization of Linear Multidimensional Systems, Multidim. Syst. Sign. Process., 26, 405-437 2015).  \\
DOI 10.1007/s11045-013-0265-0   \\
\noindent
[53] Pommaret,J.-F.:From Thermodynamics to Gauge Theory: the Virial Theorem Revisited, pp. 1-46 in "Gauge Theories and Differential geometry,", NOVA Science Publisher (2015).  \\
\noindent
[54] Pommaret, J.-F.: Airy, Beltrami, Maxwell, Einstein and Lanczos Potentials revisited, Journal of Modern Physics, 7, 699-728 (2016). \\
\noindent
http://dx.doi.org/10.4236/jmp.2016.77068   \\
\noindent
[55] Pommaret, J.-F.: Deformation Theory of Algebraic and Geometric Structures, Lambert Academic Publisher (LAP), Saarbrucken, Germany (2016). A short summary can be found in "Topics in Invariant Theory ", S\'{e}minaire P. Dubreil/M.-P. Malliavin, Springer 
Lecture Notes in Mathematics, 1478, 244-254 (1990).\\
http://arxiv.org/abs/1207.1964  \\
\noindent
[56] Pommaret, J.-F. and Quadrat, A.: Localization and Parametrization of Linear Multidimensional Control Systems, Systems \& Control Letters, 37, 247-260 (1999).  \\
\noindent
[57] Pommaret, J.-F., Quadrat, A.: Algebraic Analysis of Linear Multidimensional Control Systems, IMA Journal of Mathematical Control and Informations, 16, 275-297 (1999). \\
\noindent 
[58] Quadrat, A., Robertz, D.: Parametrizing all solutions of uncontrollable multidimensional linear systems, Proceedings of the 16th IFAC World Congress, Prague, July 4-8, 2005.  \\  
\noindent
[59] Quadrat, A.: An Introduction to Constructive Algebraic Analysis and its Applications, 
Les cours du CIRM, Journees Nationales de Calcul Formel, 1(2), 281-471 (2010).\\
\noindent
[60] Quadrat, A., Robertz, R.: A Constructive Study of the Module Structure of Rings of Partial Differential Operators, Acta Applicandae Mathematicae, 133, 187-234 (2014). \\
http://hal-supelec.archives-ouvertes.fr/hal-00925533   \\
\noindent
[61] Rotman, J.J.: An Introduction to Homological Algebra, Pure and Applied Mathematics, Academic Press (1979).  \\
\noindent
[62] Schneiders, J.-P.: An Introduction to D-Modules, Bull. Soc. Roy. Sci. Li\`{e}ge, 63, 223-295 (1994).  \\
\noindent
[63] Spencer, D.C.: Overdetermined Systems of Partial Differential Equations, Bull. Am. Math. Soc., 75 (1965) 1-114.\\
\noindent
[64] Teodorescu, P.P.: Dynamics of Linear Elastic Bodies,  Abacus Press, Tunbridge, Wells (1975) 
(Editura Academiei, Bucuresti, Romania).\\
\noindent
[65] Vessiot, E.: Sur la Th\'{e}orie des Groupes Infinis, Ann. Ec. Norm. Sup., 20, 411-451 (1903) (Can be obtained from 
http://numdam.org).  \\
\noindent
[66] Weyl, H.: Space, Time, Matter, Springer, 1918, 1958; Dover, 1952. \\
\noindent
[67] Zerz, E.: Topics in Multidimensional Linear Systems Theory, Lecture Notes in Control and Information Sciences (LNCIS) 256, 
Springer (2000).  \\
\noindent
[68] Zou, Z., Huang, P., Zang ,Y., Li, G.: Some Researches on Gauge Theories of Gravitation, Scientia Sinica, XXII, 6, 628-636 (1979).\\

\end{document}